\documentclass[aps,prf
 reprint,
 amsmath,amssymb,
 aps
]{revtex4-2}

\usepackage{graphicx}
\usepackage{dcolumn}
\usepackage{bm}

\usepackage[utf8]{inputenc}
\usepackage[T1]{fontenc}
\usepackage{mathptmx}
\usepackage{etoolbox}

\usepackage{physics}
\usepackage{float}
\usepackage{xcolor}
\usepackage{siunitx}
\usepackage{esint}
\usepackage{sidecap}
\usepackage[normalem]{ulem}

\begin{document}


\title{  }
\title[Impact of the history force on the motion of droplets in shaken liquids
]{Impact of the history force on the motion of droplets in shaken liquids
}


\author{Frederik R. Gareis}
\author{Walter Zimmermann}
\affiliation{Theoretical Physics, University of Bayreuth, 95440 Bayreuth, Germany}

\date{\today}

\begin{abstract}
Droplets, solid particles, and gas bubbles in unsteady flows experience the Basset–Boussinesq history force (BBH) in addition to steady viscous drag, added mass, and buoyancy. Although physically relevant, the BBH term is often neglected because its inclusion is analytically and numerically demanding. To assess when this approximation fails, we revisit unsteady Stokes flows around spherical droplets of finite viscosity and derive, from first principles, the velocity fields and hydrodynamic forces, including both the classical rigid-particle limit and the free-slip (zero-viscosity) bubble limit. The resulting expressions also encompass cases with time-dependent bubble radii. We further illustrate how the BBH force arises from transient, diffusion-driven vortex structures around accelerating particles. Applying these results to droplets or particles in horizontally shaken liquids (periodically accelerated flows), we find that in the transition regime between the quasi-steady Stokes limit and the inertia-dominated regime, BBH can lead to a reduction of the droplet deflection amplitude by more than 60\% compared to predictions that neglect memory effects. We also derive a characteristic scaling of the displacement amplitude in the low-frequency limit, providing an unambiguous, experimentally verifiable signature of the BBH. For light particles and gas bubbles, the BBH contribution becomes more significant (relative to the other hydrodynamic forces) compared to that o heavy particles, such as droplets in air.
\end{abstract}

\maketitle

 
\section{Introduction}

Particle-laden flows occur in a wide range of natural and industrial contexts \cite{Sirignano:2010,LothE:2023,Michaelides:2023,Balachandar:2010.1,BrandtL:2022.1,Maxey:2017.1,Elghobashi:2019.1,Mathai:2020.1,Bergougnoux2014,Balachandar:2013.1,Daitche:2015.1,LiBraggKatul2023}. Although the dynamics of suspended particles in steady viscous flows are well established, many open questions remain regarding unsteady flows, including turbulence. Time-dependent relative motion between particles and the surrounding fluid generates not only the familiar Stokes drag and added-mass forces, but also a viscous memory contribution arising from the unsteady diffusion of vorticity. This contribution is known as the classical Basset-Boussinesq history  force (BBH) \cite{Basset1888,Boussinesq1885,Vasan:2023.1}, which we consider here in the low-Reynolds number regime.

Despite its physical relevance, for example, in cloud microphysics (droplets in air), chemical processing (bubbles in liquids), microplastics, or sediment transport (solid particles in motion), the BBH term is notoriously difficult to include because it involves a convolution integral with a singular kernel, complicating both analytical treatment and numerical implementation. This difficulty has led to its frequent omission, even in simulations of turbulent flows laden with particles \cite{Haller:2019.1,Vasan:2023.1,Mathai:2020.1}. However, experimental studies, for example those involving acoustically driven micro-bubbles \cite{Lohse:2006.1,Lohse:2009.1}, translating droplets \cite{Xu_Haitao:2022.1}, or sedimenting particles in vertically shaken containers \cite{Abbad:2004.1,Abbad:2004.2} show significantly better agreement with theoretical predictions when the BBH is included. Similarly, simulations have revealed BBH-related effects on chaotic dynamics and particle dispersion in wake flows such as Kármán vortex streets \cite{Daitche:2011.1,Bagheri:2020.1}, clustering in turbulence \cite{BrandtL:2014.1} and sedimentation rates in turbulent flows \cite{MaxeyWang:1993.1,Toschi:2017.1}. BBH is often neglected when the particles are much heavier than the fluid, as with droplets in air \cite{HillR:2005.1,MaxeyWang:1993.1,Toschi:2017.1}. However, this practice has lacked quantitative validation. This raises two pressing questions:
\begin{itemize}
    \item Under what conditions does the BBH become dynamically significant and therefore must be retained in models or simulations \cite{BrandtL:2022.1,Mathai:2020.1,Elghobashi:2019.1,Prasath:2019.1}?
    \item Can its contribution be isolated and quantified directly from experiments, without requiring detailed calculations or numerical simulations?
\end{itemize}
In this work, we address these questions by examining the motion of solid particles and droplets in a periodically accelerated liquid, in the low Reynolds number regime. Motivated by earlier studies of the relative oscillatory motion of particles and fluids \cite{Odar:1964.1,MeiR:1992.1,Brady:1993.2,MeiR:1994.1,Maxey:1994.1,Abbad:2004.1,Abbad:2004.2,Coimbra:2004.1,Lampertz:2012.1,Coimbra:2024.1,LegendreD:2023.1,Barrero-Gil:2024.1}, we focus on configurations where the relative velocity is harmonic in time. In such cases, the BBH acts continuously and in a quasi-steady fashion, enabling its influence to be probed cleanly and directly, yet with results that can be generalized to more complex flows.

The basic equation for oscillatory particle or droplet dynamics is presented in Sec.\,\ref{sec:basicdropeq}, and in Sec.\,\ref{sec:periodic_sign_change}, we illustrate how the diffusion of vortices around droplets in time-dependent flows 
causes BBH. In Sec.\,\ref{sec:Horishak}, we propose an experiment with particles in a horizontally shaken liquid, which can be used to clearly identify the effects of BBH and characterize the parameter dependencies of particle deflection relative to the shaken liquid.  
Practical considerations regarding the implementation of the proposed shaking experiment are discussed in Sec.\,\ref{sec:considerations}. Conclusions and remarks can be found in Sec.\,\ref{sec:conclusions}.

In Appendix\,\ref{sec:basicvel}, we review the basic hydrodynamic description of unsteady flows around
a spherical droplet or particle \cite{Landau6eng,batchelor1967introduction,lamb1993hydrodynamics,kim1991microhydrodynamics}
as used in the main text. In Appendix\,\ref{sec:forcedropR0}, we solve these equations and determine the flow around an arbitrarily moving spherical droplet. In addition, the forces acting on the spherical particles are determined, including 
the BBH force. This appendix also revisits early treatments \cite{gorodtsov1976slow,kim1991microhydrodynamics,yang1991note}, identifies and corrects known inconsistencies \cite{galindo1993note}, and extends the analysis to compressible gas bubbles with a time-dependent radius via a coordinate transformation proposed in \cite{Magnaudet:1998.1}, as detailed in Appendix\,\ref{sec:varying_radius}.

Together, these sections provide a comprehensive analytical overview of the forces acting on spherical droplets at small Reynolds numbers, including time-dependent and history effects relevant to the BBH formulation.

\section{Equations of motion for small spherical droplets in a shaken container \label{sec:basicdropeq}}
\begin{figure}[H]
	\centering
    \begin{minipage}{.26\textwidth}
        \includegraphics[width=.8\columnwidth]{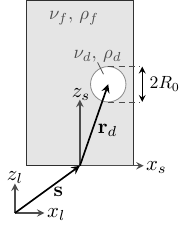}
    \end{minipage}
    \hspace{0cm}
    \begin{minipage}{.4\textwidth}
    \vspace{-.5cm}
        \caption{
        Sketch of the system. A container filled with a liquid of mass density $\rho_f$ and kinematic viscosity $\nu_f$ (gray)
        performs oscillatory motion described by the position $\vb s(t)$ of the center of its bottom. Studied is the dynamics of a small
         droplet of density $\rho_d$, kinematic viscosity $\nu_d$ and radius $R_0$ (white)  immersed in the much larger and liquid-filled container.
         The droplet position is given by $\vb r_d$ in the container-fixed coordinate system, or is
         given in the laboratory frame by $\vb r_l = \vb s + \vb r_d$.
	    }\label{fig:shaken_system}
    \end{minipage}
\end{figure}

Here we present the equations of motion for a liquid drop of mass density $\rho_d$ and kinematic viscosity $\nu_d$, which is suspended in an immiscible carrier fluid (density $\rho_f$ and viscosity $\nu_f$) contained in a shaken vessel. The dynamics of the drop are described in two coordinate systems: the laboratory system with coordinates $(x_l, y_l, z_l)$, in which the container performs oscillatory motion described by the trajectory of the center of the container bottom $\vb{s}(t)$, as shown in Fig.\,\ref{fig:shaken_system}. A second coordinate system $(x_s, y_s, z_s)$ is used, whose origin coincides with the position of the center at the bottom of the container. The axes of both systems are parallel. In the coordinate system moving with the container, the position of the center of the droplet is given by $\vb{r}_d = (x_d, y_d, z_d)$, see Fig.\,\ref{fig:shaken_system}.  
The position of the drop in the laboratory system, $\vb{r}_l$, can be described by $\vb{r}_d$ as follows:
\begin{equation}
    \vb{r}_l(t) = \vb{s}(t) + \vb{r}_d(t) \, .
    \label{eq:lab_pos}
\end{equation}
In this study, we assume small droplets so that, due to surface tension, they maintain a spherical shape of constant radius $R_0$, and we also assume large containers so that interactions between the droplets and the container walls can be neglected.

The equations of motion for the center of the droplet, either in the laboratory frame or in the moving container-fixed frame, are given by
\begin{equation}
m_d\derivative{^2 \vb{r}_l}{t^2}=\vb F -g m_d \vb e_z
\,\,\,\Longleftrightarrow\,\,\,
m_d\derivative{^2 \vb{r}_d}{t^2}=\vb F - m_d\qty[\vb{\ddot s} + g\vb e_z]
\, ,
\label{eq_eq_of_motion}
\end{equation}
where $m_d = 4\pi R_0^3 \rho_d/3$ is the mass of the droplet and $g$ is the gravitational acceleration. Central to particle dynamics are the hydrodynamic forces $\vb F$ acting on it, which also include the Basset-Boussinesq history force.
  
The theoretical understanding of the hydrodynamic forces $\vb F$ acting on particles in unsteady viscous flows is well established, but the relevant results are widely scattered in the literature \cite{gorodtsov1976slow,kim1991microhydrodynamics,yang1991note,galindo1993note,Magnaudet:1998.1,MaxeyRiley:1983.1,Basset1888,Boussinesq1885}.  
In Appendices \ref{sec:basicvel} and \ref{sec:forcedropR0}, we derive explicit and comprehensive analytical expressions for the flow around particles and droplets and the resulting hydrodynamic forces acting on them, from first principles. 
These include a summary of the analytical results for spherical droplets at low Reynolds numbers and, in preparation for future work, Appendix \ref{sec:varying_radius} also covers the case of variable particle radii such as bubbles. Consequently, in the main part of this article we focus on analyzing the dynamical equations in Eq.\,(\ref{eq_eq_of_motion}), which encompass the essence of the aforementioned fundamental hydrodynamic theory via the forces $\vb F$.

 The latter, $\vb{F} = \vb{F}_b + \vb F_a + \vb{F}_d$, consists of the buoyancy $\vb F_b$, the added mass $\vb F_a$ and the drag contribution $\vb F_d$. The buoyancy and added mass forces are given by
\begin{equation}
\label{eq:buoydrop}
	\vb F_b = m_f [\vb{\ddot s}+g\vb e_z]
    \quad \text{and} \quad 
    \vb F_a
    =
    -
    \frac{m_f}{2}\derivative{^2\vb r_d}{t^2}
 \, ,
\end{equation}
where $m_f = 4\pi R_0^3 \rho_f/3$ is the mass of the displaced carrier fluid.  The drag force acting on a spherical droplet in an unsteady viscous flow is given by (see Refs.\,\cite{galindo1993note,yang1991note},  Appendices \ref{sec:basicvel} and  \ref{sec:forcedropR0}),
\begin{align}
    \vb F_d =
    -
    6\pi R_0\mu_f\frac{2+3\kappa}{3+3\kappa}
    \derivative{\vb r_d}{t}
    +
    3 R_0\mu_f
    \int \text ds\, 
    \derivative{^2 \vb r_d (s)}{s^2}
    G(t-s,\kappa)
    \, ,
    \label{eq:Fd_formula}
\end{align}
where the first term represents the instantaneous Hadamard–Rybczynski drag on droplets, while the second term captures the unsteady contribution due to BBH. Here, $\mu_f = \rho_f \nu_f$ is the dynamic viscosity of the surrounding fluid and
\begin{align}
    \kappa = \mu_d / \mu_f
    \, 
\end{align}
denotes the viscosity ratio between the liquid in the droplet and the carrier fluid. 
The function $G(t,\kappa)$ is a memory kernel that accounts for the unsteady diffusion of vortical structures around a particle (see Sec.\,\ref{sec:periodic_sign_change}) and is defined as
\begin{align}
    G(t,\kappa )
    =
    2\,\text{Im}\qty(\int_0^\infty 
    \text d\omega
    \frac{\bar G(\omega,\kappa )}{\omega}\,
    \,
    e^{-i \omega t}
    )
    \, .
\end{align}
The Fourier transform $\bar G$ is given by
\begin{equation}
	\bar G(\omega,\kappa )
	=
	\frac{4ik_f h(ik_d)+\kappa\qty[3(1+ik_f)h(ik_d)+\chi -3ik_f(1+\kappa)\chi]}{3i(3i+k_f)h(ik_d)+3\kappa\qty[i(3i+k_f)h(ik_d)+\chi +\kappa\chi]}
	\label{eq:kernel_1}
	,
\end{equation}
with
\begin{align}
	h(x)&=(x^2+3)\tanh(x)-3x\,,
	\qquad \qquad 
	\chi=2h(ik_d)+i k_d^2(k_d-\tan k_d)
    \,,
\end{align}
and
\begin{equation}
k_d=\sqrt{i\omega \frac{R_0^2}{\nu_d} }
\,,\quad\quad
k_f=\sqrt{i\omega \frac{R_0^2}{\nu_f}}
\,.
\end{equation}
An analytical expression for $G(t,\kappa)$ can be obtained in the limiting cases $\kappa \to 0$ and $\kappa \gg 1$, corresponding to the free- and non-slip boundary conditions on the droplet surface.  
For $\kappa \gg 1$, one recovers the classical Basset–Boussinesq history kernel for a solid sphere (see Refs.\,\cite{Landau6eng,Basset1888,Boussinesq1885} and the Appendices of the present work)
\begin{align}
G_s(t)
&=
G(t,\kappa \to \infty)
=
-2\sqrt{\tau_f}
\begin{cases}
\sqrt{\pi/t }
&t>0
\\
0
&t<0
\end{cases}
\, ,
\label{eq:classic_BBH_kernel}
\end{align}
while in the limit $\kappa \to 0$, the corresponding expression for a 'bubble' with free-slip boundary conditions is (\cite{yang1991note}, App.\,\ref{sec:arb_motion_small_kappa})
\begin{equation}
G_b(t)
=
G(t,\kappa \to 0)
=
-\frac{8\pi }{3}
\begin{cases}
\exp\qty(\frac{9t}{\tau_f}) \text{Erfc}\qty(\sqrt{\frac{9t}{\tau_f}})
& t>0
\\
0
&t<0
\end{cases}
\, ,
\end{equation}
where in both cases
\begin{equation}
    \tau_f = \frac{R_0^2}{\nu_f}
    \, 
    \label{eq:tau_f_def}
\end{equation}
defines the viscous relaxation time of the carrier fluid, representing the characteristic timescale over which the flow field adjusts to a sudden change in the droplet velocity. 

Combining Eqs.\,\eqref{eq_eq_of_motion} and \eqref{eq:buoydrop}, we can write down the equation for the droplet motion,
\begin{equation}
\qty( m_d + \frac{m_f}{2} )\derivative{^2 \vb{r}_d}{t^2}
=
    \vb F_d
    + g\big(m_f-m_d\big)\vb e_z 
    + \vb{\ddot s} \big( m_f- m_d\big)
\, .
\label{eq_eq_of_motion_2}
\end{equation}
With the coordinate transformation $\vb r_l = \vb r_d + \vb s$, the corresponding expression in the laboratory frame becomes 
\begin{align*}
    m_d\derivative{^2 \vb{r}_l}{t^2}
    =
    -
    6\pi R_0\mu_f\frac{2+3\kappa}{3+3\kappa}
    \derivative{}{t}\,(\vb r_l-\vb s)
    +
    3 R_0\mu_f
    \int \text d\tau\, 
    \derivative{^2 (\vb r_l-\vb s) (\tau)}{\tau^2}
    G(t-\tau,\kappa)
    -
    \frac{m_f}{2}\derivative{^2(\vb r_l-\vb s)}{t^2}
    +
    g(m_f-m_d) \vb e_z
    +
    m_f \vb{\ddot s}
    \, .
\end{align*}
If a particle is so small that spatial variations of the flow velocity on the scale of a particle can be neglected, spatial fluctuations in the fluid velocity are irrelevant to the force balance, and this equation reduces, for $\kappa \to \infty$, to the equation for a solid particle introduced by Maxey and Riley \cite{MaxeyRiley:1983.1}.

       For  general finite values of $\kappa$, the complex frequency dependence of the memory kernel $\bar{G}(\omega,\kappa)$ prevents a closed analytical representation of $G(t,\kappa)$ or a simple formula for $\vb F_d$. Nevertheless, the kernel $G(t,\kappa)$ can still be numerically evaluated for any given $\kappa$. If the droplet undergoes a periodic harmonic motion relative to the carrier fluid at a single frequency $\omega$, the response $\bar{G}(\omega,\kappa)$ also depends solely on that frequency, allowing both $G(t,\kappa)$ and $\vb F_d$ to be determined analytically, as detailed in Sec.\,\ref{sec:periodic_sign_change}. This single-frequency simplification not only facilitates the theoretical analysis of viscous memory effects, but also provides a direct route for experimental validation through the controlled application of oscillatory forces to droplets and particles in the proposed shaken-container experiment, as described in Sec.\,\ref{sec:Horishak}.

\section{Generation of the History Force by Vortex Dynamics \label{sec:periodic_sign_change}}
Here we demonstrate that the memory contribution in Eq.\,\eqref{eq:Fd_formula} enhances the viscous drag on a droplet or particle. This enhancement arises because the vortex detachment around a droplet moves unsteadily.  

To illustrate this effect, consider a droplet undergoing harmonic motion relative to the surrounding liquid in a container at rest:
\begin{align}
    \vb r_d(t) = \vb r_l(t) = - X_0 \sin(\omega t + \varphi) \vb e_z
    \, .
    \label{eq:prescribed_harmonic_motion}
\end{align}
During such oscillatory motion, the droplet experiences a time-dependent drag force of the form (see Eq.\,\eqref{eq:harmonic_force}):  
\begin{align}
    \vb F_d =
    -
    6\pi R_0\mu_f\frac{2+3\kappa}{3+3\kappa}
    \derivative{\vb r_d}{t}
    +
    6\pi R_0\mu_f X_0 \omega 
    \Big( 
    \text{Re}\qty[\bar G(\omega,\kappa)]
    \cos(\omega t+\varphi)
    +
    \text{Im}\qty[\bar G(\omega,\kappa)]
    \sin(\omega t+\varphi)
    \Big)
    \vb e_z
    \, .
    \label{eq_harmonic_drag}
\end{align}
The first term corresponds to the instantaneous Hadamard–Rybczynski drag \cite{Leal:2007}, while the second term captures the unsteady contribution due to the BBH.  

\begin{figure}[H]
    \centering
    \begin{minipage}{.6\textwidth}
        \includegraphics[width=.87\columnwidth]{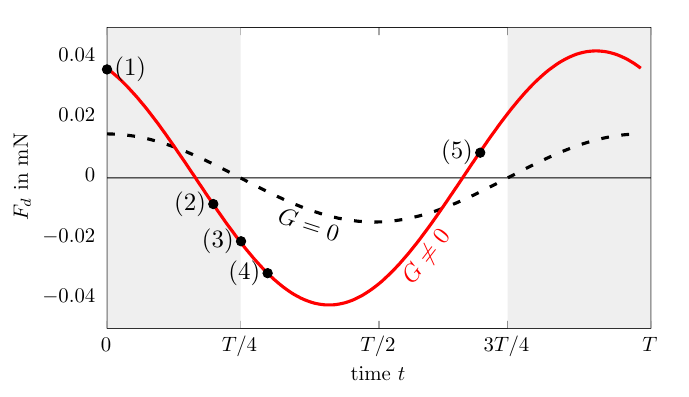}%
    \end{minipage}%
    \hspace{-.2cm}
    \begin{minipage}{.4\textwidth}%
    \vspace{-.5cm}
        \caption{
        Drag force acting on a solid particle executing oscillations of the form $\vb r_d=-X_0\sin(\omega t)\vb e_z$ with respect to a  quiescent liquid. The solid red line represents the drag force including the BBH force as calculated using Eq.\,\eqref{eq_harmonic_drag}. The dashed line shows the instantaneous Hadamard-Rybczynski viscous drag alone. The black dots at (1)-(5) indicate the times at which the flow profiles around a particle are shown in Fig.\,\ref{fig:flow_field_illustration}. Other parameters include: $\omega=4\si{\radian/\second},\, R_0=1\si{\milli\meter},\,X_0=0.4\si{\meter},\,\mu_f=0.001\si{\pascal\second},\,\mu_d=0.01\si{\pascal\second},\,\rho_f=\rho_d=1000\si{\kilo\gram\per\cubic\meter}$.
        }
        \label{fig:illustration_force}
    \end{minipage}
\end{figure}

In Fig.\,\ref{fig:illustration_force}, we compare the total viscous drag $F_d$ [solid red line, Eq.\,\eqref{eq_harmonic_drag}] with the instantaneous drag obtained by setting $G=0$ (black dashed line). The total drag exhibits a larger amplitude and a pronounced phase lead, showing that the BBH force enhances the viscous drag and can dominate in certain parameter ranges.

\begin{figure}[htb!]
    \centering%
    \begin{minipage}{.325\textwidth}%
        \includegraphics[width=\columnwidth]{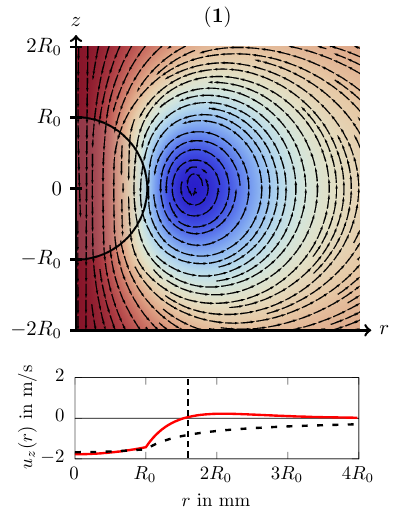}%
    \end{minipage}%
    \begin{minipage}{.325\textwidth}%
        \includegraphics[width=\columnwidth]{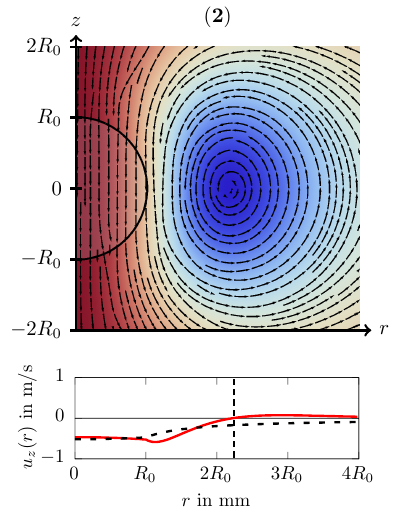}%
    \end{minipage}%
    \begin{minipage}{.325\textwidth}%
        \includegraphics[width=\columnwidth]{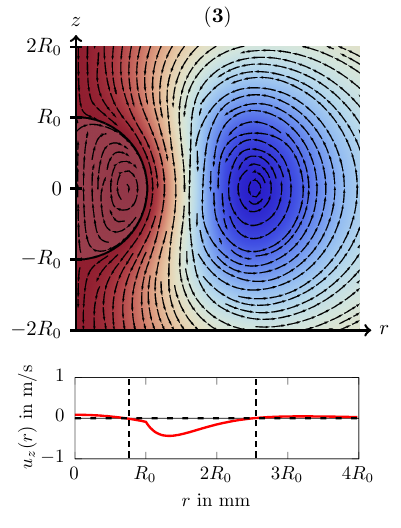}%
    \end{minipage}%
    \\%
    \begin{minipage}{.325\textwidth}%
        \includegraphics[width=\columnwidth]{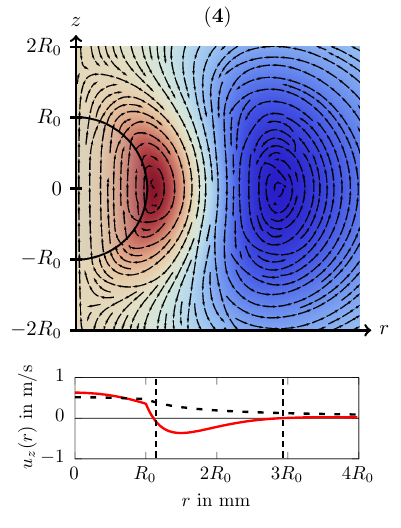}%
    \end{minipage}%
    \begin{minipage}{.325\textwidth}%
        \includegraphics[width=\columnwidth]{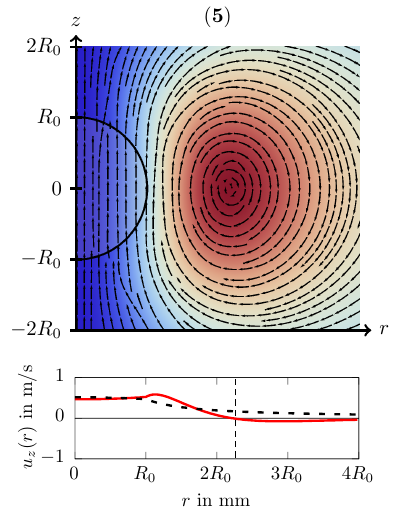}%
    \end{minipage}%
    \begin{minipage}{.325\textwidth}%
        \vspace{-.35cm}
        \caption{
        Snapshots of the fluid flow in the laboratory system (arrows), and qualitative heat maps of the stream function, around a solid particle at different phases of the oscillation cycle in Fig.\,\ref{fig:illustration_force} at times (1)\,-\,(5). The upper panels show the velocity field in a radial cross-section, while the lower panels display the unsteady profile of the vertical velocity component $u_z(r)$ in the equatorial plane (solid lines), compared with the corresponding steady velocity (dashed lines). The quasi-steady \textit{Stokes} profile assumes instantaneous flow relaxation, valid only for $\delta \ll 1$. The unsteady profiles illustrate how a finite $\delta$ causes vortices, increasing the viscous drag force. The parameters were chosen as in Fig.\,\ref{fig:illustration_force}. Times: $t=0$ in (1), $t=0.2T$ in (2), $t=0.25T$ in (3) $t=0.3T$ in (4) and $t=0.7T$ in (5).
        }
    \label{fig:flow_field_illustration}%
    \end{minipage}%
\end{figure}

To understand the physical origin of this frequency-dependent contribution, it is instructive to examine the flow fields near the droplet at times (1)\,-\,(5) marked in Fig.\,\ref{fig:illustration_force}. These flow fields are given analytically in Appendix\,\ref{sec:flow_harm_osz} and are illustrated in Fig.\,\ref{fig:flow_field_illustration}.   
%
For a droplet moving uniformly through a viscous fluid, the resulting 'Stokes flow' is stationary: the vertical fluid velocity $u_z(r)$ in the equatorial plane decreases monotonically with distance $r$, as indicated by the dashed profiles in the lower panels of Fig.\,\ref{fig:flow_field_illustration}. If the velocity of the droplets changes with time, the fluid near the particle responds almost instantaneously, while the fluid farther away adjusts more gradually. The timescale of this adaptation is determined by the viscous relaxation time $\tau_f$.  

For harmonic oscillations, the extent of this velocity adaptation is characterized by the dimensionless frequency
\begin{equation}
    \delta = \omega \tau_f = \frac{\omega R_0^2}{\nu_f}
    \, .
    \label{eq:delta_def}
\end{equation}
As $\delta$ increases, the particle changes its direction of motion faster than the carrier fluid can relax after a sudden change in velocity. Consequently, farther from the particle, the vertical velocity component has not yet adjusted to the new particle motion. This mismatch causes the fluid velocity to exhibit periodic reversals with distance, giving rise to spatial structures of counter-rotating vortices, as illustrated in Fig.\,\ref{fig:flow_field_illustration}. Such vortices have recently been observed experimentally \cite{Jaroslawski2025_StokesianSettling} and in numerical simulations \cite{LegendreD:2023.1}.

In panel (3) of Fig.\,\ref{fig:flow_field_illustration}, the particle has just reversed its velocity. However, the tangential shear of the fluid velocity at its surface, and hence the drag force $F_d$, has already changed sign between times (1) and (2). By contrast, for the case without memory effects ($\delta \ll 1$), both the gradient and the instantaneous drag change sign simultaneously 
with the velocity of the particles.  

As a result, the extrema of $F_d(t)$ occur earlier than those of instantaneous drag, as seen in Fig.\,\ref{fig:illustration_force}.   
Hence, the observed phase shift and the increased magnitude of the drag are direct manifestations of the BBH force, which encapsulates the diffusive vortex dynamics of unsteady Stokes flows. This effect is further investigated for a particle in a horizontally shaken container in the next section.

\section{Effects of the History force on the droplet response in horizontally shaken liquids \label{sec:Horishak}}
  
\begin{figure}[H]
	\centering
    \begin{minipage}{.44\textwidth}
        \includegraphics[width=.93\columnwidth]{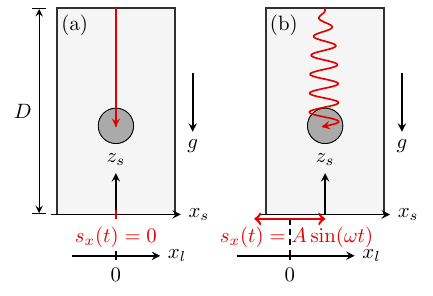}
    \end{minipage}
    \hspace{.1cm}
    \begin{minipage}{.54\textwidth}
        \vspace{-.4cm}
        \caption{
        Sketch of an experimental setup to study the effects of the added mass and the BBH force on a droplet (gray) in a horizontally shaken container filled with an immiscible carrier liquid. Part (a) shows a spherical droplet sedimenting along the red arrow in a stationary liquid container of height $D$. Part (b) depicts the container undergoing horizontal harmonic shaking with amplitude $A$ and frequency $\omega$. The periodic horizontal motion of the liquid induces a corresponding periodic acceleration of the droplet via viscous and buoyancy forces, deflecting it from its straight sedimentation path. The dependence of the droplet deflection on shaking amplitude and frequency provides insight into the forces acting on the particle in unsteady flows. The position of the center of the shaken container in the laboratory frame is indicated by the red arrow at the bottom of part (b), $s_x(t) = A \sin(\omega t)$.
	    }\label{fig:coordinates}
    \end{minipage}
\end{figure}

In this section, we analyze how the deflection of droplets and solid particles depends on the parameters of a horizontally shaken liquid. We identify parameter regimes in which BBH effects are significant and cannot be neglected, and we derive a characteristic power law that uniquely reflects these effects and can be directly tested experimentally. Fig.\,\ref{fig:coordinates} illustrates an experimental setup suitable for verifying the predicted behavior.

The center of the bottom of the container in Fig.\,\ref{fig:coordinates}, denoted by $\vb s(t)$, undergoes a sinusoidal horizontal motion,
\begin{equation}
    \vb s (t) = A \sin(\omega t)\vb e_x
    \, ,
\label{eq:shaking_motion}
\end{equation}
where $A$ and $\omega$ are the shaking amplitude and the angular frequency, respectively. The corresponding container acceleration is
\begin{equation}
\vb {\ddot s}= -A\omega^2 \sin(\omega t)\vb e_x
\,,
\label{eq:acceleration_shaking_experiment}
\end{equation}
such that Eq.\,\eqref{eq_eq_of_motion_2} takes the form
\begin{equation}
\qty( \frac{1}{2} + \frac{\rho_d}{\rho_f} )\derivative{^2 \vb{r}_d}{t^2}
=
    \frac{\vb F_d}{m_f}
    + g\qty(1-\frac{\rho_d}{\rho_f} )\vb e_z 
    -A\omega^2 \sin(\omega t) \qty( 1 - \frac{\rho_d}{\rho_f} ) \vb e_x
\, .
\label{eq_eq_of_motion_3}
\end{equation}
During sedimentation or rise, horizontal shaking of the carrier fluid induces a lateral oscillatory motion of the particle whenever $\rho_d \neq \rho_f$.

The horizontal particle response depends on the viscosity ratio $\kappa$, the density ratio $\rho_d/\rho_f$, and the dimensionless frequency parameter $\delta$, which compares the oscillation period $1/\omega$ with the viscous relaxation time $\tau_f$ of the surrounding fluid, see Eq.\,\eqref{eq:delta_def}. By systematically varying these parameters, the proposed experiment provides direct access to the frequency-dependent effects of the BBH force. In particular, this approach enables an experimental determination of the response kernel $\bar G(\omega,\kappa)$.

In the following, we derive theoretical predictions for the amplitude and phase of the horizontal droplet motion, which can be easily tested in experiments.

\subsection{Sedimentation and rise of droplets without container shaking}

The vertical motion of a particle, $z_d(t)$, reaches its terminal settling or rising velocity on a short time scale of order $\sim\tau_p$. The Basset–Boussinesq history force contributes only during this initial transient, as recently investigated in Ref.\,\cite{Jaroslawski2025_StokesianSettling}, and becomes negligible once steady motion is established.

Consequently, beyond this transient we approximate the vertical component of the drag force by retaining only the instantaneous Hadamard–Rybczynski contribution and neglecting both memory effects and particle acceleration in Eq.\,\eqref{eq_eq_of_motion_3}. This yields the steady-state sedimentation (or rise) velocity,
\begin{equation}
 \frac{1}{\tau_p}
    \derivative{z_d}{t}
=
g\qty(\frac{\rho_f}{\rho_d}-1 ) 
\, ,
\label{eq:basic_shake_z}
\end{equation}
with the abbreviations 
\begin{equation}
    \tau_p = \frac{2}{9}\frac{R_0^2}{\nu_f }\frac{\rho_d}{\rho_f}M
    \,,\qquad
    M=\frac{3+3\kappa}{2+3\kappa}
    \, .
\end{equation}
The sedimentation velocity $\dot z_d$ vanishes for $\rho_d = \rho_f$, is directed upward for $\rho_d < \rho_f$, and downward for $\rho_d > \rho_f$. In the limit of very small Reynolds numbers, this stationary vertical motion does not couple to the horizontal droplet dynamics induced by container shaking. This separation of vertical and horizontal motion is assumed throughout the remainder of this section.

\subsection {Horizontal droplet dynamics in a shaken container}
\begin{SCfigure}[][htb]
	\centering
	\includegraphics[width=.67\textwidth]{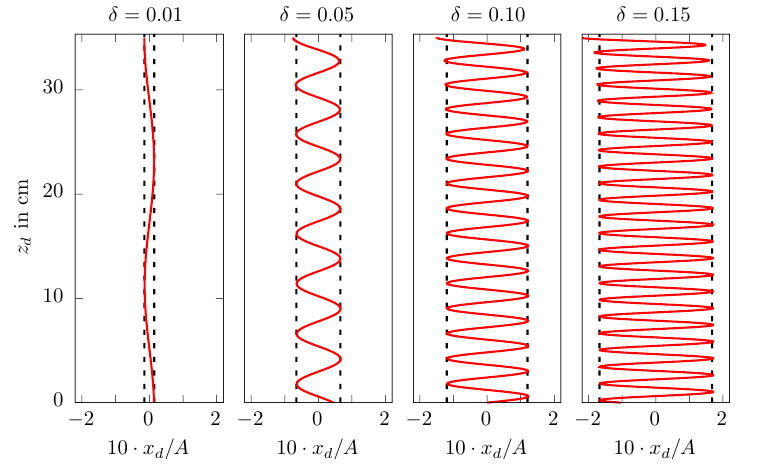}
    \caption{
    Shown are the vertical coordinate $z_d(t)$ and the enlarged relative horizontal motion $10\, x_d(t)/A$ of numerically simulated sedimentation trajectories of rigid particles in a horizontally shaken container for four different dimensionless shaking frequencies: $\delta= 0.01$, 
    $\delta= 0.05$,
    $\delta= 0.10$,
    and $\delta= 0.15$,  
    whereby the Reynolds number is always small, $\text{Re} \lesssim 1$.
    After a short settling time, the trajectories show sinusoidal behavior of $x_d(t)$ with a constant oscillation amplitude $X_0$
    ($\propto A$) indicated by the dashed line, where the ratio  $X_0/A$ is given analytically by Eq.\,\eqref{eq:Sigma2} below.
    Further parameters: $\rho_d/\rho_f=8$, $\rho_f=1000 \si{\kilo\gram\per\cubic\meter}$, $\nu_f=0.001\si{\square\meter\per\second}$, $g=9.81\si{\meter\per\square\second}$ and $R_0=2.25\si{\milli\meter}.$
    \vspace{.7cm}
	}
	\label{fig:traject_difffreq}
\end{SCfigure}

The horizontal dynamics of a droplet in a horizontally shaken container, described by the coordinate $x_d(t)$, is governed by Eqs.\,\eqref{eq:Fd_formula} and \eqref{eq_eq_of_motion_3}, which we combine to obtain
\begin{equation}
    \qty( \frac{m_f}{2} + m_d )\derivative{^2 x_d}{t^2}
    =
    -
    6\pi R_0\mu_f\frac{2+3\kappa}{3+3\kappa}
    \derivative{x_d}{t}
    +
    3 R_0\mu_f
    \int \text ds\, 
    \derivative{^2 x_d (s)}{s^2}
    G(t-s,\kappa)
    -A\omega^2 \sin(\omega t) \qty( m_f - m_d )  
\, .
\label{eq:basic_shake_x}
\end{equation}
Fig.\,\ref{fig:traject_difffreq} shows exemplary sedimentation trajectories $(x_d(t), z_d(t))$ obtained by numerical integration of Eq.\,\eqref{eq:basic_shake_z} and Eq.\,\eqref{eq:basic_shake_x} for a rigid particle, i.e., within the limit $\kappa \gg 1$, using the solid-sphere kernel $G_s(t)$ from Eq.\,\eqref{eq:classic_BBH_kernel}. The results are shown for several dimensionless shaking frequencies $\delta$ with otherwise identical parameters.
After a brief transient, the horizontal motion of the particle settles into sinusoidal oscillations with the shaking frequency $\omega$ and constant amplitude $X_0$:
\begin{equation}
    x_d(t) = - X_0 \sin(\omega t + \varphi)
    \, .
    \label{eq:x_dansatz}
\end{equation}
The amplitude calculated from this asymptotic behavior is indicated by the dashed lines in Fig.\,\ref{fig:traject_difffreq}. The oscillation amplitude increases systematically with $\delta$ and depends characteristically on viscous memory effects at higher frequencies.

For harmonic motion, the memory integral in Eq.\,\eqref{eq:basic_shake_x} reduces to a single-frequency contribution and can be expressed in terms of the response kernel $\bar G(\omega,\kappa)$, as in Eq.\,\eqref{eq_harmonic_drag}. Substituting the ansatz \eqref{eq:x_dansatz} into Eq.\,\eqref{eq:basic_shake_x} then yields the following equation for the horizontal droplet motion:
\begin{align}
\qty( \frac{m_f}{2} + m_d )\derivative{^2 x_d}{t^2}
 &=
 -(m_f-m_d) A \omega^2 \sin(\omega t)
- 6\pi R_0 \mu_f \frac{2+3\kappa}{3+3\kappa} \derivative{x_d}{t}
 \nonumber\\
 &\quad 
 +
 6\omega X_0 \pi R_0 \mu_f \Big(\text{Re}\qty[\bar G(\omega,\kappa)]\cos(\omega t+\varphi) 
    +  \text{Im}\qty[\bar G(\omega,\kappa)]\sin(\omega t+\varphi)
    \Big)
 \label{eq:basic_shake_x2}
 \,.
\end{align}
Using standard trigonometric identities to expand $\sin(\omega t+\varphi)$ and $\cos(\omega t+\varphi)$, and substituting the ansatz \eqref{eq:x_dansatz} into Eq.\,\eqref{eq:basic_shake_x2}, one can collect the linearly independent terms proportional to $\sin(\omega t)$ and $\cos(\omega t)$. Requiring each of these contributions to vanish separately yields the following system of algebraic equations for the oscillation amplitude $X_0$ and phase $\varphi$:
\begin{subequations}
\label{eq:coeff_eq}
	\begin{align}
	0
	&=
    c_1 \cos(\varphi)
	+ \frac{c_2}{\kappa + 1}\sin(\varphi)-\frac{2 A\omega (m_d - m_f)}{X_0}
	\,,
	\\
	0
	&=
	c_1\sin(\varphi) 
	-
    \frac{c_2}{\kappa + 1}\cos(\varphi)\,,
	\end{align}
	\end{subequations}
	with the abbreviations
	\begin{subequations}
    \label{eq:c_abbreviations}
	\begin{align}
	c_1 
	&=
	\omega (2m_d + m_f) - 12\pi \text{Im}\left[\bar{G}(\omega,\kappa)\right] \mu_f R_0\,,
    \label{eq:c_1_abbreviation}
	\\
	c_2
	&=
	4\pi \mu_f R_0 \left(2 + 3\kappa + 3(\kappa + 1)\text{Re}\left[\bar{G}(\omega,\kappa)\right]\right)
	.
    \label{eq:c_2_abbreviation}
	\end{align}	
\end{subequations}
The analytical solution of Eqs.\,\eqref{eq:coeff_eq} reads 
\begin{subequations}
\label{eq:ampl_phi_gen}
\begin{align}
\label{eq:ampl_phi_gen_X}
X_0 
&=
\,
2A\omega  \,\frac{(m_d - m_f)(1 + \kappa)}{\sqrt{c_2^2 + c_1^2(1 + \kappa)^2}}
\,,
\\
\cot \varphi &=
\frac{c_1 (1+\kappa)}{c_2}
\,.
\label{eq:ampl_phi_gen_phi}
\end{align}
\end{subequations}
For equal mass densities, $\rho_d=\rho_f$, the droplet is advected synchronously with the carrier liquid and therefore remains in its initial horizontal position in the frame fixed to the container, giving $X_0=0$. For $\rho_d\neq\rho_f$, the interplay of inertia and hydrodynamic forces gives rise to a characteristic frequency-dependent response described by the amplitude $X_0(\omega)$ and the phase $\varphi(\omega)$.

The validity of the analytical results is demonstrated in Fig.\,\ref{fig:traject_difffreq}. The dashed lines show the oscillation amplitudes $X_0$ predicted by Eq.\,\eqref{eq:ampl_phi_gen_X}, which are in excellent agreement with the amplitudes obtained from direct numerical integration of Eqs.\,\eqref{eq:basic_shake_z} and \eqref{eq:basic_shake_x}. This quantitative agreement confirms the harmonic ansatz \eqref{eq:x_dansatz} as an accurate description of horizontal dynamics.

Beyond validating the theory, the harmonic response also provides a direct route to reconstruct the frequency-dependent kernel of the history force, $\bar G(\omega,\kappa)$, for arbitrary viscosity ratios $\kappa$. By experimental measurement of the response functions $X_0(\omega)$ and $\varphi(\omega)$, the real and imaginary parts of the kernel, $\text{Re}\!\left[\bar{G}(\omega,\kappa)\right]$ and $\text{Im}\!\left[\bar{G}(\omega,\kappa)\right]$, can be determined explicitly. Solving Eqs.\,\eqref{eq:c_abbreviations} and \eqref{eq:ampl_phi_gen} for $\bar G(\omega,\kappa)$ yields
\begin{align}
    \text{Re}\left[\bar{G}(\omega,\kappa)\right]
    &=
    -\frac{1}{6}
    \qty(\frac{4+6\kappa}{1+\kappa} - \frac{A\omega(m_d-m_f)}{\pi R_0X_0\mu_f}\sin\varphi)\,,
    \\
    \text{Im}\left[\bar{G}(\omega,\kappa)\right]
    &=
    \frac{\omega}{12\pi R_0X_0\mu_f}
    \Big(
    (2m_d+m_f)X_0-2A(m_d-m_f)\cos\varphi
    \Big)
    \, .
\end{align}
In Sec.\,\ref{sec:rigidpart} we focus on the effects of added mass and the BBH force on the deflection amplitude $X_0$ and phase $\varphi$ as functions of the driving frequency $\omega$ for the limiting case $\kappa \to \infty$ of a rigid particle, generating experimentally testable predictions. In Sec.\,\ref{sec:dropletpart}, we show that the response of droplets with finite viscosity ratios $\kappa$ is qualitatively similar to that of solid particles.

\subsection{Response of a solid spherical particle \label{sec:rigidpart}}
%
A particularly relevant limit is that of a solid particle subject to a no-slip boundary condition at its surface. This situation is obtained from the droplet theory in the regime $\kappa \gg 1$. In this limit, the kernel reduces to $\bar G(\omega,\kappa) = -i\sqrt{i\omega \tau_f}$, and the relative displacement amplitude $X_0/A$ from Eq.\,\eqref{eq:ampl_phi_gen_X} becomes
\begin{align}
\label{eq:Sigma2}
\frac{X_0}{A} &= 
 \frac{4\delta}{9}
 \frac{\rho_d/\rho_f-1}
{\sqrt{\left(2+H\sqrt{2\delta}\right)^2+\left[ H\sqrt{2\delta} +\frac{2}{9} \delta \left(\alpha+2\frac{\rho_d}{\rho_f}\right)\right]^2}}
\,,
\end{align}
while the phase from Eq.\,\eqref{eq:ampl_phi_gen_phi} reads
\begin{align}
    \cot\varphi
&=
\frac{H\sqrt{2\delta}+2\delta (\alpha+2\rho_d/\rho_f)/9}{2+H\sqrt{2\delta}}
\label{eq:cot_phi_rigid}
\, .
\end{align}
These expressions involve the parameters $\alpha$ and $H$, which allow one to include distinct physical contributions selectively. The choice $\alpha = 1$ accounts for the added mass of the displaced fluid, while $\alpha = 0$ suppresses it. Similarly, $H = 1$ includes the BBH, and $H = 0$ omits it. This flexibility makes it possible to isolate the dynamical impact of either effect. Note, the relative displacement
$X_0/A$ and the phase shift $\varphi$ depend only on the dimensionless frequency $\delta$ and the ratio between the mass densities $\rho_d/\rho_f$.

\subsubsection{Discussion of the displacement amplitude of particles}

\begin{SCfigure}[][htb]
    \centering
    \includegraphics[width=8.25cm]{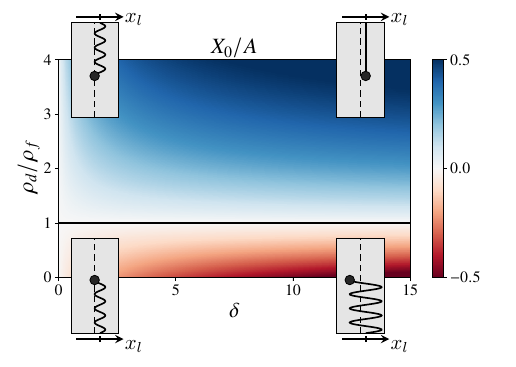}
    \caption{ Heat map of the amplitude ratio $X_0/A$ predicted by Eq.\,\eqref{eq:Sigma2} as a function of the density ratio $\rho_d/\rho_f$ and the frequency $\delta$. The horizontal black line separates heavy (sedimenting) particles from light (rising) ones. The corner sketches qualitatively illustrate the expected laboratory-frame dynamics in the limits $\delta\ll 1$ (left panels), $\delta\gg 1$ with $\rho_d/\rho_f\ll 1$ (bottom right), and $\delta\gg 1$ with $\rho_d/\rho_f\gg 1$ (top right). These cases are discussed in the main text. \vspace{1.75cm}}
    \label{fig:X_0_heatmap}
\end{SCfigure}

The amplitude of the relative deflection with respect to the shaken liquid, given by Eq.\,\eqref{eq:Sigma2}, depends only on the two parameters $\delta$ and $\rho_f/\rho_d$ and therefore can also be clearly represented graphically, 
as shown in Fig.\,\ref{fig:X_0_heatmap}.
The black line in Fig.\,\ref{fig:X_0_heatmap} shows that  $X_0/A$ changes sign with $(\rho_d/\rho_f - 1)$, as indicated by Eq.\,\eqref{eq:Sigma2}. This reflects the fact that the buoyancy force reverses when the particle becomes lighter than the displaced fluid, or vise versa.  The functional dependence of $X_0/A$ on $\delta$ is similar for all density ratios: $X_0/A$ disappears for $\delta\ll 1$ and reaches a plateau value depending on the density ratio for $\delta\gg 1$,  as can be seen in Eq.\,\eqref{eq:plateau_delta}. These two limiting ranges of $X_0/A$ are discussed in more detail.

(1) Low-frequency limit $\delta \ll 1$. This corresponds to particles with $R_0 \ll \sqrt{\nu_f/\omega}$, for which the vorticity generated at the surface diffuses away rapidly compared to one oscillation period, and the surrounding flow relaxes quasi-steadily to the Stokes profile.  
In this limit, the dominant hydrodynamic contribution in Eq.\,\eqref{eq:basic_shake_x2} is the instantaneous Stokes drag, which scales linearly with $R_0$. For solid particles ($\kappa\gg 1$), the contribution of BBH scales as $R_0^2$ (see Eq.\,\eqref{eq:Fd_formula} 
and Eq.\,\eqref{eq:classic_BBH_kernel}) which is subdominant. This results in $X_0/A = 0$, i.e., the particle does not undergo displacement relative to the locally oscillating fluid and strictly follows the motion of the container. This regime is central in a number of physical contexts, including the dynamics of small droplets in turbulent clouds \cite{pumir_wilkinson,bec_gustavsson_mehlig}. It is common (and often consistent) in this regime to retain only the Stokes drag and neglect the contribution of the BBH, especially since for heavy particles ($\rho_d/\rho_f\gg 1$), additional inertial terms proportional to $\ddot{\vb s}$ are also subdominant. In clouds, the potential relevance of the BBH force and added mass in strongly intermittent regions of high acceleration is not fully understood \cite{Shaw2003}.

(2) High-frequency limit $\delta \gg 1$. 
In this regime, the relative displacement approaches
\begin{align}
\label{eq:plateau_delta}
\frac{X_0}{A} (\delta \gg 1)
\;\to\;
\frac{2}{\alpha + 2\rho_d/\rho_f}
\Bigl(\rho_d/\rho_f - 1\Bigr)
\, ,
\end{align}
and the phase lag tends to zero $\varphi \to 0$, i.e., the particle oscillates in the laboratory system in anti-phase to the container.
This plateau is governed by a balance between inertia and buoyancy and depends only on the density ratio. Consequently, 
the dependence of the displacement amplitude on the driving frequency, viscosity, and the contribution of BBH becomes negligible for large $\delta$. The added mass ($\alpha=1$) becomes especially relevant for light particles with $\rho_d/\rho_f\ll 1$, for which the inertial mass is dominated by the displaced fluid. In the limit $\rho_d/\rho_f\to 0$, Eq.\,\eqref{eq:plateau_delta} gives $X_0\to -2A$. 
In this case, the motion of the particle in the laboratory-frame at large $\delta$ is given via Eqs.\,\eqref{eq:lab_pos} and \eqref{eq:x_dansatz} by $x_l(t) = 3A\,\sin(\omega t)$ (bottom right panel of Fig.\,\ref{fig:X_0_heatmap}).

For heavy particles with $\rho_d/\rho_f\gg 1$, such as iron particles in water ($\rho_d/\rho_f\sim 8$) or water droplets in the air, the plateau approaches asymptotically $X_0/A \to 1$.  
The trajectory in the laboratory frame then tends to $x_l(t)=0$ for $\delta\gg 1$, meaning that the particle is essentially unaffected by the container oscillation (top right panel of Fig.\,\ref{fig:X_0_heatmap}).

\subsubsection{Relative importance of the history force}

Stokes drag represents a quasi-steady response and therefore cannot capture the viscous drag accurately as the forcing frequency increases. As discussed above, the leading-order correction in $\delta$ is provided by BBH, which becomes significant before inertial effects eventually dominate at very large $\delta$.

\begin{SCfigure}[][htb]
	\centering
    \begin{minipage}{.5\textwidth}%
        \includegraphics[width=.9\columnwidth]{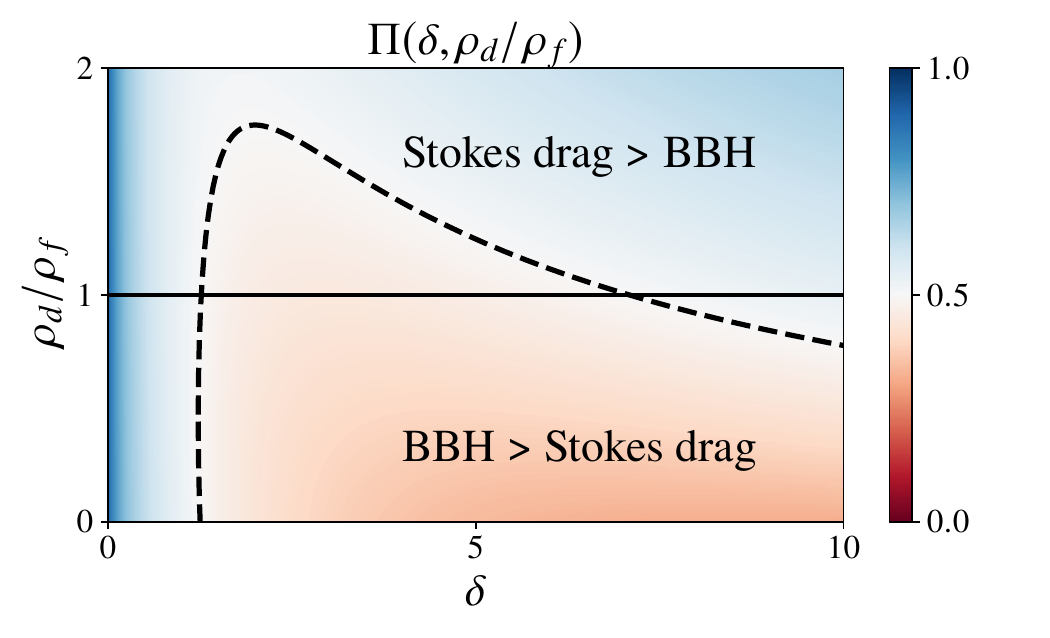}
    \end{minipage}
    \caption{
Heat map of the ratio between the particle displacement amplitude with and without BBH, $\Pi(\delta,\rho_d/\rho_f)$, quantifying the reduction of the displacement amplitude due to the BBH contribution. Red regions indicate strong suppression ($\Pi<0.5$), whereas blue regions correspond to weaker BBH effects. The axes show the dimensionless frequency $\delta$ and the density ratio $\rho_d/\rho_f$. The dashed curve marks $\Pi=0.5$ and thus separates the regime dominated by the history force from the regime in which Stokes drag provides the leading contribution. The visualization highlights that the BBH force is most relevant for harmonically forced solid particles with $\rho_d/\rho_f<1$ and intermediate $\delta$. For small $\delta$, $\Pi\approx 1$ as Stokes drag dominates, while at large $\delta$ or large density ratios the particle dynamics becomes inertia-dominated, again yielding $\Pi\approx 1$ because drag becomes irrelevant. \vspace{0cm}
}
	\label{fig:elongation_ratio_density}
\end{SCfigure}

To quantify the relevance of the history force at intermediate values of $\delta$, we consider the reduction of the displacement amplitude $X_0$ when the BBH term is included. This is characterized by the ratio
\begin{align}
\label{eq:ratio_RH}
\Pi(\delta, \rho_d/\rho_f)&=\frac{X_0(H=1,\delta, \rho_d/\rho_f)}{X_0(H=0,\delta, \rho_d/\rho_f)}
\,,
\end{align}
where $H=1$ corresponds to the case with BBH and $H=0$ excludes the contribution of BBH. The heat map in Fig.\,\ref{fig:elongation_ratio_density} shows $\Pi(\delta, \rho_d/\rho_f)$ as a function of the dimensionless frequency and density ratio. The red regions indicate a strong suppression of the particle displacement amplitude due to the contribution of BBH  ($\Pi<0.5$), while the blue regions correspond to weaker effects ($\Pi>0.5$).

The heat map demonstrates that the relative influence of the history force decreases as $\rho_d/\rho_f$ increases. This trend reflects the increasing importance of particle inertia and buoyancy compared to viscous drag. For small density ratios, by contrast, the history force becomes increasingly dominant: for example  for $\rho_d/\rho_f \sim 0.5$, the BBH contribution 
reduces the displacement amplitude by more than $60\%$, corresponding to the red regions in Fig.\,\ref{fig:elongation_ratio_density}.

Moreover, the effect of the BBH force is most pronounced at intermediate values of $\delta$, consistent with the behavior seen
in Ref.\,\cite{Coimbra:2001.1}. In this regime, the contribution of BBH substantially decreases the amplitude $X_0$, indicating that the particle tends to follow the fluid motion more closely. Physically, this results from the fact that the history force penalizes rapid accelerations of the particle relative to the surrounding fluid.

\subsubsection{Unique (measurable) properties of BBH  at small $\delta$}

Beyond its quantitative impact on amplitude, the BBH force also modifies the characteristic behavior of the response $X_0/A$, especially in the 
small-$\delta$ regime, where the viscous effects are strongest. A particularly sensitive diagnostic is the quantity $X_0/(A\delta)$. In Fig.\,\ref{fig:Ampli_small_delt}, this dependence is shown with ($H=1$) and without ($H=0$) the history force. Figure\,\ref{fig:Ampli_small_delt}(a) displays the behavior throughout the entire range of $\delta$, while Fig.\,\label{fig:Ampli_small_delt}(b) magnifies the small$-\delta$ region.
  \begin{figure}[H]	
      \begin{minipage}{.5\textwidth}
          \includegraphics[width=\columnwidth]{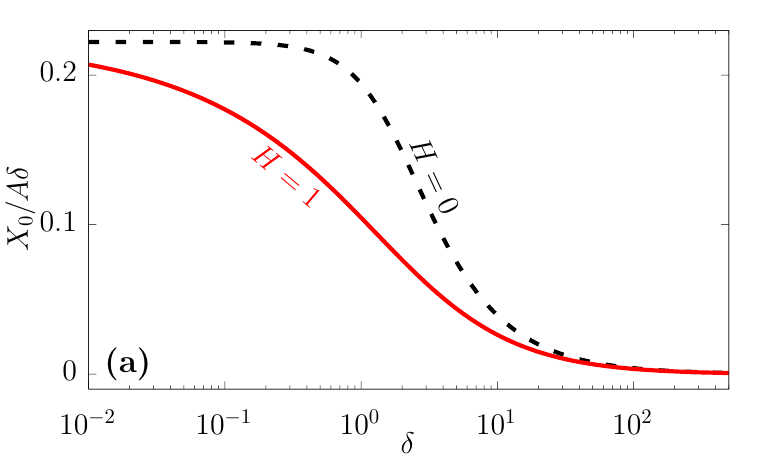}
      \end{minipage}%
      \begin{minipage}{.5\textwidth}%
        \vspace{-.1cm} 
          \includegraphics[width=\columnwidth]{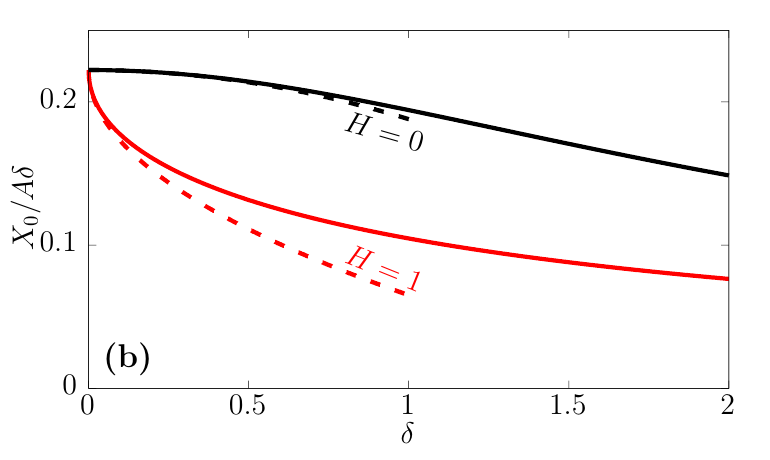}
      \end{minipage}
      \caption{
    $\delta$-dependence of $X_0/(A\delta)$ with ($H=1$, red) and without ($H=0$, black) the BBH force.  
    Panel (a) shows the overall suppression of the amplitude by the BBH force across a wide frequency range.  
    Panel (b) focuses on the small-$\delta$ regime, comparing the asymptotic results in Eqs.\,\eqref{eq:X0whist} and \eqref{eq:X0ohist} (dashed) with the full expression in Eq.\,\eqref{eq:Sigma2} (solid).  
    Here $\rho_d/\rho_f=2$.
    }
    \label{fig:Ampli_small_delt}
\end{figure}
Using Eq.\,\eqref{eq:Sigma2}, we expand $X_0/(A\delta)$ for $\delta\ll 1$ with and without the BBH.  
When the history force is included ($H=1$), the expansion yields
\begin{align}
\label{eq:X0whist}
\frac{X_0}{ A \delta}  \sim  \frac{2}{9} \left(\frac{\rho_d}{\rho_f}-1\right)
 \left(1-H\sqrt{\frac{\delta}{2}}\right)
+ O(\delta)\,,
\end{align}
demonstrating that the departure from the leading-order constant scales as $\delta^{1/2}$.  
In contrast, neglecting the BBH ($H=0$) gives
\begin{align}
\label{eq:X0ohist}
 \frac{X_0}{ A \delta} \sim  \frac{2}{9} \left(\frac{\rho_d}{\rho_f}-1\right)
 \left(1- \frac{\delta^2}{2}\, \qty[\frac{\alpha+2\rho_d/\rho_f}{9}]^2\right)+ O(\delta^4 )
 \, ,
\end{align}
where the deviation now scales as $\delta^{2}$.  
These distinct power-law behaviors lead to opposite curvature signs in Fig.\,\ref{fig:Ampli_small_delt}(b), and thus constitute a clear experimental signature of the BBH force.  
Because inertial effects become negligible as $\delta\to 0$, the observed scaling $X_0/(A\delta)\propto\delta^{1/2}$ is directly attributable to the BBH contribution.  
This provides an experimentally accessible means to identify the history force without recourse to numerical simulations in an experiment as proposed here or in Ref.\,\cite{Coimbra:2004.1}.

The structure of these power laws remains unchanged when the mass ratio crosses $\rho_d/\rho_f=1$.  
However, they cannot be well measured for $\rho_d/\rho_f\simeq 1$, where $X_0$ vanishes, nor for $\rho_d\gg\rho_f$, where $X_0\simeq A$ becomes basically independent of $\delta$.

\begin{figure}[H]
	\centering
    \begin{minipage}{.47\textwidth}
            \includegraphics[width=\columnwidth]{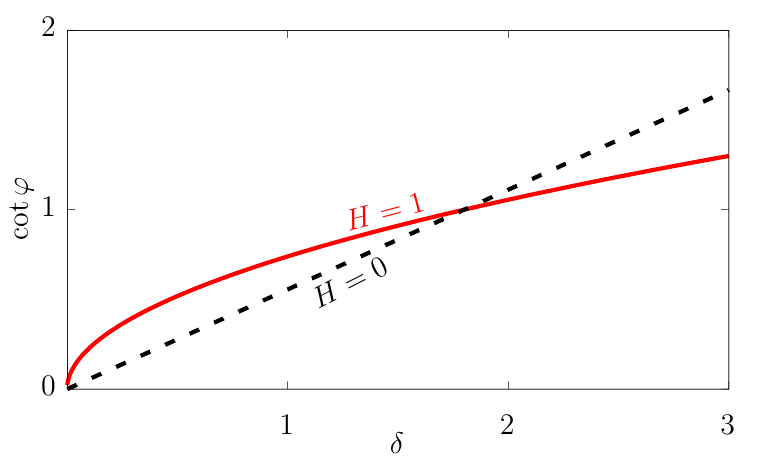}
    \end{minipage}
    \begin{minipage}{.4\textwidth}
            \caption{
        $\delta$-dependence of $\cot(\varphi)$ from Eq.\,\eqref{eq:cot_phi_rigid} with ($H=1$, solid red line) and without ($H=0$, dashed black line) the BBH force.  
        Here $\rho_d/\rho_f=2$.
        }
        	\label{fig:varphi_history}
    \end{minipage}
\end{figure}

A complementary diagnostic is the phase quantity $\cot(\varphi)$ defined in Eq.\,\eqref{eq:cot_phi_rigid} and illustrated in Fig.\,\ref{fig:varphi_history}.  
For $H=0$ it varies linearly with $\delta$.  
For small density ratios and large $\delta$, the phase behavior is dominated by added-mass effects.  
However, with the BBH included the dependence becomes non-linear and exhibits a $\delta^{1/2}$ scaling at small $\delta$, providing another experimentally accessible signature of the history force.  
In all cases, the phase shift vanishes as $\delta\to 0$ in the quasi-static limit.

\subsection{Response of spherical droplets and air bubbles\label{sec:dropletpart}}

\begin{figure}[H]	
    \includegraphics[width=.47\textwidth]{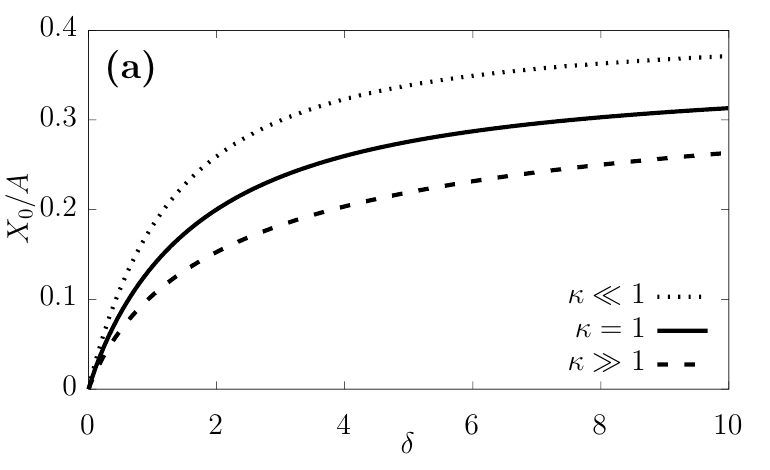}
	  \includegraphics[width=.47\textwidth]{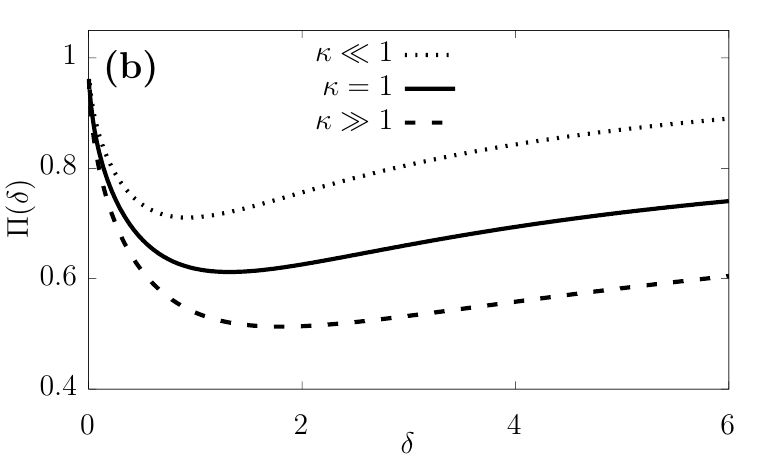}
    \begin{minipage}{.47\textwidth}
        \includegraphics[width=\columnwidth]{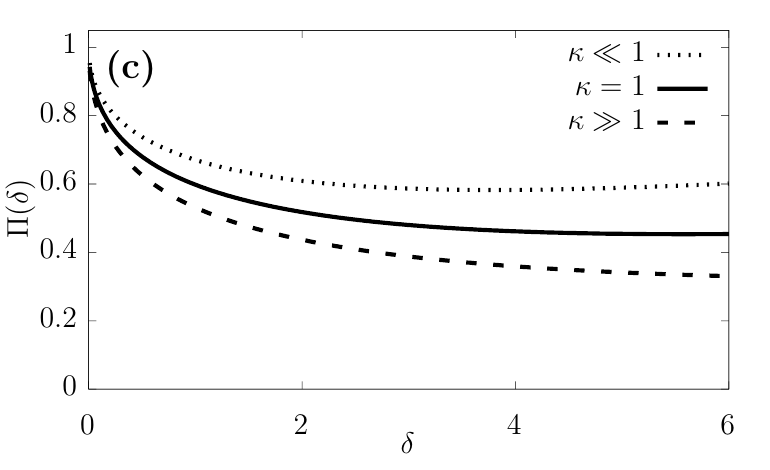}
    \end{minipage}
    \hspace{1.7cm}
    \begin{minipage}{.4\textwidth}
    \vspace{-.5cm}
        \caption{Panel (a) shows the relative horizontal displacement $X_0/A$ as a function of $\delta$ for a rigid particle (dashed line), a droplet with a viscosity ratio of $\kappa = 1$ (solid line), and a droplet with very low viscosity (dotted line). All cases have the same mass density ratio of $\rho_d/\rho_f = 2$. In panel (b), the ratio $\Pi(\delta)$, as given by Eq.\,\eqref{eq:ratio_RH}, is plotted to compare the displacement with and without the BBH effect for the same parameters as in panel (a). Finally, panel (c) presents the ratio $\Pi(\delta)$ for the same three viscosity ratios but for a very small mass density ratio, $\rho_d/\rho_f \to 0$.}
        \label{fig:Amplidelt_droplet}
    \end{minipage}
    \hspace{0cm}
\end{figure}

%
For droplets or gas bubbles in liquids, the assumption $\kappa \gg 1$ (appropriate for solid particles) and the associated non-slip boundary condition no longer hold. When a droplet moves relative to the surrounding fluid, viscous stresses induce internal circulation, as illustrated in Fig.\,\ref{fig:flow_field_illustration} panel (3). This internal flow reduces the shear stress at the droplet interface, thereby decreasing the effective viscous drag compared to that experienced by a rigid particle. 
The relative displacement amplitude $X_0/A$ remains the highest for droplets in the inertia-dominated regime at higher values of $\delta$, which can be achieved by increasing the shaking frequency $\omega$ or decreasing the kinematic viscosity $\nu_f$. For droplets with low internal viscosity, the velocity difference between the droplet interior and the surrounding fluid becomes smaller, resulting in a reduced drag force and, consequently, a larger displacement amplitude. Thus, the relative amplitude $X_0/A$ increases as the viscosity ratio $\kappa$ decreases.
This expectation is confirmed by the results shown in Fig.\,\ref{fig:Amplidelt_droplet}\,(a), where $X_0/A$ is plotted for a rigid particle ($\kappa \gg 1$), a droplet with intermediate viscosity contrast ($\kappa = 1$) and a droplet with very low internal viscosity ($\kappa \ll 1$), all for a fixed density ratio $\rho_d/\rho_f = 2$.
The effects  of BBH  are characterized by the ratio $\Pi(\delta,\kappa)$, defined in Eq.\,\eqref{eq:ratio_RH}, which quantifies the reduction of the droplet displacement relative to the liquid due to memory effects at the respective value of $\kappa$. In Fig.\,\ref{fig:Amplidelt_droplet}\,(b), $\Pi(\delta)$ is plotted for the same density ratio $\rho_d/\rho_f = 2$ and three different viscosity ratios. The displacement reduction becomes more pronounced as $\kappa$ increases. This trend is also observed in Fig.\,\ref{fig:Amplidelt_droplet}\,(c) in the limit of a very small density ratio, $\rho_d/\rho_f\to 0$. Then, case $\kappa \ll 1$ represents gas bubbles with free-slip boundary conditions, while $\kappa \gg 1$ approximates bubbles with non-slip boundary conditions, due to surfactants \cite{Grace_Weber:1978}. For the small mass densities of gas bubbles, the relative importance of the history force becomes more pronounced, as also observed for solid particles in Fig.\,\ref{fig:elongation_ratio_density}\,(b) at similarly low density ratios.
\begin{figure}[H]
	\centering
    \begin{minipage}{.47\textwidth}
            \includegraphics[width=\columnwidth]{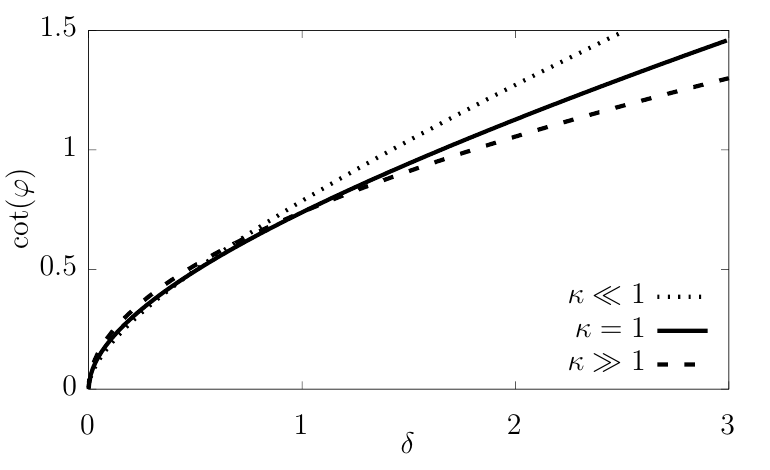}
    \end{minipage}
    \begin{minipage}{.4\textwidth}
            \caption{ The function $\cot\varphi$ of the relative phase shift $\varphi$ for the three values of $\kappa$ as in Fig.\,\ref{fig:Amplidelt_droplet}(a) and the mass ratio $\rho_d/\rho_f=2$.}
	          \label{fig:phase_droplet}
    \end{minipage}
\end{figure}
In Fig.\,\ref{fig:phase_droplet}, we compare the function $\cot(\varphi(\delta))$ for particles with three different viscosity ratios: droplets with $\kappa = 1$ (solid line), gas bubbles with $\kappa \ll 1$ (dotted line) and solid particles with $\kappa \gg 1$ (dashed line). For rigid particles, we previously found that $\cot(\varphi(\delta)) \propto \delta^{1/2}$ in the limit of small $\delta$, see Eq.\,\eqref{eq:cot_phi_rigid}. The curves for droplets ($\kappa = 1$) and gas bubbles ($\kappa \ll 1$) in Fig.\,\ref{fig:Amplidelt_droplet}\,(c) show a similar qualitative square-root-like scaling behavior.
These comparisons indicate that the qualitative features of the Basset–Boussinesq history force described for solid particles in Sec.\,\ref{sec:rigidpart} also apply to incompressible spherical droplets. Moreover, both $X_0/A$ and $\cot(\varphi)$ exhibit comparable power-law scaling in the small-$\delta$ regime, as found for solid particles, although the magnitude of the effects is reduced for droplets.
In realistic shaken liquids, gas bubbles typically undergo volume changes as a result of pressure oscillations. In the asymptotic limits $\kappa \ll 1$ and $\kappa \gg 1$, the hydrodynamic forces acting on the bubbles can still be calculated analytically, as discussed in Ref.\,\cite{Magnaudet:1998.1} and Appendix\,\ref{sec:bubble_history_var_R}. In a shaking experiment, as described in the present work, this analogy could be tested experimentally for droplets and gas bubbles, especially to probe the effects of a variable bubble radius.

Note, coupling between bubble compressibility and container acceleration can introduce a time-averaged buoyancy force \cite{Buchanan1962, JAMESON196635, Ellenberger:2007.1, Sorokin:2012.1, Elbing:2016.1}, causing the bubble to migrate slowly in the horizontal direction towards a container wall.
Qualitatively, if a bubble is initially displaced to the left of the container center, then during rightward (positive) acceleration, the increased fluid pressure compresses the bubble more strongly than during leftward (negative) acceleration. Since buoyancy is proportional to the product of bubble volume and acceleration, the force is stronger when the bubble is less compressed, i.e., during negative acceleration. This asymmetry produces a net buoyant force directed toward the wall at negative $x$, leading to horizontal migration from an unstable equilibrium at the center of the container.

\section{Considerations for conducting the proposed shaking experiment\label{sec:considerations}}
     
In this section, we summarize some estimates and considerations for conducting the experiment described in the previous section to analyze the effects of BBH on the dynamics of particles in unsteady fluid flows. Accurate measurements of particle dynamics require several observable cycles of oscillation of a particle during its sedimentation or ascent before it hits the upper or lower container wall. This can be achieved by choosing small differences between the mass density of the particles and the liquid. However, a low density contrast leads to small particle amplitudes $X_0$ that may be challenging experimentally.

\subsection{Amplitude limitations}
The oscillation amplitude of a particle $X_0$ with respect to the carrier fluid in response 
to shaking of the carrier liquid decreases with decreasing density contrast $\abs{\rho_d/\rho_f-1}$. 
According to Eq.\,\eqref{eq:Sigma2}, a small particle displacement $X_0$ can be amplified by a larger container oscillation amplitude $A$, but $A$ is limited by the mechanical constraint that the whole container must be accelerated in every cycle. The following estimate focuses on rigid particles, but also applies to droplets qualitatively. According to Fig.\,\ref{fig:elongation_ratio_density}, the region of interest corresponds to $\rho_d/\rho_f \lesssim 2$ and $\delta \lesssim 10$, where the history force has its largest relative impact. The main constraints are the minimum detectable particle amplitude $X_0^*$ and the maximum achievable container amplitude $A^*$. For given $X_0^*$ and $A^*$, Eq.\,\eqref{eq:Sigma2} defines the accessible part of the parameter space in the $(\delta,\rho_d/\rho_f)$ plane, corresponding to the range with $X_0/A \gtrsim X_0^*/A^*$. As an example, consider silicone oil with kinematic viscosity
$\nu_f \sim 10^{-4}\,\si{\square\meter\per\second}$ and particles of radius $R_0 \sim 1\,\si{\milli\meter}$. A shaking frequency in the range $0 < \omega/2\pi \lesssim 150\,\si{\hertz}$
corresponds for this example to a dimensionless frequency $0 < \delta \lesssim 10$ with $\delta = \omega R_0^2/\nu_f$. 

\begin{SCfigure}[][htb]
	\centering
    \begin{minipage}{.5\textwidth}%
        \includegraphics[width=.92\columnwidth]{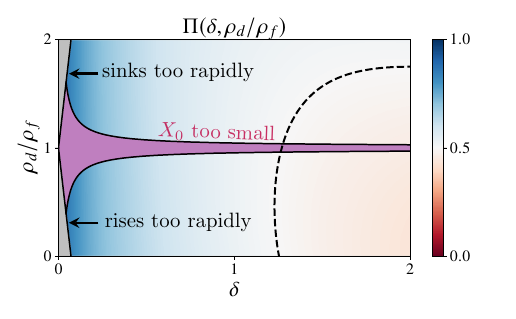}
    \end{minipage}
    \caption{Experimental accessibility in the $\delta$-$\rho_d/\rho_f$ parameter space. The purple region corresponds to $X_0/A < X_0^*/A^* = 5\cdot 10^{-3}$, where particle displacements are too small to resolve. The gray regions indicate where particles rise or sediment too quickly to complete at least $N=10$ oscillation cycles in a container of height $D = 20\,\si{\centi\meter}$. Other parameters are $R_0 = 1\,\si{\milli\meter}$, $g = 9.81\,\si{\meter\per\second\squared}$, and $\nu_f = 10^{-4}\,\si{\meter\squared\per\second}$. Under these estimates, most of the region where history effects dominate (reddish shading beyond the dashed line, as in Fig.\,\ref{fig:elongation_ratio_density}) remains accessible. \label{fig:experimental_limits} \vspace{1.2cm}}
    \end{SCfigure} 

In this regime, shaking amplitudes up to $A^* \sim 2\,\si{\milli\meter}$ are plausible, while modern equipment can likely detect particle displacements as small as $X_0^* \sim 10\,\si{\micro\meter}$. According to this estimate, the mentioned amplitude limitations lead to a small experimentally inaccessible parameter space, as indicated by the purple region in
Fig.\,\ref{fig:experimental_limits}. Most of the relevant parameter space for characterizing the effects of BBH is experimentally accessible, including most of the region where $\Pi < 0.5$ and the history force is the dominant part of the viscous drag.

\subsection{Number of observable oscillation cycles during sedimentation} 
To accurately measure the horizontal deflection amplitude $X_0$ and the phase $\varphi$ of a particle in the shaken carrier liquid, several measurable oscillation cycles of a particle in the volume of the container are required before it hits the bottom or top of the container. Using the steady-state sedimentation velocity from Eq.\,\eqref{eq:basic_shake_z}, for the above parameters and $\rho_d/\rho_f \lesssim 2$, we estimate:
\begin{align}
    \abs{\dot z_d} &= \frac{2R_0^2}{9\nu_f} g \abs{1-\frac{\rho_d}{\rho_f}} \lesssim 0.02\,\si{\meter\per\second} 
    \qquad \text{and}\qquad 
    \tau_p          = \frac{2}{9}\frac{R_0^2}{\nu_f}\frac{\rho_d}{\rho_f} \lesssim 0.005\,\si{\second}.
    \label{eq:vertical_settling_estimation}
\end{align}
For a container of height $D = 20\,\si{\centi\meter}$, sedimentation or rise occurs over $t_s = D / \abs{\dot z_d} \sim 10\,\si{\second} \gg \tau_p$, making transient effects negligible. The number of horizontal oscillation cycles is then
\begin{align}
    N = \frac{\omega t_s}{2\pi} = \frac{9 \nu_f^2 D}{4\pi g R_0^4} \frac{\delta}{\abs{1-\frac{\rho_d}{\rho_f}}} \sim 140 \frac{\delta}{\abs{1-\frac{\rho_d}{\rho_f}}}.
    \label{eq:shakeoscill}
\end{align}

To accurately measure the amplitude and phase of the horizontal particle oscillation, several oscillation cycles are necessary, e.g., more than $N \gtrsim 10$  cycles. With the parameters mentioned for silicone oil, this condition represents only a minor limitation, affecting a very small region at small $\delta$ as indicated by the gray areas in Fig.\,\ref{fig:experimental_limits}.

    \subsection{Nonzero Reynolds number}
Our analysis focuses on the regime of small values of the Reynolds number $\mathrm{Re} = R_0 |\dot{\mathbf r}_d|/\nu_f$. In the context of the shaking experiment, it is convenient to express $\mathrm{Re}$ in terms of the horizontal peak velocity. In addition, using the estimate for the vertical velocity from Eq.\,\eqref{eq:vertical_settling_estimation}, the horizontal contribution dominates, and we obtain the following
\begin{equation}
    \mathrm{Re} \sim \frac{X_0}{R_0}\,\delta \lesssim 1 
    \, .
\end{equation}
For parameters $X_0^* \sim 10\,\si{\micro\meter}$ and $R_0 \sim 1\,\si{\milli\meter}$, $\mathrm{Re}$ thus remains within the Stokes limit up to $\delta \sim 100$. In less viscous fluids or with larger particles, this threshold is reached significantly earlier, making the proposed experiment particularly well-suited for systematically investigating the onset of finite Reynolds number effects.

Within the Stokes limit ($\mathrm{Re}= 0$), the ratio $X_0/A$ is independent of the driving amplitude $A$. At finite $\mathrm{Re}$, this breaks down, i.e., increasing $A$  the magnitude of $X_0/A$ and the phase may change because nonlinear contributions to the Navier-Stokes equations become relevant \cite{MeiR:1992.1,MeiR:1994.1,Berlemont:1990.1,Sirignano:1998.1}. Vertical sedimentation or ascent also becomes coupled with horizontal motion. With the proposed experiment, such coupling can be detected directly through variations in vertical velocity, changes in amplitude response, or phase shift as $A$ and $\omega$ vary. Any measurable deviation from the Stokes prediction directly indicates nonlinear effects. This would supply valuable information for extending BBH-type models to finite Reynolds numbers, e.g.\ along the lines of Refs.\,\cite{MeiR:1994.1,MeiR:1992.1}, and help establish when these linearized models (widely used in particle-laden turbulence) cease to be reliable \cite{Vasan:2023.1,Daitche:2011.1,Daitche:2015.1}.

A related open question concerns the role of particle rotation. Although it is known that a constantly translating sphere for $\mathrm{Re}=0$ experiences the same drag whether it rotates or not \cite{Gladkov2022}, at small but finite $\mathrm{Re}$ a rotating sphere experiences an enhanced drag \cite{Ovseenko1968}. How this rotational-translational coupling manifests itself in unsteady flows remains unresolved. The shaking experiment therefore offers an opportunity to probe an aspect of BBH physics that has received little attention but may be relevant, for example, to run-and-tumble dynamics of micro-swimmers and to inertial particle modeling in turbulent environments.

\subsection{Deformability and compressibility}
\begin{SCfigure}{}{htb!}
    \centering
    \includegraphics[width=0.5\linewidth]{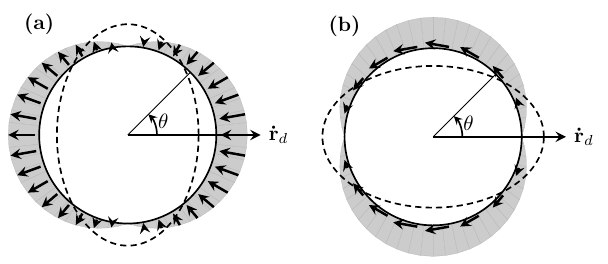}
    \caption{%
    (a) Pressure distribution around a steadily translating droplet with a highly
    viscous interior. The radial envelope scales with $\cos\theta$, with
    $\theta=0$ aligned to the direction of motion, illustrating the upstream
    overpressure and downstream reduction.  
    (b) Corresponding shear–stress distribution. The gray halo varies as
    $|\sin\theta|$, capturing the angular dependence of tangential viscous
    stresses, which vanish at the poles and reach a maximum at the equator.
    Simplified arrows indicate the direction of surface shear, and the large
    arrow at $\theta=0$ marks the direction of droplet motion.
    }
    \label{fig:deformation}
\end{SCfigure}

A further limitation concerns particle deformability. For an incompressible droplet with a highly viscous interior
($\kappa\gg1$), the changes in shape are governed by the competition between the surface–tension forces and the viscous
and pressure stresses imposed by the exterior flow. According to common textbooks \cite{kim1991microhydrodynamics,Leal:2007}
or as described in Appendix\,\ref{sec:forcedropR0}, the disturbance pressure around a steadily
translating sphere scales as $p_{\text{flow}}\sim 3\mu_f U/(2R_0)\cos\theta$, while the corresponding tangential viscous stress
scales as $\sigma_{r\theta}\sim 3 \mu_f U/(2R_0)\sin\theta$, where $U=|\dot{\mathbf r}_d|$ and $\theta$ is the polar angle measured from $\theta=0$ in the
direction of motion. The situation is illustrated in Fig.\,\ref{fig:deformation}. Marangoni stresses arising from surface tension gradients are assumed to be absent.
Both stresses scale as $\mu_f U/R_0$, while the surface tension opposes deformation via the Laplace pressure $p_L=2\sigma/R_0$. Their ratio defines the capillary number,
\begin{equation}
    \mathrm{Ca}=\frac{\mu_f U}{\sigma}\sim \frac{\mu_f X_0\omega}{\sigma},
\end{equation}
which quantifies the relevance of deformation \cite{stone1994dynamics}. With
$\mathrm{Ca}\ll1$ the shape remains close to spherical.

A second source of deformation is the acceleration-induced hydrostatic pressure difference across the droplet during
shaking. Its scale is $\Delta p_{\text{stat}}\sim |\rho_f-\rho_d|\,a\,R_0$, where $a=A\omega^2$ is the driving acceleration.
Comparing this with the Laplace pressure yields a Bond number
\begin{equation}
    \mathrm{Bo}=\frac{|\rho_f-\rho_d|\,A\omega^2\,R_0^2}{\sigma}.
\end{equation}
For liquid-liquid systems with small density differences $|\rho_f-\rho_d|$, and $\mathrm{Bo}\ll1$ for the typical parameter range considered here. Using representative values ($\sigma\sim 50\cdot 10^{-3}\,\si{\newton\per\meter}$, $X_0\sim100\,\mu\si{\meter}$,
$f\simeq150\,\si{\hertz}$, $\mu_f\sim 100\,\si{cSt}\times\rho_f$, $\rho_f\sim 1000\si{\kilo\gram\per\cubic\meter}$) gives $\mathrm{Ca}\lesssim 0.2$, so the deformation remains
weak. Although such effects could be investigated systematically, they can be made subdominant under the conditions explored.

\subsection{Interactions between droplets and boundary effects}
 Hydrodynamic interactions cause two adjacent particles in a stationary liquid to sediment faster than a single particle  \cite{Russel:89}. 
 The proposed experimental setup allows investigations of how the sedimentation rate of neighboring particles changes in time-dependent fluid motions compared to stationary fluids by systematically varying the shaking frequency and amplitude. This enables the study of the influence of hydrodynamic memory effects on the particles. This experimental approach can therefore serve as a guide for investigating more complex dynamics with multiple interacting particles or walls.

\section{Remarks and Conclusions \label{sec:conclusions}}

In this work, we have revisited the hydrodynamic forces acting on droplets, solid particles, and gas bubbles in unsteady Stokes flows, with particular emphasis on the BBH. Starting from the fundamental hydrodynamic equations, we re-derived the complete velocity field and force expressions for spherical droplets of arbitrary viscosity ratio $\kappa$ in 
Appendix\,\ref{sec:forcedropR0}, recovering as limiting cases both the classical rigid-particle and the free-slip bubble results. The theoretical framework also accommodates time-dependent droplet radii, as detailed in \cite{Magnaudet:1998.1} and App.\,\ref{sec:varying_radius}, thus providing a unified description applicable to a wide class of particle–fluid systems.

It may seem unusual to many, but this project was inspired by the 2019 “International Young Physicists’ Tournament” (IYPT), an international competition for high school students in which F.G. had participated while still in school. The phenomenon of gas bubbles sinking in vertically shaken liquids was one of the 17 problems internationally announced for the 2019 IYPT competition. During this competition, the authors noted that existing theoretical studies on gas bubbles in shaken liquids did not account for the classical Basset-Boussinesq history force. This observation formed the starting point for the present study on solid particles, droplets, and gas bubbles in shaken liquids, and the work for this publication was subsequently carried out in parallel with F.G.'s physics studies.

Beyond our analytical derivations in this work, we have clarified the physical origin of the BBH force in terms of transient, diffusion-driven vortex structures that arise when the droplet accelerates and persists over a range proportional to $\tau_f$. These unsteady vortices near the particles generate, both an enhanced drag amplitude and a characteristic phase lead, effects that become prominent when the dimensionless frequency $\delta=\omega\tau_f$ approaches unity and when the density contrast $\abs{1-\rho_d/\rho_f}$ is small.
Our analysis shows that incomplete relaxation of the surrounding flow during successive cycles of particle oscillations relative to the carrier fluid is directly responsible for the viscous history force, regardless of whether the particle is a solid sphere, a droplet, or a bubble.

Building on these findings, we propose an experimental setup to investigate the dynamics of sedimenting (rising) droplets in a horizontally shaken and thus accelerated liquid. In this experiment, the lateral (horizontal) displacement of the droplet results from the interplay of inertia, buoyancy, added mass, and the contribution of the BBH. We derived closed-form expressions for the particle displacement amplitude $X_0$ and the phase $\varphi$ as functions of the shaking frequency, viscosity ratio and density ratio. These results show that in the transition region between quasi-stationary Stokes flow and inertia-dominated particle dynamics, BBH can reduce the lateral particle displacement amplitude by more than $60\%$ compared to predictions that neglect memory effects.
In the low-frequency limit, we identified a characteristic scaling of $X_0(\omega)$ that provides a clear and experimentally verifiable signature of the BBH force. Our formulation further enables direct reconstruction of the history kernel $\bar{G}(\omega,\kappa)$ from the measured amplitude and phase of the droplet displacement relative to the carrier liquid.

The present analysis applies to small spherical droplets at low Reynolds number and our results provide quantitative criteria for when the BBH term must be retained in modeling and simulations, particularly for light particles or bubbles in liquids, where memory effects become comparable to or exceed added-mass contributions.

Our findings suggest several avenues for future research. The extension of the present framework to droplets or bubbles with time-dependent radii represents a natural next step, especially in contexts involving volume oscillations or acoustic forcing. Further work may also explore weakly inertial regimes with small Reynolds numbers or interactions among several droplets.

\section*{Acknowledgments}

We gratefully acknowledge the Wilhelm and Else Heraeus Foundation (Hanau, Germany) for supporting FG's participation in the  International Young Physicists' Tournament (IYPT)  and the German Young Physicists' Tournament (GYPT), as well as for supporting WZ with a WE-Heraeus Senior-Professorship. The authors also thank Jacques Magnaudet and Michael Wilczek for valuable discussions.

\appendix

\section{Basic equations for the flow field around a moving spherical droplet \label{sec:basicvel}}

The fundamental equations that govern the flow field around a spherical droplet that moves unsteadily without mixing with the surrounding liquid are well known from classical textbooks \cite{Landau6eng, batchelor1967introduction, lamb1993hydrodynamics,kim1991microhydrodynamics} and are summarized here for completeness. The properties of the incompressible carrier fluid are given by the mass density $\rho_f$ and the kinematic viscosity $\nu_f$ and those of the incompressible droplet by $\rho_d$ and $\nu_d$, whereby 
the special case of a rigid particle is described by the limiting case of high droplet viscosity $\nu_d$.   

\begin{figure}[htb]
	\centering
    \begin{minipage}{.45\textwidth}
        \includegraphics[width=\columnwidth]{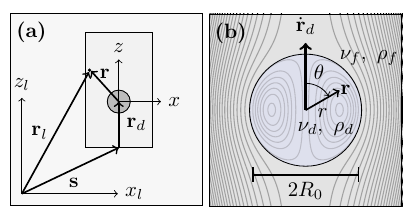}
    \end{minipage}%
    \hspace{.5cm}
    \begin{minipage}{.46\textwidth}%
        \caption{Part (a) shows how  $\vb r_l$ in the laboratory system is represented by  $\vb s$ at the center of the container bottom, the droplet coordinates $\vb r_d$ in the frame moving with the container and the relative position $\vb r$ to the droplet.
        Part (b) shows a spherical particle moving with $\vb{\dot r}_d$ relative to the carrier liquid (gray) of viscosity $\nu_f$ and mass density $\rho_f$. The particle is a liquid droplet of kinematic viscosity $\nu_d$, mass density $\rho_d$ (light blue), and constant radius $R_0$. In the coordinate system attached to the droplet, each  $\vb r$ is expressed by the angle $\theta$ to the velocity axis, the radial distance $\norm{\vb r}=r$ from the center of the droplet and an azimuthal angle $\varphi$.
	    }
	\label{fig:flow_arounddrop}
    \end{minipage}
\end{figure}

The flow field ${\vb v}$ around a moving droplet is represented in a coordinate system $\vb r=(x,y,z)$ whose origin is co-moving with the center of the droplet, as indicated in Fig.\,\ref{fig:flow_arounddrop}\,(a). 
In terms of $\vb r$, the position of a point $\vb r_l$ in the laboratory system is given by 
\begin{equation}
	\vb r_l =\vb r+\vb s(t) + \vb r_d(t)
	\,,
\end{equation}
and the velocity field ${\vb v}$ in the co-moving frame can be expressed by that in the laboratory system $\vb u(\vb r_l,t)$ as follows:
\begin{equation}
\vb v(\vb r,t)= \vb u(\vb r+\vb s(t)+\vb r_d(t),t) - \vb{\dot s}(t)-\vb{\dot r}_d(t)\,.
\label{eq:velocity_transformation}
\end{equation}
In large containers, far from moving droplets, for $\norm{\vb r}\gg R_0$, there is no flow of the carrier liquid relative to the container. In this case, the velocity field of the liquid in the container is given by $\vb{u}=\vb{\dot s}$. This leads to the following condition:
\begin{equation}
\vb u(\norm{\vb r}\gg  R_0,t)\to \vb{\dot s}
\,
\Rightarrow
\,
\vb{v}(\norm{\vb{r}}\gg R_0,t) \to  -\vb{\dot r}_d(t) = \vb{v}_\infty(t)
\, .
\label{eq:condition_at_infinity}
\end{equation}
In the coordinate system moving with the droplet, the upstream velocity $\vb{v}_\infty$ in Eq.\,\eqref{eq:condition_at_infinity} has the same magnitude but opposite direction as the droplet velocity $\vb{\dot r}_d$.

An advantageous property of the described coordinate transformation is that it preserves the distances between points and that the spatial derivatives remain invariant. This property is utilized in further calculations and also means that the continuity equation for incompressible fluids has the same form in both systems:
\begin{equation}
\grad_l\cdot  \vb{u}(\vb r_l,t)=0 \quad \Longleftrightarrow \quad \grad\cdot \vb{v}(\vb r,t)=0
\label{eq:incompress}
\, .
\end{equation}
Here, $\nabla_l$ with the subscript $"l"$ indicates the divergence in the laboratory frame, while $\grad$ represents derivatives with respect to $x$, $y$, and $z$ in the comoving frame. In the following, the mass density and kinematic viscosity are sometimes abbreviated as $\rho=\rho_d$ and $\nu=\nu_d$ within the droplet ($\norm{\vb{r}}<R_0$), and as $\rho=\rho_f$ and $\nu=\nu_f$ outside the droplet ($\norm{\vb{r}}>R_0$), respectively. In this study, we have a length scale $R_0$, an external time scale, given by the inverse of the shaking frequency $\omega$, and we consider incompressible fluid flows in the range of small Reynolds numbers $\mbox{Re}$  but finite values of the frequency parameter $\delta$:
\begin{align}
\label{eq:ReStokes}
\mbox{Re}=
\frac{\norm{\vb{v}_\infty}R_0}{\nu} \ll 1 
\,,\qquad 
\delta =\frac{\omega R_0^2}{\nu} 
\, .
\end{align}
The velocity field is determined by the Navier-Stokes equations,
\begin{align}
\partialderivative{\vb{u}(\vb r_l,t)}{t}
+
\big( \vb{u}(\vb r_l,t)\cdot \grad_l \big) \vb{u}(\vb r_l,t)
=
-\frac{ \grad_l p_l(\vb r_l,t)}{\rho} + \nu \nabla^2_l\vb{u}(\vb r_l,t) -\grad_l \Phi_l(\vb r_l,t)
\, 
\label{eq:stokes_general}
,
\end{align}
which includes a general, but conservative, body force $\vb{f}(\vb{r}_l,t) = -\grad_l\Phi_l(\vb{r}_l,t)$. $\Phi_l$ describes, for example, the gravitational potential along the $z_l$-direction. Using Eq.\,\eqref{eq:velocity_transformation} with $\vb v_\infty(t) = -\dot{\vb r}_d(t)$, we transform Eq.\,\eqref{eq:stokes_general} into a reference frame moving with the droplet. In the creeping flow approximation at small values of the Reynolds number, we can then neglect the convection term $\vb v\cdot \grad \vb v$. This results in the time-dependent Stokes equation:
\begin{align}
\partialderivative{\vb{v}(\vb r,t)}{t}
&=
-\grad\qty(\frac{p(\vb r,t)}{\rho}+\Phi(\vb r,t)-\vb r\cdot\qty[\derivative{\vb{v}_\infty}{t}-\derivative{\vb{\dot s}}{t}]) 
+ \nu \nabla^2\vb{v}(\vb r,t)
\, ,
\label{eq:Stokes_equation}
\end{align}
where the definitions are
\begin{subequations}
	\begin{align}
	p(\vb r,t)&=p_l(\vb r+\vb s(t)+\vb r_d(t),t)\,,
    \qquad 
	\Phi(\vb r,t)=\Phi_l(\vb r+\vb s(t)+\vb r_d(t),t)
    \,,
	\end{align}
\end{subequations}
ensure that the scalar fields $p$ and $\Phi$ are evaluated at the correct points in the space.

The solutions of these equations inside and outside the droplet must be matched at the droplet surface using the boundary conditions. Changing to spherical coordinates as shown in Fig.\,2\,(b), we can describe the velocity field $\mathbf{v}$ in terms of the radial distance $r$ from the center of the droplet, the polar angle $\theta$ relative to the direction of motion, and the azimuthal angle $\varphi$. For motion along a fixed direction without rotation, the problem exhibits rotational symmetry around the velocity axis $\dot{\mathbf{r}}_d$, and the velocity field becomes independent of $\varphi$. This allows us to write the velocity field as
\begin{equation}
\mathbf{v}(\mathbf{r}, t) = \mathbf{v}(r, \theta, t) = v_r(r, \theta, t)\,\mathbf{e}_r + v_\theta(r, \theta, t)\,\mathbf{e}_\theta. \label{eq:generalVelocity_split}
\end{equation}
The general case of arbitrarily directed unsteady motion can be constructed by decomposing the velocity into three orthogonal components, each aligned with a Cartesian axis. Due to the linearity of the Stokes equations, the corresponding flow and force contributions can be calculated separately and superimposed. 

This argument implicitly assumes that the droplet’s orientation does not vary over time. In fact, even if the droplet undergoes a finite angular velocity or reorientation, the translational hydrodynamic force should remain unaffected in the Stokes regime \cite{Ovseenko1968,Gladkov2022}. 

Due to the rotational symmetry around the direction of the droplet velocity $\dot{\vb r}_d$, the velocity field remains independent of $\varphi$. In the absence of rotational motion, $v_\varphi=0$. It should be noted that the droplet is assumed to be undeformable and that no liquid exchange takes place across the boundary, which leads to the following condition for the radial velocity component $v_r$:  
\begin{equation}
v_r(r=R_0,\theta,t) = 0
\label{eq:normalBC}
\, .
\end{equation}
The constant radius $R_0$ also implies that normal stresses on the surface (including the Young-Laplace pressure induced by surface tension) are not relevant to flow dynamics. For a time-dependent radius, e.g., in bubble studies, these stresses must be considered as they couple to the radius dynamics. The resulting compressible dynamics are interesting and will be addressed in future work.

Regarding the tangential velocities inside and outside the spherical droplet, they are identical on its surface at $r=R_0$, and the tangential stresses must also be similar at this interface. 
Regarding stresses, the force $\vb{\text{d}F}$ exerted by the flow $\vb{v}$ on a surface element d$S$ of the droplet with a normal vector $\vb{n}=\vb{e}_r$ is usually expressed as a contraction with the hydrodynamic stress tensor $\sigma_{ij}$. Using the summation convention, its components take the form,
\begin{equation}
\text{d}F_j = \sigma_{ij} n_i \text{d}S
=\sigma_{rj} \text dS
\label{eq:force_general}
\, ,
\end{equation}
and are a function of the spatial derivatives of the flow velocity field $\vb v$ at $\norm{\vb r}=R_0$
in the co-moving reference system of the droplet.  
The spatial derivatives remain unchanged during the transformation to the laboratory system according to Eq.\,\eqref{eq:velocity_transformation} and therefore do not cause an additional shear force.
 Consequently, the total surface force d$ {\vb F}$ expressed by ${\bf v}$ is identical to that calculated in terms of $\vb u$ at the time-dependent position of the boundary in the laboratory system. The tangential force $\vb {\text{d}t}_f$ exerted by the flow from outside on the droplet per surface element ${\text d}S$ is then given by the $\varphi$- and $\theta$-components of $\vb {\text{d}F}$, i.e., it is with $\epsilon>0$ given by:
\begin{equation}
	\vb {\text{d}t}_f
	= \lim_{\epsilon\rightarrow 0}\Big[\sigma_{\varphi r}(r=R_0+\epsilon)\vb e_\varphi + \sigma_{\theta r}(r=R_0+\epsilon)\vb e_\theta\Big]\text dS
	\, .
	\label{eq:stress_from_outside}
\end{equation}
The counterpart of the circulation inside the droplet is given by:
\begin{equation}
	\vb {\text{d}t}_d
	= \lim_{\epsilon\rightarrow 0}\Big[\sigma_{\varphi r}(r=R_0-\epsilon)\vb e_\varphi + \sigma_{\theta r}(r=R_0-\epsilon)\vb e_\theta\Big]\text dS
    \, .
\end{equation}
The tangential stress changes continuously and is therefore identical on both sides of the droplet surface:
\begin{equation}
	\vb {\text{d}t}_d=\vb {\text{d}t}_f
	.
	\label{eq:tangcont}
\end{equation}
With the velocity defined in Eq.\,\eqref{eq:generalVelocity_split}, the tensor component $\sigma_{\varphi r}$ vanishes. Since we have $\partial_\theta v_r=0$ on the surface of the spherical drop, we can deduce from the definition of $\sigma_{\theta r}$ and
Eq.\,\eqref{eq:tangcont} the following boundary condition for the tangential velocity component $v_\theta$,
\begin{align}
	\partialderivative{}{r}\qty(\frac{\mu_f v_\theta}{r})_{r=R_0 }
	=
	\partialderivative{}{r}\qty(\frac{\mu_d v_\theta}{r})_{r=R_0 }
	\label{eq:tangentialBC}
,
\end{align}
with dynamic viscosities $\mu_f=\rho_f\nu_f$ and $\mu_d=\rho_d\nu_d$. The limits with respect to $\epsilon$ have been removed in this formula
for simplicity.
Eqs.\,\eqref{eq:condition_at_infinity}, \eqref{eq:incompress}, \eqref{eq:Stokes_equation}, \eqref{eq:normalBC} and \eqref{eq:tangentialBC} combined with the continuity of the velocity field across the droplet interface form a comprehensive set of equations for the velocity field $\vb v$ and pressure $p$.

If the droplet is at rest in the shaken coordinate system, with $\vb{v}_\infty = 0$, we can immediately see that the solution corresponds to the hydrostatic equilibrium of the fluid where the velocity vanishes ($\vb{v}=0$) and the pressure in the fluid reads:
\begin{equation}
p_\text{static}(\vb{r}, t) = p_0(t) - \rho_f \qty[\Phi(\vb{r}, t) + \vb{r} \cdot \derivative{\vb{\dot{s}}}{t}]
.
\label{eq:static_solution}
\end{equation}
The function $p_0(t)$ depends on the boundary conditions along the surface of the carrier liquid and disappears when integrating all the stresses on the spherical surface of a droplet. Therefore, equation \eqref{eq:static_solution} is sufficient to calculate the force acting on a droplet. The fluid inside the droplet is also in hydrostatic equilibrium. However, the pressure jumps across the boundary as result of the surface-tension-induced Laplace pressure. This difference is spatially uniform within the droplet and does not affect the velocity profile. The form of $p_\text{static}$ also illustrates that the droplet experiences a buoyancy force independent of its relative motion through the container, caused by an external potential $\Phi$ or vibration $\ddot{\vb s}$ of the container.

We can then consider the general case of a droplet with $\vb{v}_\infty(t) \neq 0$ as a perturbation to this hydrostatic equilibrium. Therefore, we split the total pressure into its static contribution and a yet-to-be-determined part due to flow $p_\text{flow}(\vb{r}, t)$:
\begin{subequations}
	\begin{align}
	p(\vb r,t)&= p_\text{static}(\vb r,t) +p_\text{flow}(\vb r,t)
	\label{eq:split_pressure}
	.
	\end{align}
\end{subequations}
Since the static part does not cause fluid flow, $\vb v$ is caused by the relative motion of the droplet alone. In addition to the continuity of the velocity at $r=R_0$, the complete set of equations that govern this velocity perturbation $\vb v$ and the pressure induced by the flow can be expressed as follows:
\begin{subequations}
	\begin{align}
	\partialderivative{\vb{v}}{t}
	+\grad\qty(\frac{p_\text{flow}}{\rho_d}-\vb r\cdot\derivative{\vb{v}_\infty}{t})
	-\nu_d \nabla^2\vb{v}	
	&=0
	\quad (\text{for}~
	r<R_0)\,,
	\label{eq:interior_Stokes_raw}
	\\
	\partialderivative{\vb{v}}{t}
	+\grad\qty(\frac{p_\text{flow}}{\rho_f}-\vb r\cdot\derivative{\vb{v}_\infty}{t})
	-\nu_f \nabla^2\vb{v}
	&=0
	\quad (\text{for}~
	r>R_0)\,,
	\label{eq:exterior_Stokes_raw}
	\\
	\partialderivative{}{r}\qty(\frac{\mu_f  v_\theta}{r})_{r=R_0}
	-\partialderivative{}{r}\qty(\frac{\mu_d v_\theta}{r})_{r=R_0 }
	&=
	0,
	\label{eq:tangential_BC_vflow}
	\\
	v_r(r=R_0,\theta,t) &= 0\,,
	\label{eq:radial_BC_vflow}
	\\
	\vb{v}(r \gg R_0,\theta,t) &= \vb{v}_\infty(t)\,,
	\label{eq:far_field_vflow}
	\\
	\div \vb{v}&=0
	\label{eq:incompressibility}
	.
	\end{align}
	\label{eq:governing_set}
\end{subequations}
In the next section, we will solve these equations for the velocity field $\vb v$ and for $p_\text{flow}$, and from this we will calculate the force acting on the droplet. When $r\to 0$ we demand regular, symmetric solutions.

\section{Velocity field around a moving spherical droplet and the force acting on it  
\label{sec:forcedropR0}}
  
Equations\,\eqref{eq:governing_set} form a complete set to determine the flow field $\vb{v}(\vb r,t)$ and the pressure $p_\text{flow}$ around a moving droplet. For their solution, we were inspired by §24 of Ref.\,\cite{Landau6eng} and structured our calculations as follows. In Appendix\,\ref{sec:flow_harm_osz}, we first calculate the flow field $\vb{v}(\vb{r},t)$ in the vicinity of a spherical droplet undergoing harmonic oscillations through the carrier fluid. The resulting viscous drag forces acting on the droplet due to its relative motion through the carrier fluid are calculated in Appendix\,\ref{sec:force_harm_osz}.
  
In Appendix\,\ref{sec:drag_arbit_t}, the forces are formulated for a general time-dependent relative velocity $\vb v_\infty(t)$. Therefore, after a Fourier decomposition of $\vb v_\infty(t)$, the resulting viscous drag acting on a droplet in the range of low Reynolds numbers is calculated by a linear superposition of the solutions from Appendix\,\ref{sec:flow_harm_osz} and Appendix\,\ref{sec:force_harm_osz}. The results depend on the following ratio:
    \begin{equation}
    \kappa =\frac{\mu_d}{\mu_f}
    \, .
    \end{equation}
   
   In Appendix\,\ref{sec:arb_motion_large_kappa} the Fourier transform is inverted and the general drag force is explicitly calculated in the limit $\kappa \gg 1$, which corresponds to solid beads, and in section \ref{sec:arb_motion_small_kappa} in the limit $\kappa \ll 1$, as for fixed radius spherical gas bubbles. 
   
   In Appendix\,\ref{sec:comparememory}, we discuss the typical time scales over which the memory kernel decays in the limiting cases of $\kappa \gg 1$ and $\kappa \ll 1$. Illustrations of the flow field around a harmonically oscillating droplet are found in the main text.

\subsection{Determination of ${\vb v}({\bf r},t)$ for a harmonic $\vb v_\infty(t)$ \label{sec:flow_harm_osz}}

Here, we determine the flow field \(\mathbf{v}(\mathbf{r}, t)\) around a spherical droplet with a constant radius \(R_0\), where the carrier fluid at large distances from the droplet moves with a velocity in the \(z\)-direction:
\begin{equation}
\vb{v}_\infty(t) = v_\infty(t) \vb{e}_z\,.
\label{eq:direction_of_vinfty}
\end{equation}
The choice of \(\vb{v}_\infty\) directed along the \(z\)-axis is somewhat arbitrary and will be generalized later by decomposing an arbitrary \(\vb{v}_\infty\) into its components along the \(x\)-, \(y\)- and \(z\)-axes. In the linear regime of small Reynolds numbers, the dynamics in orthogonal directions decouple, allowing us to superimpose the solutions.

By taking  the curl of Eq.\,\eqref{eq:interior_Stokes_raw} and Eq.\,\eqref{eq:exterior_Stokes_raw} the pressure and the liquid
velocity $\vb{v}_\infty$ far from the droplet, both drop out for incompressible liquids inside and outside of the droplet, and we find,
	\begin{align}
	\partialderivative{\qty(\curl\vb{v})}{t}&=\nu \nabla^2\qty(\curl\vb{v})
	\,
,
	\label{eq:curl_stokesEquation_inside}
	\end{align}	
%
with kinematic viscosities $\nu=\nu_d$ for $r<R_0$ and $\nu=\nu_f$ for $r>R_0$. Since rotational symmetry around the direction of ${\vb {\dot r}_d}\propto \vb e_z$, the flow field $\vb{v}(\vb{r},t)$ can be represented by a stream function \cite{batchelor1967introduction,lamb1993hydrodynamics} $\psi(r,\theta,t)$, where the spherical coordinates $(r,\theta,\varphi)$ correspond to those introduced in Fig.\,\ref{fig:flow_arounddrop}\,(b). This leads to the following representation of the velocity field:
\begin{align}
\vb{v} 
= 
\curl \qty(\frac{\psi}{r\sin\theta}\vb{e}_\varphi)
=
\frac{1}{r^2\sin\theta}\partialderivative{\psi}{\theta}\vb e_r
-
\frac{1}{r\sin\theta}\partialderivative{\psi}{r}\vb e_\theta
.
\label{eq:stream_f_def}
\end{align}
The form of Eq.\,\eqref{eq:stream_f_def} ensures that the liquid described by $\psi$ is incompressible. Taking the curl of $\vb v$ gives in terms of the stream function $\psi$:
\begin{align}
\curl\vb{v} &= \curl\qty(\frac{1}{r^2\sin\theta}\partialderivative{\psi}{\theta}\vb{e}_r-\frac{1}{r\sin\theta}\partialderivative{\psi}{r}\vb{e}_\theta) 
= -\frac{1}{r\sin\theta}\qty[\partialderivative{^2\psi}{r^2}+\frac{\sin\theta}{r^2}\partialderivative{}{\theta}\qty(\frac{1}{\sin\theta}\partialderivative{\psi}{\theta})]\vb{e}_\varphi
= -\frac{\hat{E}\psi}{r\sin\theta}\vb{e}_\varphi
\, .
\label{eq:E_def}
\end{align}
With the operator $\hat E$ defined in this way, we can reformulate Eq.\,\eqref{eq:curl_stokesEquation_inside} as follows:
\begin{align}
\partialderivative{\qty(\hat{E}\psi)}{t} = - \nu   r\sin\theta\nabla^2 \qty(-\frac{\hat{E}\psi}{r\sin\theta}\vb{e}_\varphi)\cdot \vb e_\varphi
= \nu \hat{E}\qty(\hat{E} \psi)
.
\end{align}
Accordingly, the linear Navier-Stokes equations for the stream function inside and outside the droplet take the following forms:
\begin{subequations}
\label{eq:stokesEquation_final}
	\begin{align}
	\qty(\hat{E}-\frac{1}{\nu_d}\partialderivative{}{t} )\hat{E}\psi=0
	\quad \mbox{for} \quad
	r<  R_0\,,
	\label{eq:stokesEquation_final_1}
	\\
	\qty(\hat{E}-\frac{1}{\nu_f}\partialderivative{}{t} )\hat{E}\psi=0
	\quad  \mbox{for} \quad
	r> R_0 \,.
	\label{eq:stokesEquation_final_2}
	\end{align}
\end{subequations}
The solution $\psi(r,\theta,t)$ to these equations has to satisfy the boundary conditions given by Eq.\,\eqref{eq:tangential_BC_vflow} and Eq.\,\eqref{eq:radial_BC_vflow}, as well as the far-field condition in Eq.\,\eqref{eq:far_field_vflow}. Using Eq.\,\eqref{eq:direction_of_vinfty}, $\vb{v}_\infty$ can be expressed in the following form:
\begin{equation}
\vb{v}_\infty(t)
= v_\infty(t) \vb{e}_z
= v_\infty (t) \qty[\cos\theta \vb{e}_r-\sin\theta \vb{e}_\theta].
\end{equation}
Hence, by comparison to Eq.\,\eqref{eq:stream_f_def}, the stream function \(\psi\) must match the following expression at large distances:
\begin{align}
\psi(r \gg R_0,\theta,t)  =
v_\infty(t) \frac{r^2}{2}\sin^2\theta +F(t,\theta)
\label{eq:far_field_psi}
.
\end{align}
Note that the arbitrary function $F(t,\theta)$ is only relevant for particles with a time-dependent radius $R(t)$, as is the case for compressible bubbles. In this scenario, $F$ describes a radially time-dependent flow that decays with the distance $r$ from the particle as $r^{-2}$. However, for droplets with a constant $R_0$, one can set $F=0$. The boundary condition, which ensures zero normal flux, as stated in Eq.\,\eqref{eq:radial_BC_vflow}, can be given in terms of the stream function by:
\begin{equation}
\partialderivative{\psi}{\theta}\Big|_{r=R_0}
=
0
\,.
\label{eq:normalBC_streamFunction_bubble}
\end{equation}
Similarly, the continuity of the stress gives according to Eq.\,\eqref{eq:tangential_BC_vflow}:
\begin{equation}
\partialderivative{}{r}\qty(\frac{\mu_f}{r^2}\partialderivative{\psi}{r})\Big|_{r=R_0}
=
\partialderivative{}{r}\qty(\frac{\mu_d}{r^2}\partialderivative{\psi}{r})\Big|_{r=R_0}
\label{eq:tangentialBC_streamFunction}
.
\end{equation}
Following the strategy outlined at the beginning of the section, we first consider a single-harmonic oscillation with
\begin{equation}
v_\infty(t) = 
v_\omega \cos(\omega t+\phi) = 
\frac{v_\omega e^{i\phi}}{2}
e^{i\omega t} 
+   
\mbox{c.c.}
\, ,
\label{vinftyomeg}
\end{equation}
where $v_\omega$ is a real constant and $v_\infty$ is represented as a linear combination of two complex Fourier modes $e^{i\omega t}$ and $e^{-i\omega t}$. Due to the linearity of the equations, we also expect $\psi$ to be a linear combination of the form,
\begin{align}
\psi(r,\theta,t) =
\frac{v_\omega e^{-i\phi}}{2} 
\psi_\omega (r,\theta,t)
+ 
\frac{v_\omega e^{i\phi}}{2}
\psi_{-\omega}(r,\theta,t)\,
,
\label{eq:psiharm}
\end{align}
where the complex function $\psi_\omega$ satisfies Eqs.\,\eqref{eq:stokesEquation_final} and describes the flow profile for a single Fourier mode in $v_\infty$. 
Using Eq.\,\eqref{eq:far_field_psi} (and $F=0$) we obtain the stream function in the far field:
\begin{align}
\psi_\omega (r \gg R_0,\theta,t)  
=
e^{-i\omega t}
\sin^2\theta
\frac{r^2}{2}
\label{eq:far_field_psi_omega}
.
\end{align}
At smaller radii, we can use the ansatz,
\begin{align}
\psi_\omega(r,\theta,t)
 &=
e^{-i\omega t}  
\sin^2\theta
R_0^2
\begin{cases}
g_d(r) & r<R_0
\\
g_f(r) & r>R_0
\end{cases},
\label{eq:stream_function_ansatz}
\end{align}
to which we can apply the operator $\hat E$, as defined in Eq.\,\eqref{eq:E_def}. Then we find the following:
\begin{equation}
\hat E\psi_\omega =\qty(\partialderivative{^2}{r^2}-\frac{2}{r^2})\psi_\omega 
= \hat L \psi_\omega\,.
\end{equation}
In the case of one frequency of $v_\infty(t)$ in \eqref{vinftyomeg}, with these relations we obtain
from the partial differential equations \eqref{eq:stokesEquation_final_1} and \eqref{eq:stokesEquation_final_2} for $\psi(t)$ two ordinary differential equations for $g_d(r)$ and $g_f(r)$:
\begin{subequations}
\begin{align}
\hat L\hat L g_d+\frac{k_d^2}{R_0^2}\hat L g_d=0
\quad\quad
(r<R_0)\,,
\label{eq:gd_equation}
\\
\hat L\hat L g_f+\frac{k_f^2}{R_0^2}\hat L g_f=0
\quad\quad
(r>R_0)
.
\label{eq:gf_equation}
\end{align}
\end{subequations}
Here, we have introduced the parameters $k_d$ and $k_f$,
\begin{equation}
k_d=\sqrt{i\omega \tau_d }
\,,\quad\quad
k_f=\sqrt{i\omega \tau_f}\,,
\label{eq:k_definition}
\end{equation}
including the two viscous diffusion times,
\begin{equation}
\tau_f=\frac{R_0^2}{\nu_f}\,,\qquad \tau_d=\frac{R_0^2}{\nu_d}
\,.
\label{eq:def_taudff}
\end{equation}
The boundary condition for $\psi(t)$ for a vanishing normal flow on the drop surface in Eq.\,\eqref{eq:normalBC_streamFunction_bubble} is transformed into the boundary condition for $\psi_\omega$:
\begin{equation}
\partialderivative{\psi_\omega}{\theta}\Big|_{r=R_0}
=
0
\label{eq:normalBC_streamFunction_bubble_fourier}
\,
.
\end{equation}
According to 
Eq.\,\eqref{eq:normalBC_streamFunction_bubble_fourier}, the solution $g_d(r)$ of Eq.\,\eqref{eq:gd_equation} must meet  the boundary condition $g_d(R_0)=0$.   In addition, $g_d$ is also symmetrical and non-singular at $r=0$ and this results in    
\begin{equation}
g_d(r)=
\alpha^d \qty(\xi^2 -\frac{\cos(k_d \xi)-\sin(k_d \xi)/(k_d \xi)}{\cos(k_d)-\sin(k_d)/k_d})
\label{eq:ansatz_for_gd}
,
\end{equation}
with dimensionless length $\xi=r/R_0$.
The parameter $\alpha^d$ is yet unknown. The general solution  $g_f(r)$ of Eq.\,\eqref{eq:gf_equation} reads:
\begin{align}
g_f(r)
&=
\frac{\alpha^f_1}{\xi}+\alpha^f_2\xi^2 +\alpha^f_3 e^{i k_f\xi }\qty(i-\frac{1}{k_f \xi}) 
+
\alpha^f_4 e^{-i k_f\xi }\qty(i+\frac{1}{k_f \xi})
\label{eq:ansatz_for_gf}
.
\end{align}
Eq.\,\eqref{eq:far_field_psi_omega} requires that this expression approaches $\xi^2/2$ for large values of $\xi$. The term proportional to $\alpha_1^f$ vanishes for large $\xi$, while the contribution of $\alpha_4^f$ diverges as long as $\omega >0$ (the case with $\omega <0$ gives the complex conjugate quantity $\psi_{-\omega}=\psi_\omega^*$). We can therefore implement the correct far-field behavior by setting $\alpha^f_2=1/2$ and $\alpha^f_4=0$. For Eq.\,\eqref{eq:normalBC_streamFunction_bubble_fourier} to be fulfilled, $g_f$ must satisfy the condition,
\begin{equation}
g_f(R_0)=0 \, ,
\label{eq:condition_for_gf}
\end{equation}
while the continuity of the flow velocity tangential to the droplet surface yields the constraint:
\begin{equation}
g_d'(R_0)=g_f'(R_0)
\label{eq:continousVelocity}
.
\end{equation}
In addition, Eq.\,\eqref{eq:tangentialBC_streamFunction} provides:
\begin{equation}
\kappa\qty[2 g_d'(R_0)-R_0g_d''(R_0)]
=
2 g_f'(R_0)-R_0 g_f''(R_0)
\label{eq:stress_continuity}
.
\end{equation}
The equations \eqref{eq:condition_for_gf}, \eqref{eq:continousVelocity} and \eqref{eq:stress_continuity} form a closed set of equations that determine  the remaining unknowns $\alpha^d$, $\alpha^f_1$ and $\alpha^f_3$. With the abbreviations 
\begin{subequations}
	\begin{align}
	h(x)&=(x^2+3)\tanh(x)-3x\,
	\qquad\text{and}\qquad 
	\chi=2h(ik_d)+i k_d^2(k_d-\tan k_d)\,,
\end{align}
\end{subequations}
one obtains:
\begin{subequations}
  \begin{align}
	\alpha^d 
	&=
	-\frac{3}{2}\frac{(i+k_f)(k_d-\tan k_d)}{(3-ik_f)h(i k_d)-\chi \kappa}\,,
    \\
	\alpha^f_1
	&=
	\frac{1}{2}-\frac{3}{2i k_f^2}
    \frac{\qty(i+k_f-\frac{2}{3}ik_f^2)\chi \kappa - 2\qty(i+k_f-ik_f^2-\frac{1}{3}k_f^3) h(ik_d)}{   (3-ik_f)h(i k_d)-\chi\kappa    }\, ,
	\\
	\alpha^f_3
	&=
	-\frac{3}{2k_f}\frac{\chi \kappa -2h(ik_d)}{(3-ik_f)h(i k_d)-\chi \kappa}e^{-ik_f}
    \,.
	\end{align}
	\label{eq:solutions_alpha}
\end{subequations}
With these coefficients, $\psi_\omega$ is fully determined and the resulting analytic expression for the stream function $\psi(r,\theta,t)$ can be found using Eq.\,\eqref{eq:psiharm}. The resulting long expressions for $\psi(r,\theta,t)$ are not explicitly given but can be easily obtained using computer algebra.
 \begin{figure*}
	\includegraphics[width=.75\textwidth]{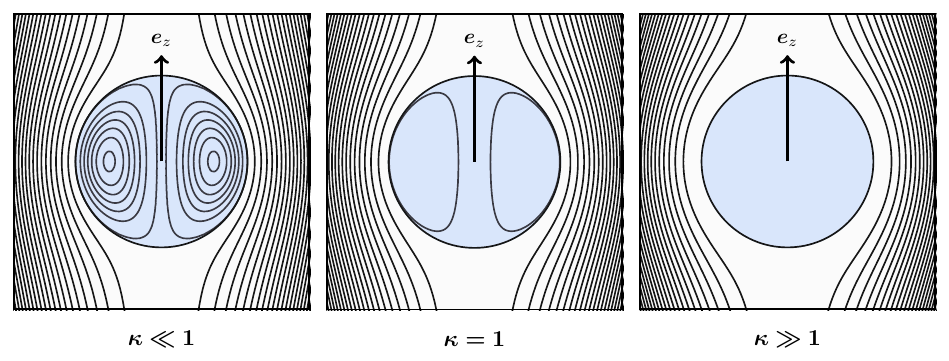}
	\caption{This figure schematically presents snapshots of the stream function, $\psi(r,\theta,t)$, at $t = \pi/(2\omega)$ for different viscosity ratios $\kappa$ between the interior and exterior of the droplet, assuming a density ratio of $\rho_d/\rho_f = 1$.
    For a small viscosity ratio, $\kappa \ll 1$ (here: $\kappa = 10^{-3}$), even a slight stress at the surface induces significant circulation within the droplet.
    When $\kappa = 1$, the strength of the internal circulation is notably reduced.
    For a large viscosity ratio, such as $\kappa = 1000$, the system approaches the behavior of rigid particles, with circulation virtually vanishing.
    The increments of the stream function are the same across all three images, meaning that the reduction in circulation strength is directly due to the varying viscosity ratio.
    A small frequency $\omega$ is used, with the streamlines closely following the steady-state Stokes velocity profile.
	\label{fig:streamfunction}}
\end{figure*}
In Fig.\,\ref{fig:streamfunction},  $\psi(r,\theta,t)$ is visualized in the co-moving frame for three different scenarios characterized by different viscosity ratios $\kappa$. The contour plots provide a visual representation of the stream function within and around the droplet.

For relatively small values of $\kappa$, there is a pronounced circulation flow inside the droplet. This behavior can be explained by Eq.\,\eqref{eq:tangential_BC_vflow}, as small values of $\kappa$ effectively result in stress-free boundary conditions. This limiting case is applicable, for example, to gas bubbles in water, where the viscosity ratio $\mu_\text{air}/\mu_\text{water}\approx 0.01$ is significantly smaller than unity.

For $\kappa=1$, the strength of the circulation inside a droplet decreases noticeably, but remains finite. This scenario arises for liquid droplets surrounded by another immiscible liquid. Mathematically, this is the most complex case, as the analytical expressions do not allow for further simplifications.

In the regime of high viscosity ratios $\kappa$, the circulation within the droplet practically disappears. Considering Eq.\,\eqref{eq:tangential_BC_vflow} for this scenario reveals that due to the high viscosity, even small circulations in the droplet are associated with enormous stresses that cannot be balanced unless the outer tangential flows vanish at the surface. This case corresponds to the non-slip boundary conditions found on the surface of solid particles.

\subsection{Drag force in terms of the stream function \label{sec:force_harm_osz}  }%
During the movement of a droplet relative to the surrounding liquid, pressure and viscous frictional forces act on it, which are determined here using the stream function calculated in the previous section. 

The force $\vb{\text{d}F}$ acting on a surface element d$S$ of a droplet with a normal vector $\vb{n}=\vb{e}_r$ has, according to Eq.\,\eqref{eq:force_general}, the following form in spherical coordinates:
\begin{align}
\vb{\text{d}F} &= 
\qty[\sigma_{rr}\vb e_r+\sigma_{r\theta}\vb e_\theta+\sigma_{r\varphi}\vb e_\varphi]_{r=R_0} \, \text{d}S
=
\qty[\qty(\tau_{rr}-p)\vb e_r+\sigma_{r\theta}\vb e_\theta]_{r=R_0} \, \text{d}S
\label{eq:direction_of_force}
\, .
\end{align}
Herein, $\sigma_{r\varphi}$ vanishes due to symmetry reasons, as mentioned above, and the two other coefficients are \cite{Landau6eng,batchelor1967introduction}:
\begin{align}
\tau_{rr}&=\sigma_{rr}+p
=2\mu_f \partialderivative{v_r}{r}
=2\mu_f \partialderivative{}{r}\qty(\frac{1}{r^2\sin\theta} \partialderivative{\psi}{\theta})\,,
\label{eq:shear_coff_1_def}
\\
\sigma_{r\theta}&=
\mu_f\qty(\frac{1}{r}\partialderivative{v_r}{\theta}+\partialderivative{v_\theta}{r}-\frac{v_\theta}{r})
=
-\frac{\mu_f}{r\sin\theta}\qty(\partialderivative{^2\psi}{r^2}-\frac{2}{r}\partialderivative{\psi}{r} - \frac{\sin\theta}{r^2}\partialderivative{}{\theta}\qty(\frac{1}{\sin\theta}\partialderivative{\psi}{\theta}))
\,.
\label{eq:shear_coff_2_def}
\end{align}
If the pressure $p$ outside the drop is decomposed as in Eq.\,\eqref{eq:split_pressure}, we can use Eq.\,\eqref{eq:static_solution} and integrate the contribution of $p_\text{static}$, which gives the buoyancy force:
\begin{align}
	\vb F_b 
	&=
	-\oiint_{r=R_0}  p_\text{static}  \,\,\vb e_r \, \text{d}S
    =
	-\iiint_{r\leq R_0}  \big(\grad p_\text{static}\big) \text{d}^3r
    =
	\rho_f\iiint_{r\leq R_0}  \grad\qty(\Phi(\vb r,t)+\vb r\cdot \derivative{\vb{\dot s}}{t}) \text{d}^3r
    =
	m_f \vb{\ddot s}+\rho_f\iiint_{r\leq R_0}  \,\grad\Phi(\vb r,t)\, \text{d}^3r
	\label{eq:buoyancy}
	\, .
\end{align}
Here, we have used the divergence theorem and
\begin{align}
      m_f=\frac{4}{3}\pi R_0^3\rho_f
      \, 
 \label{eq:mf_definition}     
\end{align}
is the mass of the liquid displaced by the drop. Equation \eqref{eq:buoyancy} illustrates that when the container is shaken, the buoyancy force, arising from the external potential, is modified by an effective gravitational acceleration induced by the shaking.

The flow resulting from the movement of the droplet causes a dynamic force contribution:
\begin{align}
	\vb F_d
	&=
	\oiint_{r=R_0}\qty[\qty(\tau_{rr}-p_\text{flow})\vb e_r+\sigma_{r\theta}\vb e_\theta]\, \text{d}S
	\label{eq:forceEquation}
	.
\end{align}
Together, the total hydrodynamic force $\vb F$ on the droplet is given by:
\begin{equation}
	\vb F= \vb F_b+\vb F_d
 \label{eq:force_split}
	\, .
\end{equation}
 To calculate the force in Eq.\,\eqref{eq:forceEquation}, the pressure $p_\text{flow}$ caused by the flow is required. It is determined as a function of $\vb v$ by Eq.\,\eqref{eq:exterior_Stokes_raw} as follows:
\begin{align}
\grad\qty(\frac{p_\text{flow}}{\rho_f}-\vb r\cdot\derivative{\vb{v}_\infty}{t})&= \nu_f \nabla^2\vb{v} - \partialderivative{\vb{v}}{t}
\label{eq:stokes_for_pressure}
.
\end{align}
Applying the Grassmann identity and the incompressibility condition (with some intermediate steps omitted), the Laplacian becomes:
\begin{align}
\nabla^2\vb v
&= 
\grad\qty(\div\vb v)-\curl\qty(\curl\vb v)
=
-\curl\qty(\curl\curl\qty(\frac{\psi}{r\sin\theta}\vb e_\varphi))
=-\frac{\vb e_\varphi}{r\sin\theta}\cross \grad\qty(\hat E\psi).
\end{align}
Consequently, Eq.\,\eqref{eq:stokes_for_pressure} can be reformulated, and the following equation can be reformulated for the determination of the flow contribution to the pressure:
\begin{align}
\grad\qty(\frac{p_\text{flow}}{\rho_f}-\vb r\cdot\derivative{\vb{v}_\infty}{t})
 &=
 -\nu_f\frac{\vb e_\varphi}{r\sin\theta}\cross \grad\qty(\hat E\psi)
 -\curl\qty(\frac{1}{r\sin\theta}\partialderivative{\psi}{t}\vb e_\varphi).
\label{eq:pressure_equation}
\end{align}
With the decomposition of the stream function $\psi$ in Eq.\,\eqref{eq:psiharm}, the previous equation can be further rewritten using the expressions in Eq.\,\eqref{eq:stream_function_ansatz} for $\psi_\omega$ and in Eq.\,\eqref{eq:ansatz_for_gf} for $g_f$:
\begin{align}
\label{eq:pflow_1}
\grad\qty(\frac{p_\text{flow}}{\rho_f}-z\derivative{v_\infty}{t})
&=
\grad\qty(
\frac{v_\omega e^{-i\phi-i\omega t}}{2} i\omega
\qty[1-\frac{R_0^3}{r^3}\alpha^f_1]r\cos\theta )
+
\mbox{c.c.}
\end{align}
The first contribution on the right-hand side is caused by $\psi_\omega$, while the second and complex conjugate contribution arises due to $\psi_{-\omega}=\psi_\omega^*$. To find this equation, we have also used the relationship between $\omega$ and $k_f$ given in Eq.\,\eqref{eq:k_definition}.  Equation \eqref{eq:pflow_1} determines the pressure outside a droplet  and we obtain with $z=r\cos\theta$ the flow-induced pressure
\begin{equation}
p_\text{flow}=-\rho_f i\omega \frac{v_\omega e^{-i\phi}}{2} e^{-i\omega t} \frac{R_0^3}{r^3}\alpha^f_1 r\cos\theta 
+ \mbox{c.c.}\,,
\label{eq:pflow_coeff_form}
\end{equation}
whereby the integration constant is absorbed in $p_0$. Again, the first term on the right is caused by $\psi_\omega$, as introduced in Eq.\,\eqref{eq:psiharm}.

Using the definitions of the components $\tau_{rr}$ and $\sigma_{r\theta}$ from Eqs.\,\eqref{eq:shear_coff_1_def} and \eqref{eq:shear_coff_2_def}, these take the following form at the drop surface with $r=R_0$:
\begin{align}
\tau_{rr}(r=R_0)
&=
-\frac{12 \mu_f}{R_0}\qty(\alpha^f_1-\alpha^f_3\qty[\frac{1-ik_f -\frac{1}{3}k_f^2}{k_f}]e^{ik_f})
\frac{v_\omega e^{-i\phi}}{2}
e^{-i\omega t}\cos\theta
+ \mbox{c.c.}\,,
\label{eq:rr_shear_component_coeff_form}
\\
\sigma_{r\theta}(r=R_0)
&=
-\frac{6 \mu_f}{R_0}\qty(\alpha^f_1-\alpha^f_3\qty[\frac{1-ik_f -\frac{1}{2}k_f^2+\frac{1}{6}ik_f^3}{k_f}]e^{ik_f})
\frac{v_\omega e^{-i\phi}}{2}
e^{-i\omega t}\sin\theta
+ \mbox{c.c.}
\label{eq:rtehta_shear_component_coeff_form}
\end{align}
With the expressions in Eq.\,\eqref{eq:pflow_coeff_form}, Eq.\,\eqref{eq:rr_shear_component_coeff_form} and Eq.\,\eqref{eq:rtehta_shear_component_coeff_form} the drag force onto the droplet follows by evaluating the integral in Eq.\,\eqref{eq:forceEquation}. Since $\tau_{rr}, \, \sigma_{r\theta}$ and $p_\text{flow}$ are independent of $\varphi$, the components $x$ and $y$ of $\vb F_d$ vanish and the drag force in the $z$-direction is given by 
\begin{align}
\vb F_d
&= 
2\pi R_0^2\vb e_z\int_0^\pi\qty[\qty(\tau_{rr}-p_\text{flow})\cos\theta-\sigma_{r\theta}\sin\theta]\sin\theta \, \text{d}\theta
= 
3 \pi R_0\mu_f\qty[\frac{2}{9}\alpha^f_1k_f^2  +\frac{4}{9} \alpha^f_3(1-ik_f)k_fe^{ik_f}]
v_\omega e^{-i\omega t-i\phi}\vb e_z
+
\mbox{c.c.}
\end{align}
Inserting the expressions for $\alpha^f_1$ and $\alpha^f_3$ from Eqs.\,\eqref{eq:solutions_alpha}, we obtain after some algebra the compact expression for the drag force:
\begin{align}
\vb F_d
&=
6\pi R_0\mu_f\qty[
\frac{2+3\kappa}{3+3\kappa}-\frac{k_f^2}{9}+\bar G(\omega)]
\frac{v_\omega}{2}e^{-i\omega t-i\phi}
\vb e_z
+
\mbox{c.c.}\,,
\label{eq:fourier_component_force}
\end{align}
with frequency dependent $\bar G(\omega)$ given by
\begin{equation}
	\bar G(\omega)
	=
	\frac{4ik_f h(ik_d)+\kappa\qty[3(1+ik_f)h(ik_d)+\chi -3ik_f(1+\kappa)\chi]}{3i(3i+k_f)h(ik_d)+3\kappa\qty[i(3i+k_f)h(ik_d)+\chi +\kappa\chi]}
	\label{eq:kernel}
	.
\end{equation}
It should be noted that $\vb v_\infty$ in $x$ or in $y$ direction results in the same magnitude of the force in the $x$ or $y$ direction, i.e., the drag force for relative velocities in $x$ and $y$ direction is also determined. Furthermore, 
 Eq.\,\eqref{eq:force_split} takes, by using Eq.\,\eqref{eq:fourier_component_force},
the following form:
\begin{align}
\vb F
&=
\vb F_b+\frac{m_f}{2}\derivative{\vb v_\infty}{t}
+
6\pi R_0 \mu_f v_\omega \bigg[
\text{Im}\qty[\bar G(\omega)]\sin(\omega t+\phi)
+  
\qty(\frac{2+3\kappa}{3+3\kappa}+\text{Re}\qty[\bar G(\omega)])\cos(\omega t+\phi)  
\bigg]\vb e_z
.
\label{eq:harmonic_force}
\end{align}
Although the explicit expressions for the real and imaginary parts of \(\bar G(\omega)\) can be quite long, they can be easily evaluated numerically. 

If $\omega$ approaches zero, the two expressions $k_d$ and $k_f$ disappear. By successive application of the L'Hospitals rule, the limit $\omega \to 0 $ can be evaluated and it gives the Hadamard-Rybczynski formula \cite{Rybczynski:1911,Hadamard:1911,Leal:2007,batchelor1967introduction} for steady droplet motion through viscous liquids with constant velocity ${\vb v}_\infty=\hat {\vb v}_\infty$:
\begin{equation}
\vb F=
\vb F_b
+
6\pi R_0\mu_f\frac{2+3\kappa}{3+3\kappa}
\hat {\vb v}_{\infty}
\label{eq:hadamard_rybczynski}
.
\end{equation}
In an inviscid carrier fluid ($\mu_f = 0$), apart from buoyancy, only the second term in Eq.\,\eqref{eq:harmonic_force} contributes to the total force $\vb F$, representing the added mass effect, which affects the droplet's inertia. 

\subsection{Drag forces for a general time-dependence of $\vb v_\infty(t)$
\label{sec:drag_arbit_t}}
In general, the velocity $\vb v_\infty(t)$ is neither harmonic in time nor aligned along $\vb e_z$ but may be represented by the following Fourier decomposition,
\begin{equation}
\label{eq:vgeninfty}
\vb v_\infty(t)
=
\sum_{j=x,y,z}v_j(t) \vb e_j
=
\sum_{j=x,y,z}\qty(\int\frac{\text d\omega}{\sqrt{2\pi}} \bar v_j(\omega )  e^{-i\omega t}) \vb e_j
\, ,
\end{equation}
with
\begin{equation}
\bar v_j(\omega)=\int\frac{\text dt}{\sqrt{2\pi}} v_j(t)  e^{i\omega t}
\label{eq_velocity_spectrum}
.
\end{equation}
In the low Reynolds number range, the flow velocity $\vb v$ is described by the linear Navier-Stokes equation, Eq.\,\eqref{eq:Stokes_equation}. With a general excitation $\bf v_\infty$ as in Eq.\,\eqref{eq:vgeninfty}, the resulting flow velocity and also the total viscous frictional force $\vb F_d$ are linear superpositions of the contributions from Eq.\,\eqref{eq:fourier_component_force} induced by the different excitation frequencies and all  orientations. Hence, the total drag force acting on the droplet is given by
\begin{align}
\vb F_d
&=
6\pi R_0\mu_f
\sum_{j=x,y,z}\int \frac{\text d\omega}{\sqrt{2\pi}}\qty[
\frac{2+3\kappa}{3+3\kappa}-\frac{k_f^2}{9}+\bar G(\omega)]
\bar v_j e^{-i\omega t}\vb e_j
\nonumber\\
&=
6\pi R_0\mu_f\frac{2+3\kappa}{3+3\kappa}
\vb v_\infty(t)
+
\frac{m_f}{2}\derivative{\vb v_\infty(t)}{t}
-
3 R_0\mu_f
\int \text dt'
\derivative{\vb v_\infty (t')}{t'}
G(t-t')
\, ,
\label{eq:Fddetail}
\end{align}
with the kernel
\begin{align}
G(t)
&=
\int_{-\infty}^\infty\text d\omega
\frac{\bar G(\omega)}{i\omega}
e^{-i\omega t}
\, 
.
\end{align}
However, $\bar G$ in Eq.\,\eqref{eq:kernel} is only valid for $\omega>0$ due to the choice of $\alpha_4^f=0$ in Eq.\,\eqref{eq:ansatz_for_gf}. For $\omega <0$ we receive the complex conjugated expression, i.e., we have the relationship $\bar G(-\omega) = \bar G^*(\omega)$ and therefore
\begin{align}
\label{eq:Ggeneral}
G(t)
&=
-\int_0^\infty \text d\omega\qty(i
\frac{\bar G(\omega)}{\omega}
e^{-i\omega t}
-
i\frac{\bar G(\omega)^*}{\omega}
e^{i\omega t}
)
=
2\,\text{Im}\qty(\int_0^\infty \text d\omega
\frac{\bar G(\omega)}{\omega}
e^{-i\omega t}
)
\, .
\end{align}
The complicated frequency dependence through $k_f$ and $k_d$ still allows a numerical evaluation of $\vb F_d$ but analytical representations of $\vb F_d$ for general $\vb v_\infty(t)$ are only tractable in the regimes of $\kappa\ll 1$ and $\kappa\gg 1$, as described in the following two subsections. 

\subsection{Forces on a particle for $\kappa \gg 1$ \label{sec:arb_motion_large_kappa}}

We first examine the limiting case $\kappa \to \infty$, which enforces rigid boundary conditions on the droplet surface, analogous to those for solid particles in practical applications. In this scenario, the kernel in Eq.\,\eqref{eq:kernel} simplifies to the following form:
\begin{equation}
\bar G(\omega)=-ik_f=-i\sqrt{i\omega \tau_f}.
\label{eq:G_kernel_solid}
\end{equation}
Then, the integral in Eq.\,\eqref{eq:Ggeneral} can be evaluated using contour integration.
Generally speaking, we make the following two substitutions: for $t>0$, we choose $\xi=\sqrt{i\omega}$, and for $t<0$, we use $\xi=\sqrt{-i\omega}$. These substitutions transform the integral into a standard Gaussian integral for easier computation and we obtain the following:
\begin{align}
G(t)
&=
2\sqrt{\tau_f}\, \text{Im}\qty(\int_0^\infty \text d\omega
\frac{e^{-i\omega t}}{\sqrt{i\omega}}
)
=
2\sqrt{\tau_f}
\begin{cases}
\text{Im}\qty(-2i\int_0^\infty \text d\xi e^{-t\xi^2}
)
&t>0
\\ 
\text{Im}\qty(2\int_0^\infty \text d\xi e^{-\abs{t}\xi^2}
)
&t<0
\end{cases}
=
-2\sqrt{\tau_f}
\begin{cases}
\sqrt{\pi/t }
&t>0
\\
0
&t<0
\end{cases}.
\end{align}
As expected, due to causality, the kernel vanishes for $t<0$.

The force acting on such a solid, spherical particle has the following form according to Eq.\,\eqref{eq:force_split} and Eq.\,\eqref{eq:Fddetail}:
\begin{align}
\vb F
&=
\vb F_b
+  6\pi R_0\mu_f
\vb v_\infty
+
\frac{m_f}{2}\derivative{\vb v_\infty}{t}
+
6 R_0^2
\sqrt{\pi\mu_f\rho_f}
\int_{-\infty}^t \text dt'
\qty(\frac{1}{\sqrt{t-t'}}
\derivative{\vb v_\infty (t')}{t'}
\label{eq:force_solid_object}
)
.
\end{align}
This force includes the buoyancy force, the added mass effect, the
instantaneous Stokes drag on the particle, and the classical contribution of the history force \cite{Basset1888,Boussinesq1885}.

\subsection{Forces on a particle for $\kappa\ll 1$ \label{sec:arb_motion_small_kappa}}

As a further limiting case, we consider spherical air bubbles of constant radius moving through fluids with constant ambient pressure.
 In this scenario, the ratio \(\kappa \ll 1\) is much smaller than one, and the Fourier transform of the kernel in Eq.\,\eqref{eq:Ggeneral} simplifies for positive angular frequencies \(\omega > 0\) as follows:
\begin{equation}
\bar G(\omega)
=-\frac{4}{3}\frac{ik_f/3}{1-ik_f/3}
=-\frac{4}{3}\frac{i\sqrt{i\omega \tau_f/9}}{1-i\sqrt{i\omega \tau_f/9}}
.
\end{equation}
Thus, we need to compute according to \eqref{eq:Ggeneral},
the following integral:
\begin{align}
G(t)
&=
\frac{8}{3}\,\text{Im}\qty(\int_0^\infty \text d\omega
\frac{\sqrt{i\omega \tau_f/9}}{i\omega\qty[1-i\sqrt{i\omega \tau_f/9}]}
e^{-i\omega t}
)
=
\frac{8}{3}
\begin{cases}
\text{Im}\qty(2\int_0^\infty 
\frac{\exp\qty(-9\xi^2t/\tau_f)}{i+\xi}\text d\xi)
& t>0\,,
\\
\text{Im}\qty(2\int_0^\infty \frac{\exp\qty(-9 \xi^2\abs{t}/\tau_f)}{1+\xi}\text d\xi)
&t<0\,,
\end{cases}
\end{align}
where we have introduced the substitutions $\xi=\sqrt{i\omega \tau_f/9}$ for $t>0$ and $\xi=-i\sqrt{i\omega \tau_f /9}$ for $t<0$. The integral for $t<0$, again, results in a real value, thus making its imaginary component vanish. For $t>0$ we bring the integral back into a Gaussian form and make the denominator real, obtaining the imaginary part of the integral that is essential for further calculation:
\begin{align}
&\text{Im}\left(\int_0^\infty \frac{\exp\left(-9\xi^2  t/\tau_f\right)}{i+\xi}\text d\xi\right)
=
- \exp\qty(\frac{9 t}{\tau_f})\int_0^\infty \frac{\exp\left(-9 \left(1+\xi^2\right)t/\tau_f\right)}{1+\xi^2}\text d\xi
.
\end{align}
The resulting integral can be solved 
using the following identity for a positive real number $\zeta$:
\begin{align}
\derivative{}{\zeta}
&\int_0^\infty \frac{\exp\qty(-\zeta \qty[1+\xi^2])}{1+\xi^2}\text d\xi
=
-\int_0^\infty \exp\qty(-\zeta \qty[1+\xi^2])\text d\xi
=
-\frac{\sqrt{\pi}}{2}\frac{e^{-\zeta}}{\sqrt{\zeta }}
=
\frac{\pi}{2}\derivative{}{\zeta}\qty[1-\text{Erf}\qty(\sqrt{\zeta})]
.
\end{align}
The integration constant with respect to the $\zeta$-integration in our particular scenario is determined by the fact that the kernel $G(t)$ tends to zero at large $t$, i.e., $G(t\rightarrow \infty)=0$. Finally, we obtain the following:
\begin{equation}
G(t)=
-\frac{8\pi }{3}
\begin{cases}
\exp\qty(\frac{9t}{\tau_f}) \text{Erfc}\qty(\sqrt{\frac{9t}{\tau_f}})
& t>0
\\
0
&t<0
\end{cases}
.
\end{equation}
Therefore, the  force on a very low viscous incompressible liquid droplet is given by:
\begin{align}
&\vb F=
\vb F_b
+
\frac{m_f}{2}\derivative{\vb v_\infty}{t}
+  4\pi R_0\mu_f
\vb v_\infty(t)
+
8\pi R_0\mu_f
\int_{-\infty}^t\text dt'
\qty[
\derivative{\vb v_\infty (t')}{t'}
\exp\qty(\frac{9(t-t')}{\tau_f}) \text{Erfc}\qty(\sqrt{\frac{9(t-t')}{\tau_f}})
]\, .
\nonumber\\
\label{eq:force_bubble}
\end{align}
Similarly to small rigid spheres, this formula includes the buoyancy forces, the added mass effect, and the stationary Stokes drag with a factor '4' instead of '6'.  However, the form of the contribution of the history force in the case of incompressible droplets with very small viscosity, more precisely for vanishing viscosity, differs significantly from the contribution in Eq.\,\eqref{eq:force_solid_object} for solid particles with no-slip boundary conditions. The formulas so far apply to droplets of fixed radius $R_0$, and their extension to  a time-dependent $R(t)$ radius for gas bubbles is discussed in Appendix\,\ref{sec:varying_radius} and in further work.

It is instructive to consider the general expressions for the drag force on a solid particle in Eq.\,\eqref{eq:force_solid_object} and a bubble of constant radius in Eq.\,\eqref{eq:force_bubble} in two simple special cases: In Appendix \ref{sec:comparememory} we study how the flow around the bubble relaxes to the Stokes profile after a sudden onset of a relative velocity. This illustrates the diffusion time scale by Eq.\,\eqref{eq:def_taudff}, which is required until the flow can be considered in steady state. The second example is the harmonic time dependence of $\vb v_\infty(t)$ in Eq.\,\eqref{vinftyomeg}, used to construct the more general expressions for the velocity profile and the drag force. Here, the ratio between the oscillation period and the diffusion timescale is a key parameter for the relevance of the history force. Since this type of periodic motion also occurs in the quantitative description of the horizontal shaking experiment, it is presented in the main text.

\subsection{Relaxation of the force for a rapid acceleration to constant velocity
\label{sec:comparememory}}
%
%
The simplest nontrivial situation considers the force on a particle after the sudden acceleration to a stationary relative velocity. An example of a fast continuous start of fluid motion at $t=0$ is $\vb v_\infty(t)=-\frac{1}{2}\vb v_0[1 + \tanh(t/\tau_{\mbox{on}})]$, where the onset time $\tau_{on}$ should be chosen to be significantly smaller than the viscous diffusion time $\tau_f$.
In this case, we approximate this switch-on process by the Heaveside function,
\begin{equation}
\vb v_\infty (t)
=
\begin{cases}
0 & t < 0 \\
-\vb v_0 & t \geq 0
\end{cases}
\label{eq:illustrative_motion}
,
\end{equation}
and we ignore the range $0<t<\tau_{\mbox{on}}$ in the following. With this choice, we have $\text d\vb v_\infty/\text dt=-\vb v_0\delta(t)$ and the force acting on a non-moving droplet with $\kappa \gg 1$ (limit of a solid particle) for $t>\tau_{\mbox{on}}$ is obtained via Eq.\,\eqref{eq:force_solid_object} as follows:  
\begin{align}
\vb F = \vb F_b - 6\pi R_0 \mu_f \vb v_0 \left[1 + \sqrt{\frac{\tau_f}{\pi t}}\right]\,.
\label{eq:historydecay_rigid}
\end{align}
For droplets with very low viscosity, that is, $\kappa\ll 1$ as for gas bubbles with a constant radius, the relative velocity of Eq.\,\eqref{eq:force_bubble} gives the following time-dependent viscous drag force: 
\begin{align}
\vb F
&=
\vb F_b
-
4\pi R_0\mu_f
\vb v_0\qty[
1
+
2\exp\qty(\frac{9 t}{\tau_f})\text{Erfc}\qty(\frac{9 t}{\tau_f})]
.
\label{eq:historydecay_lowv}
\end{align}
Equations\,\eqref{eq:historydecay_rigid} and \,\eqref{eq:historydecay_lowv} show that after a sudden onset of relative velocity in both cases with $\kappa \gg 1$ and $\kappa \ll 1$, the effective viscous friction force on a droplet greatly exceeds the Stokes friction value due to the BBH. The duration of this aftereffect is determined by the viscous diffusion time \(\tau_f\). If \(t \gg \tau_f \), the force converges to the classical Stokes friction value at constant relative velocity \(\vb{v}_0\). Hence, a change in the velocity of the liquid that occurs over a short period of time \(\tau_{on}\) influences the viscous frictional force on a particle even then, when the relative velocity of the particle does not change anymore. 

A qualitative interpretation of these results is as follows: A stationary droplet in a flow causes a perturbation of the fluid-velocity field, which decreases with the inverse of the distance according to Stokes' law. If the relative velocity between a droplet and the carrier fluid is suddenly switched on as in Eq.\,\eqref{eq:illustrative_motion}, the stationary Stokes velocity profile around the droplet is established only on a time scale $t>\tau_f$. During the transition phase, the fluid parts farther away from the droplet have not yet reached their stationary velocity. This temporarily increases the shear gradients close to the surface of the particle. Consequently, a higher viscous friction force is observed, as described by the formulas in Eqs.\,\eqref{eq:historydecay_rigid} and \eqref{eq:historydecay_lowv} 
until a stationary Stokes profile is obtained.

\section{Forces on gas bubbles with time-dependent radius \label{sec:varying_radius}}

In this final Appendix, we extend the theory from Appendices \ref{sec:basicvel} and \ref{sec:forcedropR0} - which applies to droplets of constant radius in accelerating fluids - to spherical gas bubbles with a time-dependent radius $R(t)$ \cite{Magnaudet:1998.1,Magnaudet:2004.1,Magnaudet:2000.1}. As with the liquid droplets discussed in the main text, the surface tension at the air-liquid interface determines up to what size the gas bubbles remain spherical. Our calculations closely follow those of Ref.\,\cite{Magnaudet:1998.1}.

To determine the viscous friction force acting on a gas bubble, it is sufficient to know the velocity and pressure fields outside the bubble. As before, even for spherical gas bubbles with variable radius $R(t)$, the carrier liquid does not cross the gas-liquid interface and this leads to the following boundary condition for the normal component of the carrier liquid velocity:
\begin{align}
    v_r(r=R(t),\theta,t)&=\derivative{R(t)}{t}
    \label{eq:radial_BC_Rt}
    \,.
\end{align}
As seen in Sec.\,\ref{sec:basicvel} the tangential velocity at the surface of a drop follows from the continuity of the stresses across the drop interface. Because of the negligibly low viscosity of gases, the movement of the gas inside the bubble is insignificant for the flow of the surrounding liquid. Therefore, free-slip boundary conditions apply to the flow of the surrounding fluid on the bubble surface. However, for gas bubbles in a liquid with a high concentration of surfactant, no-slip boundary conditions are a good approximation \cite{Grace_Weber:1978}, corresponding to case $\kappa\gg 1$ in Appendix\,\ref{sec:arb_motion_large_kappa}. In both special cases, the equations for the flow around the bubble are closed, and we can calculate the drag force onto the bubble without resolving its (compressible) interior dynamics.

In the following, we choose a time-dependent bubble radius with the initial condition $R(t=0)=R_0$ and we perform a coordinate transformation as suggested in Ref.\,\cite{Magnaudet:1998.1}. Accordingly, the spatial coordinates are transformed  via the  relation
\begin{equation}
\vb x=(x,y,z) = \gamma(t) (\tilde x, \tilde y, \tilde z)= \gamma(t)\,\tilde {\vb x}
\, ,
\end{equation}
with the time-dependent stretching factor $\gamma(t)=R(t)/R_0$. This gives the relationship
\begin{equation}
\tilde{r}(r, t) = \gamma^{-1}(t)r \quad\Longleftrightarrow\quad r(\tilde{r}, t) = \gamma(t) \tilde{r}
\,
\end{equation}
between the radial coordinates. The polar angle $\theta$ remains unchanged after rescaling of the coordinate axes. In the new coordinate system, the time $\tilde{t}$ is determined as follows:
\begin{align}
\tilde t &= \int_0^t\gamma^{-2}(s)\, ds \quad\Longrightarrow\quad \frac{d\tilde t}{dt} = \gamma^{-2}(t).
\label{eq:time_transf_appendix}
\end{align}
Accordingly,  the velocity vector \(\vb{\tilde v}\) in the new coordinate system is related to the original velocity ${\vb v}$ via
\begin{align}
\vb{v} &= \frac{d\vb{x}}{dt} = \frac{d}{dt}\left(\gamma(\tilde t)\vb{\tilde x}\right) = \gamma_t \vb{\tilde x} + \gamma^{-1}\frac{d\vb{\tilde x}}{d\tilde t} = \gamma_t \vb{\tilde x} + \gamma^{-1}\vb{\tilde v},
\end{align}
where the abbreviation $\gamma_t =\text d\gamma/\text dt$ is used. This equation underscores that the rescaling of time and space coordinates modulates the velocity in the new coordinate system. In particular, it emphasizes that a point at rest in the old coordinate system has a radial velocity in the new coordinates due to the stretching by the time-dependent factor \(\gamma(t)\).

The above transformation also ensures that the boundary of the spherical bubble maintains a constant radius \(\tilde{r} = R_0\), in the transformed system, which results in the following condition for the normal velocity $\tilde{v}_r$ at the bubble's surface (cf. Eq.\,\eqref{eq:radial_BC_Rt}):
\begin{equation}
\tilde v_r(\norm{\vb{\tilde r}}=R_0)=0
\label{eq:transformed_no_flux}
\, .
\end{equation}
Next, we formulate the governing equations for the velocity field outside the droplet in terms of transformed coordinates. To do this, we use the transformation rule for the spatial derivatives:
\begin{align}
\grad 
&\to\,\gamma^{-1}\tilde\grad
\, .
\end{align}
Furthermore, the (convective) time derivative of the velocity (along a Lagrangian trajectory) becomes:
\begin{align}
    \derivative{\vb v}{t}
    &=
    \derivative{}{t}
    \qty( \gamma_t\vb{\tilde x} +\gamma^{-1}\vb{\tilde v})
    =
    \gamma_{tt}\vb{\tilde x}
    +
    \gamma_t \gamma^{-2} \vb{\tilde v}
    +
    \derivative{}{t}
    \qty( \gamma^{-1}\vb{\tilde v})
    =
    \gamma_{tt}\vb{\tilde x}
    +
    \gamma^{-3}\derivative{\vb {\tilde v}}{\tilde t}
    \, ,
\end{align}
with $\gamma_{tt}=\text d^2\gamma/\text dt^2$. In the coordinate system that moves with the droplet, where stretched coordinates are used and a rescaled time, the  Navier-Stokes equation takes the following form:
\begin{align}
\derivative{\vb{\tilde v}}{\tilde t}+\gamma^3\gamma_{tt} \vb{\tilde x}
&=
-\gamma^{2}\tilde \grad\qty(\frac{p}{\rho_f}+\Phi-\vb r\cdot\qty[ \derivative{\vb v_\infty}{t} - \derivative{\vb {\dot s}}{t}]) 
+ \nu_f \tilde \nabla^2\vb{\tilde v}
\label{eq:stokes_on_w}
\,.
\end{align}
Here, $\text d\vb{\tilde v}/\text d\tilde t$ stands for the full convective derivative (including the nonlinearity), and the continuity equation reads
\begin{align}
\tilde \div\vb{\tilde v}=-\frac{3}{R(\tilde t)}\derivative{R}{\tilde t}
\label{eq:continuity_on_w}
.
\end{align}
It is intuitively clear that the pulsating bubble generates a pulsating radial flow. This idea is further supported by the solution to Eq.\,\eqref{eq:stokes_on_w} and Eq.\,\eqref{eq:continuity_on_w} for $\vb v_\infty=\vb s=\Phi=0$, which is:
\begin{align}
\vb{\tilde v}_\text{puls}
&=\qty(\frac{R_0^3}{\tilde r^3}-1)\gamma_t \gamma \vb{\tilde x}\,,
\\
p_\text{puls}
&= \frac{\rho_f R_0^2}{2}\qty[2\frac{R_0}{\tilde r}\gamma_{tt}\gamma - \qty(\frac{R_0^4}{\tilde r^4}-\frac{4R_0}{\tilde r})\gamma_t^2]
.
\end{align}
This equation also satisfies the desired normal boundary conditions of Eq.\,\eqref{eq:transformed_no_flux}. The above solution can be used to remove the additional inertial term in Eq. \eqref{eq:stokes_on_w}, while the potential $\Phi$ and shaking acceleration $\vb{\ddot s}$ can be absorbed into a hydrostatic solution in the form of Eq.\,\eqref{eq:static_solution}. Overall, we therefore  propose the following ansatz for a solution:
\begin{equation}
\vb{\tilde v} = \vb{\tilde w} + \vb{\tilde v}_\text{puls} \quad p = p_\text{static} + p_\text{puls} + p_\text{flow}.
\label{eq:velocity_split}
\end{equation}
By inserting all of this into Eq.\,\eqref{eq:stokes_on_w}, we obtain:
\begin{align}
\partialderivative{\vb{\tilde w}}{\tilde t}
&+
\qty(\vb {\tilde w}\cdot\tilde\grad)\vb {\tilde w}
+
\qty(\vb{\tilde v}_\text{puls}\cdot\tilde\grad)\vb {\tilde w}
+
\qty(\vb {\tilde w}\cdot\tilde\grad)\vb{\tilde v}_\text{puls}
=
-\gamma^{2}\tilde \grad\qty(\frac{p_\text{flow}}{\rho_f}-\vb r\cdot \derivative{\vb v_\infty}{t}) + \nu_f \tilde \nabla^2\vb{\tilde w}
.
\end{align}
As it is mentioned in \cite{Magnaudet:1998.1}, in the limit of small Reynolds number $R\norm{\vb v_\infty}/\nu_f\ll 1$ and when the bubble expands slowly, i.e. $\abs{\dot R}/\norm{\vb v_\infty}\ll 1$, we recover the Stokes limit and find:
\begin{align}
\partialderivative{\vb{\tilde w}}{\tilde t}
=
-\gamma^{2}\tilde \grad\qty(\frac{p_\text{flow}}{\rho_f}-\vb r\cdot \derivative{\vb v_\infty}{t}) + \nu_f \tilde \nabla^2\vb{\tilde w}
\label{eq:stokes_tildew}
.
\end{align}
Transforming back into the original coordinate system, the velocity is,
\begin{equation}
\vb v = \frac{R_0}{R}\vb{\tilde w} + \frac{R^2}{r^2}\derivative{R}{t}\vb e_r
\, ,
\end{equation}
and matching of the far field condition requires:
\begin{align}
\vb v (\norm{\vb r}\gg R(t))
\to
\vb v_\infty 
+\frac{R^2}{r^2}\derivative{R}{t}\vb e_r
\,.
\end{align}
When expressed in terms of $\vb{\tilde v}$, this gives:
\begin{align}
\vb{\tilde v}\qty(\norm{ \vb {\tilde x}}\gg R_0)
\to
\gamma \vb v_\infty
+
\vb{\tilde v}_\text{puls}
\label{eq:far_field_tildev}
\, .
\end{align}
The transformed equations \eqref{eq:transformed_no_flux}, \eqref{eq:continuity_on_w}, \eqref{eq:stokes_tildew}, and \eqref{eq:far_field_tildev} can be summarized, as follows:
\begin{subequations}
	\begin{eqnarray}
	\partialderivative{\vb{\tilde w}}{\tilde t}
	&=&
	-\tilde \grad\qty(\frac{\tilde p_\text{flow}}{\rho_f}-\vb{\tilde r}\cdot \derivative{\qty(\gamma \vb v_\infty)}{\tilde t}) + \nu_f \tilde \nabla^2\vb{\tilde w} \,,\\
		&& \tilde\grad\cdot\vb{\tilde w}=0\,, \\
	&&	\vb{\tilde w}(\norm{\vb {\tilde r}}\gg R_0)
	\to
	\gamma \vb v_\infty\,,\\
	&&\tilde w_r(\norm{\vb {\tilde r}}=R_0)=0.
	\end{eqnarray}
	\label{eq:almost_complete_set}
\end{subequations}
Thereby,
\begin{equation}
p_\text{flow}=\gamma^{-2}\tilde p_\text{flow}-\rho_f\gamma_t\,  \vb{\tilde r}\cdot \vb v_\infty 
\label{eq:newPressure}
.
\end{equation}
In this way, the above equations correspond exactly to the flow around a bubble of constant radius, as discussed earlier, where the free-stream velocity $\vb v_\infty$ is modulated by $\gamma$. 

To obtain the force on the bubble, as derived in Eq.\,\eqref{eq:forceEquation}, we then need to re-express the elements of the stress tensor as:
\begin{align}
\tau_{rr}(r=R(t),\theta,t)
&=2\mu_f\partialderivative{v_r}{r}\Big|_{r=R(t)}
=2\mu_f\gamma^{-1}\partialderivative{}{\tilde r}\qty(\gamma_t \tilde r + \gamma^{-1}\tilde v_r)_{\tilde r=R_0}
=2\mu_f\gamma^{-2}\partialderivative{\tilde w_r}{\tilde r}\Big|_{\tilde r=R_0}
+\text{function}(t)
\, ,
\\
\sigma_{r\theta}(r=R(t),\theta,t)
&=\mu_f r\partialderivative{}{r}\qty(\frac{v_\theta}{r})_{r=R(t)}
=\mu_f\gamma^{-2}\tilde r\partialderivative{}{\tilde r}\qty(\frac{\tilde w_\theta}{\tilde r})_{\tilde r=R_0}
.
\end{align}
During the above calculation, we employed the no-normal-flux boundary condition to simplify the expressions. Notably, the isotropic terms (those without dependence on $\theta$) in the above formula tend to average out when integrated over the bubble's spherically symmetric surface. This is true, especially for $p_\text{puls}$. 
Furthermore, the factor $\gamma^{-2}\text dS=\gamma^{-2}R^2\sin\theta\text d\varphi\text d\theta=R_0^2\sin\theta\text d\varphi\text d\theta=\text d\tilde S$ precisely rescales the integral over $r=R(t)$ to an integral over $\tilde r=R_0$. 
This rescaling turns out to be extremely beneficial as it allows us to express the force acting on the bubble as follows,
\begin{widetext}
\begin{align}
\vb F(\tilde t)
&=-\oiint_{r=R(\tilde t)}\qty[p_\text{static} + p_\text{puls} + p_\text{flow}]\vb e_r \, \text{d}S
+\oiint_{r=R(\tilde t)}\qty[\tau_{rr}\vb e_r+\sigma_{r\theta}\vb e_\theta] \, \text{d}S
\nonumber\\
&=\vb F_b(\tilde t)+\rho_f \gamma_t \gamma^{2} \oiint_{\tilde r=R_0}\qty( \vb {\tilde r}\cdot \vb v_\infty)\vb e_r \, \text{d}\tilde S
+\gamma^{-2}\oiint_{r=R(\tilde t)}\qty[\qty(2\mu_f\partialderivative{\tilde w}{\tilde r}-\tilde p_\text{flow})\vb e_r + \mu_f\tilde r\partialderivative{}{\tilde r}\qty(\frac{\tilde w_\theta}{\tilde r})\vb e_\theta] \, \text{d}S
\nonumber\\
&=\vb F_b(\tilde t)+m_f(\tilde t) \gamma^{-1}\gamma_t\vb v_\infty
+
\oiint_{\tilde r=R_0}\qty[\qty(2\mu_f\partialderivative{\tilde w}{\tilde r}-\tilde p_\text{flow})\vb e_r + \mu_f\tilde r\partialderivative{}{\tilde r}\qty(\frac{\tilde w_\theta}{\tilde r})\vb e_\theta] \, \text{d}\tilde S
,
\end{align}
\end{widetext}
with
\begin{align}
	\vb F_b(\tilde t) 
	&=
	m_f(\tilde t) \vb{\ddot s}+\rho_f\iiint_{r\leq R(\tilde t)}  \,\grad\Phi(\vb r,\tilde t)\, \text{d}^3r
	\, ,
\end{align}
and $m_f(\tilde t)=4\pi R(\tilde t)^3\rho_f/3$. 
Although this expression might seem rather complex at first glance, it significantly simplifies when the system of equations, as detailed in Eqs.\,\eqref{eq:almost_complete_set}, is closed by choosing an appropriate tangential boundary condition. In this scenario, the last integral corresponds exclusively to the hydrodynamic force acting on a droplet with a constant radius but with a modified free-stream velocity $\vb v_\infty \to \gamma \vb v_\infty$ and with \(\vb F_b=0\). The relevant expressions for the force have already been calculated earlier in this work.

For a solid sphere, we can apply Eq.\,\eqref{eq:force_solid_object}, replacing \(\vb v_\infty\) with \(\gamma \vb v_\infty\). in contrast, for an inviscid bubble, we need to refer to the relevant part of Eq.\,\eqref{eq:force_bubble}. These calculations will be elaborated upon in the following two subsections.

\subsection{Forces for free-slip boundary conditions \label{sec:bubble_history_var_R}}
The first special case in which we can easily supplement the missing boundary condition in Eq.\,\eqref{eq:almost_complete_set} is that of an inviscid bubble. We have shown earlier that the relevant expression reads:
\begin{equation}
\partialderivative{}{r}\qty(\frac{v^\theta_\text{flow}}{r})_{r=R(t)}
=
0
\,.
\end{equation}
When expressed in transformed coordinates, in terms of $\vb{\tilde w}$, we find:
\begin{equation}
\partialderivative{}{\tilde r}\qty(\frac{\tilde w_\theta}{\tilde r})_{\tilde r=R_0}=0.
\end{equation}
Therefore, the set of equations \eqref{eq:almost_complete_set} is now complete. It corresponds to the flow around a bubble of constant radius $R_0$ and the expression for the last integral may be taken from Eq.\,\eqref{eq:force_bubble}. We find:
\begin{widetext}
\begin{align}
\vb F(\tilde t)
&=\vb F_b(\tilde t)
+  m_f(\tilde t)\gamma^{-1}\gamma_t \vb v_\infty
+  4\pi R_0 \mu_f \gamma (\tilde t) \vb v_\infty(\tilde t)
+  \frac{2}{3}\pi\rho_fR_0^3\derivative{\qty(\gamma \vb v_\infty)}{\tilde t}
\nonumber\\
&\quad
+
8\pi R_0\mu_f
\int_{-\infty}^{\tilde t}\text d\tilde t'
\qty[
\derivative{\qty(\gamma \vb v_\infty)}{\tilde t'}
\exp\qty(\frac{9\nu_f(\tilde t-\tilde t')}{R_0^2}) \text{Erfc}\qty(\sqrt{\frac{9\nu_f(\tilde t-\tilde t')}{R_0^2}})
]
\nonumber\\
&=\vb F_b(\tilde t)
+  4\pi R(\tilde t) \mu_f \vb v_\infty(\tilde t)
+  \frac{1}{2}\derivative{\qty(m_f \vb v_\infty)}{t}
+
8\pi R_0\mu_f
\int_{-\infty}^{\tilde t}\text d \tilde t'
\qty[
\derivative{\qty(\gamma \vb v_\infty)}{\tilde t'}
\exp\qty(\frac{9\nu_f(\tilde t-\tilde t')}{R_0^2}) \text{Erfc}\qty(\sqrt{\frac{9\nu_f(\tilde t-\tilde t')}{R_0^2}})
]
.
\end{align}
\end{widetext}
The time coordinate within the integral can be transformed back, via Eq.\,\eqref{eq:time_transf_appendix}, by making the substitution:
\begin{equation}
\tilde t'= \int_0^{t'}\gamma^{-2}(s)\text ds
\quad \Longleftrightarrow \quad
\text d\tilde t'
=
\gamma^{-2}(t')\text dt'
\, .
\end{equation}
This yields:
\begin{align}
\vb F(t)
&=
\vb F_b(t)
+4\pi R(t) \mu_f \vb v_\infty(t)
+\frac{1}{2}\derivative{\qty(m_f \vb v_\infty)}{t}
+8\pi\mu_f
\int_{-\infty}^{t}\text dt'\derivative{\qty(R \vb v_\infty)}{t'}\exp\qty(9\nu_f \int_{t'}^tR^{-2}(s)\text ds)
\text{Erfc}\qty(\sqrt{9\nu_f \int_{t'}^tR^{-2}(s)\text ds})
,
\end{align}
which now explicitly contains an additional inertia term (or buoyancy term) already discussed by Magnaudet\,\cite{Magnaudet:1998.1}. 
This additional contribution arises from the acceleration of the fluid far from the bubble, which is balanced by a uniform pressure gradient, and therefore results in a net force on the spherical volume of fluid occupied by the bubble.

\subsection{Forces for no-slip boundary conditions}
Repeating the procedure for no-slip boundary conditions, we start from Eq.\,\eqref{eq:force_solid_object} to obtain:
    \begin{align}
\vb F(t)
&=
\vb F_b(t)
+6\pi R(t) \mu_f \vb v_\infty(t)
+\frac{1}{2}\derivative{\qty(m_f \vb v_\infty)}{t}
+
6
\sqrt{\pi\mu_f\rho_f}
\int_{-\infty}^t \text dt'
\qty(\frac{1}{\sqrt{\int_{t'}^t R^{-2}(s)\text ds}}
\derivative{\qty(R \vb v_\infty)}{t'})
\, .
\end{align}

\nocite{*}

\begin{thebibliography}{69}%
\makeatletter
\providecommand \@ifxundefined [1]{%
 \@ifx{#1\undefined}
}%
\providecommand \@ifnum [1]{%
 \ifnum #1\expandafter \@firstoftwo
 \else \expandafter \@secondoftwo
 \fi
}%
\providecommand \@ifx [1]{%
 \ifx #1\expandafter \@firstoftwo
 \else \expandafter \@secondoftwo
 \fi
}%
\providecommand \natexlab [1]{#1}%
\providecommand \enquote  [1]{``#1''}%
\providecommand \bibnamefont  [1]{#1}%
\providecommand \bibfnamefont [1]{#1}%
\providecommand \citenamefont [1]{#1}%
\providecommand \href@noop [0]{\@secondoftwo}%
\providecommand \href [0]{\begingroup \@sanitize@url \@href}%
\providecommand \@href[1]{\@@startlink{#1}\@@href}%
\providecommand \@@href[1]{\endgroup#1\@@endlink}%
\providecommand \@sanitize@url [0]{\catcode `\\12\catcode `\$12\catcode
  `\&12\catcode `\#12\catcode `\^12\catcode `\_12\catcode `\%12\relax}%
\providecommand \@@startlink[1]{}%
\providecommand \@@endlink[0]{}%
\providecommand \url  [0]{\begingroup\@sanitize@url \@url }%
\providecommand \@url [1]{\endgroup\@href {#1}{\urlprefix }}%
\providecommand \urlprefix  [0]{URL }%
\providecommand \Eprint [0]{\href }%
\providecommand \doibase [0]{https://doi.org/}%
\providecommand \selectlanguage [0]{\@gobble}%
\providecommand \bibinfo  [0]{\@secondoftwo}%
\providecommand \bibfield  [0]{\@secondoftwo}%
\providecommand \translation [1]{[#1]}%
\providecommand \BibitemOpen [0]{}%
\providecommand \bibitemStop [0]{}%
\providecommand \bibitemNoStop [0]{.\EOS\space}%
\providecommand \EOS [0]{\spacefactor3000\relax}%
\providecommand \BibitemShut  [1]{\csname bibitem#1\endcsname}%
\let\auto@bib@innerbib\@empty
\bibitem [{\citenamefont {Sirignano}(2010)}]{Sirignano:2010}%
  \BibitemOpen
  \bibfield  {author} {\bibinfo {author} {\bibfnamefont {W.~A.}\ \bibnamefont
  {Sirignano}},\ }\href@noop {} {\emph {\bibinfo {title} {{Fluid Dynamics and
  Transport of Droplets and Sprays}}}},\ \bibinfo {edition} {2nd}\ ed.\
  (\bibinfo  {publisher} {Cambridge University Press},\ \bibinfo {address}
  {Cambridge},\ \bibinfo {year} {2010})\BibitemShut {NoStop}%
\bibitem [{\citenamefont {Loth}(2023)}]{LothE:2023}%
  \BibitemOpen
  \bibfield  {author} {\bibinfo {author} {\bibfnamefont {E.}~\bibnamefont
  {Loth}},\ }\href@noop {} {\emph {\bibinfo {title} {Fluid Dynamics of
  Particles, Drops, and Bubbles}}}\ (\bibinfo  {publisher} {Cambridge
  University Press},\ \bibinfo {address} {Cambridge},\ \bibinfo {year}
  {2023})\BibitemShut {NoStop}%
\bibitem [{\citenamefont {Michaelides}\ \emph {et~al.}(2023)\citenamefont
  {Michaelides}, \citenamefont {Sommerfeld},\ and\ \citenamefont {van
  Wachem}}]{Michaelides:2023}%
  \BibitemOpen
  \bibfield  {author} {\bibinfo {author} {\bibfnamefont {E.~E.}\ \bibnamefont
  {Michaelides}}, \bibinfo {author} {\bibfnamefont {M.}~\bibnamefont
  {Sommerfeld}},\ and\ \bibinfo {author} {\bibfnamefont {B.}~\bibnamefont {van
  Wachem}},\ }\href@noop {} {\emph {\bibinfo {title} {{Multiphase Flows with
  Droplets and Particles}}}},\ \bibinfo {edition} {3rd}\ ed.\ (\bibinfo
  {publisher} {Taylor \& Francis Inc},\ \bibinfo {address} {New York},\
  \bibinfo {year} {2023})\BibitemShut {NoStop}%
\bibitem [{\citenamefont {Balachanar}\ and\ \citenamefont
  {Eaton}(2010)}]{Balachandar:2010.1}%
  \BibitemOpen
  \bibfield  {author} {\bibinfo {author} {\bibfnamefont {S.}~\bibnamefont
  {Balachanar}}\ and\ \bibinfo {author} {\bibfnamefont {J.~K.}\ \bibnamefont
  {Eaton}},\ }\bibfield  {title} {\bibinfo {title} {Turbulent dispersed
  multiphase flow},\ }\href@noop {} {\bibfield  {journal} {\bibinfo  {journal}
  {Annu. Rev. Fluid Mech.}\ }\textbf {\bibinfo {volume} {42}},\ \bibinfo
  {pages} {111} (\bibinfo {year} {2010})}\BibitemShut {NoStop}%
\bibitem [{\citenamefont {Brandt}\ and\ \citenamefont
  {Coletti}(2022)}]{BrandtL:2022.1}%
  \BibitemOpen
  \bibfield  {author} {\bibinfo {author} {\bibfnamefont {L.}~\bibnamefont
  {Brandt}}\ and\ \bibinfo {author} {\bibfnamefont {F.}~\bibnamefont
  {Coletti}},\ }\bibfield  {title} {\bibinfo {title} {Particle-laden
  turbulence: {Progress} and perspectives},\ }\href@noop {} {\bibfield
  {journal} {\bibinfo  {journal} {Annu. Rev. Fluid Mech.}\ }\textbf {\bibinfo
  {volume} {54}},\ \bibinfo {pages} {159} (\bibinfo {year} {2022})}\BibitemShut
  {NoStop}%
\bibitem [{\citenamefont {Maxey}(2017)}]{Maxey:2017.1}%
  \BibitemOpen
  \bibfield  {author} {\bibinfo {author} {\bibfnamefont {M.}~\bibnamefont
  {Maxey}},\ }\bibfield  {title} {\bibinfo {title} {Simulation methods for
  particulate flows and concentrated suspensions},\ }\href@noop {} {\bibfield
  {journal} {\bibinfo  {journal} {Annu. Rev. Fluid Mech.}\ }\textbf {\bibinfo
  {volume} {49}},\ \bibinfo {pages} {171} (\bibinfo {year} {2017})}\BibitemShut
  {NoStop}%
\bibitem [{\citenamefont {Elghobashi}(2019)}]{Elghobashi:2019.1}%
  \BibitemOpen
  \bibfield  {author} {\bibinfo {author} {\bibfnamefont {S.}~\bibnamefont
  {Elghobashi}},\ }\bibfield  {title} {\bibinfo {title} {Direct numerical
  simulation of turbulent flows laden with droplets or bubbles},\ }\href@noop
  {} {\bibfield  {journal} {\bibinfo  {journal} {Annu. Rev. Fluid Mech.}\
  }\textbf {\bibinfo {volume} {51}},\ \bibinfo {pages} {217} (\bibinfo {year}
  {2019})}\BibitemShut {NoStop}%
\bibitem [{\citenamefont {Mathai}\ \emph {et~al.}(2020)\citenamefont {Mathai},
  \citenamefont {Lohse},\ and\ \citenamefont {Sun}}]{Mathai:2020.1}%
  \BibitemOpen
  \bibfield  {author} {\bibinfo {author} {\bibfnamefont {V.}~\bibnamefont
  {Mathai}}, \bibinfo {author} {\bibfnamefont {D.}~\bibnamefont {Lohse}},\ and\
  \bibinfo {author} {\bibfnamefont {C.}~\bibnamefont {Sun}},\ }\bibfield
  {title} {\bibinfo {title} {Bubbly and buyoant particle-laden turbulent
  flows},\ }\href@noop {} {\bibfield  {journal} {\bibinfo  {journal} {Annu.
  Rev. Condens. Matter Phys.}\ }\textbf {\bibinfo {volume} {19}},\ \bibinfo
  {pages} {529} (\bibinfo {year} {2020})}\BibitemShut {NoStop}%
\bibitem [{\citenamefont {Bergougnoux}\ \emph {et~al.}(2014)\citenamefont
  {Bergougnoux}, \citenamefont {Bouchet}, \citenamefont {Lopez},\ and\
  \citenamefont {Guazzelli}}]{Bergougnoux2014}%
  \BibitemOpen
  \bibfield  {author} {\bibinfo {author} {\bibfnamefont {L.}~\bibnamefont
  {Bergougnoux}}, \bibinfo {author} {\bibfnamefont {G.}~\bibnamefont
  {Bouchet}}, \bibinfo {author} {\bibfnamefont {D.}~\bibnamefont {Lopez}},\
  and\ \bibinfo {author} {\bibfnamefont {E.}~\bibnamefont {Guazzelli}},\
  }\bibfield  {title} {\bibinfo {title} {The motion of solid spherical
  particles falling in a cellular flow field at low {Stokes} number},\
  }\href@noop {} {\bibfield  {journal} {\bibinfo  {journal} {Phys. Fluids}\
  }\textbf {\bibinfo {volume} {26}},\ \bibinfo {pages} {093302} (\bibinfo
  {year} {2014})}\BibitemShut {NoStop}%
\bibitem [{\citenamefont {Ling}\ \emph {et~al.}(2013)\citenamefont {Ling},
  \citenamefont {Paramar},\ and\ \citenamefont
  {Balachandar}}]{Balachandar:2013.1}%
  \BibitemOpen
  \bibfield  {author} {\bibinfo {author} {\bibfnamefont {Y.}~\bibnamefont
  {Ling}}, \bibinfo {author} {\bibfnamefont {M.}~\bibnamefont {Paramar}},\ and\
  \bibinfo {author} {\bibfnamefont {S.}~\bibnamefont {Balachandar}},\
  }\bibfield  {title} {\bibinfo {title} {A scaling analysis of added-mass and
  history forces and their coupling in dispersed multiphase flows},\
  }\href@noop {} {\bibfield  {journal} {\bibinfo  {journal} {Int. J. Multiphase
  Flow}\ }\textbf {\bibinfo {volume} {57}},\ \bibinfo {pages} {102} (\bibinfo
  {year} {2013})}\BibitemShut {NoStop}%
\bibitem [{\citenamefont {Daitche}(2015)}]{Daitche:2015.1}%
  \BibitemOpen
  \bibfield  {author} {\bibinfo {author} {\bibfnamefont {A.}~\bibnamefont
  {Daitche}},\ }\bibfield  {title} {\bibinfo {title} {{On the role of the
  history force for inertial particles in turbulence}},\ }\href@noop {}
  {\bibfield  {journal} {\bibinfo  {journal} {J. Fluid Mech.}\ }\textbf
  {\bibinfo {volume} {782}},\ \bibinfo {pages} {567} (\bibinfo {year}
  {2015})}\BibitemShut {NoStop}%
\bibitem [{\citenamefont {Li}\ \emph {et~al.}(2023)\citenamefont {Li},
  \citenamefont {Bragg},\ and\ \citenamefont {Katul}}]{LiBraggKatul2023}%
  \BibitemOpen
  \bibfield  {author} {\bibinfo {author} {\bibfnamefont {S.}~\bibnamefont
  {Li}}, \bibinfo {author} {\bibfnamefont {A.~D.}\ \bibnamefont {Bragg}},\ and\
  \bibinfo {author} {\bibfnamefont {G.}~\bibnamefont {Katul}},\ }\bibfield
  {title} {\bibinfo {title} {Reduced sediment settling in turbulent flows due
  to {Basset} history and virtual mass effects},\ }\href@noop {} {\bibfield
  {journal} {\bibinfo  {journal} {Geophys. Res. Lett.}\ }\textbf {\bibinfo
  {volume} {50}},\ \bibinfo {pages} {e2023GL105810} (\bibinfo {year}
  {2023})}\BibitemShut {NoStop}%
\bibitem [{\citenamefont {Basset}(1888)}]{Basset1888}%
  \BibitemOpen
  \bibfield  {author} {\bibinfo {author} {\bibfnamefont {A.~B.}\ \bibnamefont
  {Basset}},\ }\bibfield  {title} {\bibinfo {title} {On the motion of a sphere
  in a viscous liquid},\ }\href@noop {} {\bibfield  {journal} {\bibinfo
  {journal} {Philos. Trans. R. Soc. London. Ser. A}\ }\textbf {\bibinfo
  {volume} {179}},\ \bibinfo {pages} {43} (\bibinfo {year} {1888})}\BibitemShut
  {NoStop}%
\bibitem [{\citenamefont {Boussinesq}(1885)}]{Boussinesq1885}%
  \BibitemOpen
  \bibfield  {author} {\bibinfo {author} {\bibfnamefont {J.}~\bibnamefont
  {Boussinesq}},\ }\bibfield  {title} {\bibinfo {title} {Sur la résistance
  qu'oppose un liquide indéfini en repos, sans pesanteur, au mouvement varié
  d'une sphère solide qu'il mouille sur toute sa surface, quand les vitesses
  restent bien continues et assez faibles pour que leurs carrés et produits
  soient négligeables},\ }\href@noop {} {\bibfield  {journal} {\bibinfo
  {journal} {Comptes Rendus de l'Académie des Sciences, Paris}\ }\textbf
  {\bibinfo {volume} {100}},\ \bibinfo {pages} {935} (\bibinfo {year}
  {1885})}\BibitemShut {NoStop}%
\bibitem [{\citenamefont {Jaganathan}\ \emph {et~al.}(2023)\citenamefont
  {Jaganathan}, \citenamefont {Prasath}, \citenamefont {Govindarajan},\ and\
  \citenamefont {Vasan}}]{Vasan:2023.1}%
  \BibitemOpen
  \bibfield  {author} {\bibinfo {author} {\bibfnamefont {D.}~\bibnamefont
  {Jaganathan}}, \bibinfo {author} {\bibfnamefont {S.~G.}\ \bibnamefont
  {Prasath}}, \bibinfo {author} {\bibfnamefont {R.}~\bibnamefont
  {Govindarajan}},\ and\ \bibinfo {author} {\bibfnamefont {V.}~\bibnamefont
  {Vasan}},\ }\bibfield  {title} {\bibinfo {title} {The {Basset-Boussinesq}
  history force: its neglect, validity and recent numerical developments},\
  }\href@noop {} {\bibfield  {journal} {\bibinfo  {journal} {Front. Phys.}\
  }\textbf {\bibinfo {volume} {11}},\ \bibinfo {pages} {1167338} (\bibinfo
  {year} {2023})}\BibitemShut {NoStop}%
\bibitem [{\citenamefont {Haller}(2019)}]{Haller:2019.1}%
  \BibitemOpen
  \bibfield  {author} {\bibinfo {author} {\bibfnamefont {G.}~\bibnamefont
  {Haller}},\ }\bibfield  {title} {\bibinfo {title} {{Solving the inertial
  particle equation with memory}},\ }\href@noop {} {\bibfield  {journal}
  {\bibinfo  {journal} {J. Fluid Mech.}\ }\textbf {\bibinfo {volume} {874}},\
  \bibinfo {pages} {1} (\bibinfo {year} {2019})}\BibitemShut {NoStop}%
\bibitem [{\citenamefont {Toegel}\ \emph {et~al.}(2006)\citenamefont {Toegel},
  \citenamefont {Luther},\ and\ \citenamefont {Lohse}}]{Lohse:2006.1}%
  \BibitemOpen
  \bibfield  {author} {\bibinfo {author} {\bibfnamefont {R.}~\bibnamefont
  {Toegel}}, \bibinfo {author} {\bibfnamefont {S.}~\bibnamefont {Luther}},\
  and\ \bibinfo {author} {\bibfnamefont {D.}~\bibnamefont {Lohse}},\ }\bibfield
   {title} {\bibinfo {title} {Viscosity destabilizes sonoluminescing bubbles},\
  }\href@noop {} {\bibfield  {journal} {\bibinfo  {journal} {Phys. Rev. Lett.}\
  }\textbf {\bibinfo {volume} {96}},\ \bibinfo {pages} {114301} (\bibinfo
  {year} {2006})}\BibitemShut {NoStop}%
\bibitem [{\citenamefont {Garbin}\ \emph {et~al.}(2009)\citenamefont {Garbin},
  \citenamefont {Dollet}, \citenamefont {Overvelde}, \citenamefont {Cojoc},
  \citenamefont {Fabrizio}, \citenamefont {van Wingaarden}, \citenamefont
  {Prosperetti}, \citenamefont {de~Jong},\ and\ \citenamefont
  {Lohse}}]{Lohse:2009.1}%
  \BibitemOpen
  \bibfield  {author} {\bibinfo {author} {\bibfnamefont {V.}~\bibnamefont
  {Garbin}}, \bibinfo {author} {\bibfnamefont {B.}~\bibnamefont {Dollet}},
  \bibinfo {author} {\bibfnamefont {M.}~\bibnamefont {Overvelde}}, \bibinfo
  {author} {\bibfnamefont {D.}~\bibnamefont {Cojoc}}, \bibinfo {author}
  {\bibfnamefont {E.~D.}\ \bibnamefont {Fabrizio}}, \bibinfo {author}
  {\bibfnamefont {L.}~\bibnamefont {van Wingaarden}}, \bibinfo {author}
  {\bibfnamefont {A.}~\bibnamefont {Prosperetti}}, \bibinfo {author}
  {\bibfnamefont {N.}~\bibnamefont {de~Jong}},\ and\ \bibinfo {author}
  {\bibfnamefont {D.}~\bibnamefont {Lohse}},\ }\bibfield  {title} {\bibinfo
  {title} {History force on coated microbubbles propelled by ultrasound},\
  }\href@noop {} {\bibfield  {journal} {\bibinfo  {journal} {Phys. Fluids}\
  }\textbf {\bibinfo {volume} {21}},\ \bibinfo {pages} {092003} (\bibinfo
  {year} {2009})}\BibitemShut {NoStop}%
\bibitem [{\citenamefont {Wan}\ and\ \citenamefont
  {Xu}(2022)}]{Xu_Haitao:2022.1}%
  \BibitemOpen
  \bibfield  {author} {\bibinfo {author} {\bibfnamefont {D.}~\bibnamefont
  {Wan}}\ and\ \bibinfo {author} {\bibfnamefont {H.}~\bibnamefont {Xu}},\
  }\bibfield  {title} {\bibinfo {title} {Experimental study on the motion of a
  spherical particle in a plane sound wave},\ }\href@noop {} {\bibfield
  {journal} {\bibinfo  {journal} {Acta Mech. Sin.}\ }\textbf {\bibinfo {volume}
  {38}},\ \bibinfo {pages} {721493} (\bibinfo {year} {2022})}\BibitemShut
  {NoStop}%
\bibitem [{\citenamefont {Abbad}\ and\ \citenamefont
  {Souhar}(2004{\natexlab{a}})}]{Abbad:2004.1}%
  \BibitemOpen
  \bibfield  {author} {\bibinfo {author} {\bibfnamefont {M.}~\bibnamefont
  {Abbad}}\ and\ \bibinfo {author} {\bibfnamefont {M.}~\bibnamefont {Souhar}},\
  }\bibfield  {title} {\bibinfo {title} {Effects of the history force on an
  oscillating rigid sphere at low {Reynolds} number},\ }\href@noop {}
  {\bibfield  {journal} {\bibinfo  {journal} {Exp. Fluids}\ }\textbf {\bibinfo
  {volume} {36}},\ \bibinfo {pages} {775} (\bibinfo {year}
  {2004}{\natexlab{a}})}\BibitemShut {NoStop}%
\bibitem [{\citenamefont {Abbad}\ and\ \citenamefont
  {Souhar}(2004{\natexlab{b}})}]{Abbad:2004.2}%
  \BibitemOpen
  \bibfield  {author} {\bibinfo {author} {\bibfnamefont {M.}~\bibnamefont
  {Abbad}}\ and\ \bibinfo {author} {\bibfnamefont {M.}~\bibnamefont {Souhar}},\
  }\bibfield  {title} {\bibinfo {title} {Experimental investigation on the
  history force acting on oscillating fluid spheres at low {Reynolds} number},\
  }\href {https://doi.org/10.1063/1.1779051} {\bibfield  {journal} {\bibinfo
  {journal} {Phys. Fluids}\ }\textbf {\bibinfo {volume} {16}},\ \bibinfo
  {pages} {3808} (\bibinfo {year} {2004}{\natexlab{b}})}\BibitemShut {NoStop}%
\bibitem [{\citenamefont {Daitche}\ and\ \citenamefont
  {T\'el}(2011)}]{Daitche:2011.1}%
  \BibitemOpen
  \bibfield  {author} {\bibinfo {author} {\bibfnamefont {A.}~\bibnamefont
  {Daitche}}\ and\ \bibinfo {author} {\bibfnamefont {T.}~\bibnamefont
  {T\'el}},\ }\bibfield  {title} {\bibinfo {title} {{Memory effects are
  relevant for chaotic advection of inertial particles}},\ }\href@noop {}
  {\bibfield  {journal} {\bibinfo  {journal} {Phys. Rev. Lett.}\ }\textbf
  {\bibinfo {volume} {107}},\ \bibinfo {pages} {244501} (\bibinfo {year}
  {2011})}\BibitemShut {NoStop}%
\bibitem [{\citenamefont {Bagheri}\ and\ \citenamefont
  {Sabzpooshani}(2020)}]{Bagheri:2020.1}%
  \BibitemOpen
  \bibfield  {author} {\bibinfo {author} {\bibfnamefont {M.}~\bibnamefont
  {Bagheri}}\ and\ \bibinfo {author} {\bibfnamefont {M.}~\bibnamefont
  {Sabzpooshani}},\ }\bibfield  {title} {\bibinfo {title} {{On the Importance
  of the history force in disperion of particles in von K\'arm\'an vortex
  street}},\ }\href@noop {} {\bibfield  {journal} {\bibinfo  {journal} {Adv.
  Powder Technol.}\ }\textbf {\bibinfo {volume} {31}},\ \bibinfo {pages} {3897}
  (\bibinfo {year} {2020})}\BibitemShut {NoStop}%
\bibitem [{\citenamefont {Olivieri}\ \emph {et~al.}(2014)\citenamefont
  {Olivieri}, \citenamefont {Picano}, \citenamefont {Sardina}, \citenamefont
  {Iudicone},\ and\ \citenamefont {Brandt}}]{BrandtL:2014.1}%
  \BibitemOpen
  \bibfield  {author} {\bibinfo {author} {\bibfnamefont {S.}~\bibnamefont
  {Olivieri}}, \bibinfo {author} {\bibfnamefont {F.}~\bibnamefont {Picano}},
  \bibinfo {author} {\bibfnamefont {G.}~\bibnamefont {Sardina}}, \bibinfo
  {author} {\bibfnamefont {D.}~\bibnamefont {Iudicone}},\ and\ \bibinfo
  {author} {\bibfnamefont {L.}~\bibnamefont {Brandt}},\ }\bibfield  {title}
  {\bibinfo {title} {{The effect of the Basset history force on particle
  clustering in homogeneous istropic turbulence}},\ }\href@noop {} {\bibfield
  {journal} {\bibinfo  {journal} {Phys. Fluids}\ }\textbf {\bibinfo {volume}
  {26}},\ \bibinfo {pages} {041704} (\bibinfo {year} {2014})}\BibitemShut
  {NoStop}%
\bibitem [{\citenamefont {Wang}\ and\ \citenamefont
  {Maxey}(1993)}]{MaxeyWang:1993.1}%
  \BibitemOpen
  \bibfield  {author} {\bibinfo {author} {\bibfnamefont {L.-P.}\ \bibnamefont
  {Wang}}\ and\ \bibinfo {author} {\bibfnamefont {M.~R.}\ \bibnamefont
  {Maxey}},\ }\bibfield  {title} {\bibinfo {title} {{Settling velocity and
  concentration distribution of heavy particles in homogeneous isotropic
  turbulence}},\ }\href@noop {} {\bibfield  {journal} {\bibinfo  {journal} {J.
  Fluid Mech.}\ }\textbf {\bibinfo {volume} {256}},\ \bibinfo {pages} {27}
  (\bibinfo {year} {1993})}\BibitemShut {NoStop}%
\bibitem [{\citenamefont {van Hinsberg}\ \emph {et~al.}(2017)\citenamefont {van
  Hinsberg}, \citenamefont {Clercx},\ and\ \citenamefont
  {Toschi}}]{Toschi:2017.1}%
  \BibitemOpen
  \bibfield  {author} {\bibinfo {author} {\bibfnamefont {M.~A.~T.}\
  \bibnamefont {van Hinsberg}}, \bibinfo {author} {\bibfnamefont {H.~J.~H.}\
  \bibnamefont {Clercx}},\ and\ \bibinfo {author} {\bibfnamefont
  {F.}~\bibnamefont {Toschi}},\ }\bibfield  {title} {\bibinfo {title} {{Enhance
  settling of nonheavy inertial particles in homogeneous isotropic turbulence:
  The role of the pressure gradient and the Basset history force}},\
  }\href@noop {} {\bibfield  {journal} {\bibinfo  {journal} {Phys. Rev. E}\
  }\textbf {\bibinfo {volume} {95}},\ \bibinfo {pages} {023106} (\bibinfo
  {year} {2017})}\BibitemShut {NoStop}%
\bibitem [{\citenamefont {Hill}(2005)}]{HillR:2005.1}%
  \BibitemOpen
  \bibfield  {author} {\bibinfo {author} {\bibfnamefont {R.}~\bibnamefont
  {Hill}},\ }\bibfield  {title} {\bibinfo {title} {Geometric collision rates
  and trajectories of cloud droplets falling into a {Burgers} vortex},\
  }\href@noop {} {\bibfield  {journal} {\bibinfo  {journal} {Phys. Fluids}\
  }\textbf {\bibinfo {volume} {17}},\ \bibinfo {pages} {037103} (\bibinfo
  {year} {2005})}\BibitemShut {NoStop}%
\bibitem [{\citenamefont {Prasath}\ \emph {et~al.}(2019)\citenamefont
  {Prasath}, \citenamefont {Vasan},\ and\ \citenamefont
  {Govindarajan}}]{Prasath:2019.1}%
  \BibitemOpen
  \bibfield  {author} {\bibinfo {author} {\bibfnamefont {S.~G.}\ \bibnamefont
  {Prasath}}, \bibinfo {author} {\bibfnamefont {V.}~\bibnamefont {Vasan}},\
  and\ \bibinfo {author} {\bibfnamefont {R.}~\bibnamefont {Govindarajan}},\
  }\bibfield  {title} {\bibinfo {title} {Accurate solution method for the
  {Maxey-Riley equation, and the effects of Basset} history},\ }\href@noop {}
  {\bibfield  {journal} {\bibinfo  {journal} {J. Fluid Mech.}\ }\textbf
  {\bibinfo {volume} {868}},\ \bibinfo {pages} {428} (\bibinfo {year}
  {2019})}\BibitemShut {NoStop}%
\bibitem [{\citenamefont {Odar}\ and\ \citenamefont
  {Hamilton}(1964)}]{Odar:1964.1}%
  \BibitemOpen
  \bibfield  {author} {\bibinfo {author} {\bibfnamefont {R.}~\bibnamefont
  {Odar}}\ and\ \bibinfo {author} {\bibfnamefont {W.~S.}\ \bibnamefont
  {Hamilton}},\ }\bibfield  {title} {\bibinfo {title} {Forces on a sphere
  accelerating in a viscous fluid},\ }\href@noop {} {\bibfield  {journal}
  {\bibinfo  {journal} {J. Fluid Mech.}\ }\textbf {\bibinfo {volume} {18}},\
  \bibinfo {pages} {302} (\bibinfo {year} {1964})}\BibitemShut {NoStop}%
\bibitem [{\citenamefont {Mei}\ and\ \citenamefont
  {Adrian}(1992)}]{MeiR:1992.1}%
  \BibitemOpen
  \bibfield  {author} {\bibinfo {author} {\bibfnamefont {R.}~\bibnamefont
  {Mei}}\ and\ \bibinfo {author} {\bibfnamefont {R.}~\bibnamefont {Adrian}},\
  }\bibfield  {title} {\bibinfo {title} {Flow past a sphere with an oscillation
  in the free-stream velocity and unsteady drag at finite {Reynolds} number},\
  }\href@noop {} {\bibfield  {journal} {\bibinfo  {journal} {J. Fluid Mech.}\
  }\textbf {\bibinfo {volume} {256}},\ \bibinfo {pages} {323} (\bibinfo {year}
  {1992})}\BibitemShut {NoStop}%
\bibitem [{\citenamefont {Lovalenti}\ and\ \citenamefont
  {Brady}(1993)}]{Brady:1993.2}%
  \BibitemOpen
  \bibfield  {author} {\bibinfo {author} {\bibfnamefont {P.~M.}\ \bibnamefont
  {Lovalenti}}\ and\ \bibinfo {author} {\bibfnamefont {J.~F.}\ \bibnamefont
  {Brady}},\ }\bibfield  {title} {\bibinfo {title} {The force on a sphere in a
  uniform flow with small-amplitude oscillations at finite {Reynolds number}},\
  }\href@noop {} {\bibfield  {journal} {\bibinfo  {journal} {J. Fluid Mech.}\
  }\textbf {\bibinfo {volume} {256}},\ \bibinfo {pages} {607} (\bibinfo {year}
  {1993})}\BibitemShut {NoStop}%
\bibitem [{\citenamefont {Mei}(1994)}]{MeiR:1994.1}%
  \BibitemOpen
  \bibfield  {author} {\bibinfo {author} {\bibfnamefont {R.}~\bibnamefont
  {Mei}},\ }\bibfield  {title} {\bibinfo {title} {Flow due to an oscillating
  sphere and and an expression for unsteady drag on the sphere at finite
  {Reynolds} number},\ }\href@noop {} {\bibfield  {journal} {\bibinfo
  {journal} {J. Fluid Mech.}\ }\textbf {\bibinfo {volume} {270}},\ \bibinfo
  {pages} {133} (\bibinfo {year} {1994})}\BibitemShut {NoStop}%
\bibitem [{\citenamefont {Chang}\ and\ \citenamefont
  {Maxey}(1994)}]{Maxey:1994.1}%
  \BibitemOpen
  \bibfield  {author} {\bibinfo {author} {\bibfnamefont {E.~J.}\ \bibnamefont
  {Chang}}\ and\ \bibinfo {author} {\bibfnamefont {M.~R.}\ \bibnamefont
  {Maxey}},\ }\bibfield  {title} {\bibinfo {title} {{Unsteady flow about a
  sphere at low to moderate Reynolds number. Part 1. Oscillatory motion }},\
  }\href@noop {} {\bibfield  {journal} {\bibinfo  {journal} {J. Fluid Mech.}\
  }\textbf {\bibinfo {volume} {277}},\ \bibinfo {pages} {347} (\bibinfo {year}
  {1994})}\BibitemShut {NoStop}%
\bibitem [{\citenamefont {Coimbra}\ \emph {et~al.}(2004)\citenamefont
  {Coimbra}, \citenamefont {L'Esperance}, \citenamefont {Lambert},
  \citenamefont {Trollinger},\ and\ \citenamefont {Rangel}}]{Coimbra:2004.1}%
  \BibitemOpen
  \bibfield  {author} {\bibinfo {author} {\bibfnamefont {J.~F.~M.}\
  \bibnamefont {Coimbra}}, \bibinfo {author} {\bibfnamefont {D.~L.}\
  \bibnamefont {L'Esperance}}, \bibinfo {author} {\bibfnamefont {R.~A.}\
  \bibnamefont {Lambert}}, \bibinfo {author} {\bibfnamefont {J.~D.}\
  \bibnamefont {Trollinger}},\ and\ \bibinfo {author} {\bibfnamefont {R.~H.}\
  \bibnamefont {Rangel}},\ }\bibfield  {title} {\bibinfo {title} {{An
  experimental study on stationary history effects in high-frequency Stokes
  flows}},\ }\href@noop {} {\bibfield  {journal} {\bibinfo  {journal} {J. Fluid
  Mech.}\ }\textbf {\bibinfo {volume} {504}},\ \bibinfo {pages} {252} (\bibinfo
  {year} {2004})}\BibitemShut {NoStop}%
\bibitem [{\citenamefont {Lampertz}(2012)}]{Lampertz:2012.1}%
  \BibitemOpen
  \bibfield  {author} {\bibinfo {author} {\bibfnamefont {S.}~\bibnamefont
  {Lampertz}},\ }\href@noop {} {\bibinfo {title} {Experimental measurement of
  the history forces on a rigid sphere in unsteady motion}},\ \bibinfo
  {howpublished} {Master Thesis, University of Göttingen} (\bibinfo {year}
  {2012})\BibitemShut {NoStop}%
\bibitem [{\citenamefont {AlAli}\ \emph {et~al.}(2024)\citenamefont {AlAli},
  \citenamefont {Traver},\ and\ \citenamefont {Coimbra}}]{Coimbra:2024.1}%
  \BibitemOpen
  \bibfield  {author} {\bibinfo {author} {\bibfnamefont {O.}~\bibnamefont
  {AlAli}}, \bibinfo {author} {\bibfnamefont {B.}~\bibnamefont {Traver}},\ and\
  \bibinfo {author} {\bibfnamefont {J.~F.~M.}\ \bibnamefont {Coimbra}},\
  }\bibfield  {title} {\bibinfo {title} {{Particle response to oscillatory
  flows at finite Reynolds numbers}},\ }\href@noop {} {\bibfield  {journal}
  {\bibinfo  {journal} {Phys. Fluids}\ }\textbf {\bibinfo {volume} {36}},\
  \bibinfo {pages} {103367} (\bibinfo {year} {2024})}\BibitemShut {NoStop}%
\bibitem [{\citenamefont {Gode}\ \emph {et~al.}(2023)\citenamefont {Gode},
  \citenamefont {Charton}, \citenamefont {Climent},\ and\ \citenamefont
  {Legendre}}]{LegendreD:2023.1}%
  \BibitemOpen
  \bibfield  {author} {\bibinfo {author} {\bibfnamefont {H.}~\bibnamefont
  {Gode}}, \bibinfo {author} {\bibfnamefont {S.}~\bibnamefont {Charton}},
  \bibinfo {author} {\bibfnamefont {E.}~\bibnamefont {Climent}},\ and\ \bibinfo
  {author} {\bibfnamefont {D.}~\bibnamefont {Legendre}},\ }\bibfield  {title}
  {\bibinfo {title} {{Basset-Boussinesq history force acting on a drop in an
  oscillatory flow}},\ }\href@noop {} {\bibfield  {journal} {\bibinfo
  {journal} {Phys. Rev. Fluids}\ }\textbf {\bibinfo {volume} {8}},\ \bibinfo
  {pages} {073605} (\bibinfo {year} {2023})}\BibitemShut {NoStop}%
\bibitem [{\citenamefont {Velazquez}\ and\ \citenamefont
  {Barrero-Gil}(2024)}]{Barrero-Gil:2024.1}%
  \BibitemOpen
  \bibfield  {author} {\bibinfo {author} {\bibfnamefont {A.}~\bibnamefont
  {Velazquez}}\ and\ \bibinfo {author} {\bibfnamefont {A.}~\bibnamefont
  {Barrero-Gil}},\ }\bibfield  {title} {\bibinfo {title} {{Simplified dynamics
  model of a sphere decelerating freely in a fluid}},\ }\href@noop {}
  {\bibfield  {journal} {\bibinfo  {journal} {Phys. Fluids}\ }\textbf {\bibinfo
  {volume} {36}},\ \bibinfo {pages} {023104} (\bibinfo {year}
  {2024})}\BibitemShut {NoStop}%
\bibitem [{\citenamefont {Landau}\ and\ \citenamefont
  {Lifshitz}(1987)}]{Landau6eng}%
  \BibitemOpen
  \bibfield  {author} {\bibinfo {author} {\bibfnamefont {L.~D.}\ \bibnamefont
  {Landau}}\ and\ \bibinfo {author} {\bibfnamefont {E.~M.}\ \bibnamefont
  {Lifshitz}},\ }\href@noop {} {\emph {\bibinfo {title} {Course of Theoretical
  Physics, Vol. 6 Fluid Mechanics}}}\ (\bibinfo  {publisher} {Butterworth},\
  \bibinfo {address} {Boston},\ \bibinfo {year} {1987})\BibitemShut {NoStop}%
\bibitem [{\citenamefont {Batchelor}(1967)}]{batchelor1967introduction}%
  \BibitemOpen
  \bibfield  {author} {\bibinfo {author} {\bibfnamefont {G.~K.}\ \bibnamefont
  {Batchelor}},\ }\href@noop {} {\emph {\bibinfo {title} {An Introduction to
  Fluid Dynamics}}}\ (\bibinfo  {publisher} {Cambridge University Press},\
  \bibinfo {address} {Cambridge, UK},\ \bibinfo {year} {1967})\BibitemShut
  {NoStop}%
\bibitem [{\citenamefont {Lamb}(1993)}]{lamb1993hydrodynamics}%
  \BibitemOpen
  \bibfield  {author} {\bibinfo {author} {\bibfnamefont {H.}~\bibnamefont
  {Lamb}},\ }\href@noop {} {\emph {\bibinfo {title} {Hydrodynamics}}},\
  \bibinfo {edition} {6th}\ ed.\ (\bibinfo  {publisher} {Cambridge University
  Press},\ \bibinfo {year} {1993})\BibitemShut {NoStop}%
\bibitem [{\citenamefont {Kim}\ and\ \citenamefont
  {Karrila}(1991)}]{kim1991microhydrodynamics}%
  \BibitemOpen
  \bibfield  {author} {\bibinfo {author} {\bibfnamefont {S.}~\bibnamefont
  {Kim}}\ and\ \bibinfo {author} {\bibfnamefont {S.~J.}\ \bibnamefont
  {Karrila}},\ }\href@noop {} {\emph {\bibinfo {title} {Microhydrodynamics
  Principles and Selected Applications}}}\ (\bibinfo  {publisher}
  {Butterworth-Heinemann},\ \bibinfo {address} {Boston},\ \bibinfo {year}
  {1991})\BibitemShut {NoStop}%
\bibitem [{\citenamefont {Gorodtsov}(1976)}]{gorodtsov1976slow}%
  \BibitemOpen
  \bibfield  {author} {\bibinfo {author} {\bibfnamefont {V.~A.}\ \bibnamefont
  {Gorodtsov}},\ }\bibfield  {title} {\bibinfo {title} {Slow motions of a
  liquid drop in a viscous liquid},\ }\href@noop {} {\bibfield  {journal}
  {\bibinfo  {journal} {J. Appl. Mech. Tech. Phys.}\ }\textbf {\bibinfo
  {volume} {16}},\ \bibinfo {pages} {865} (\bibinfo {year} {1976})}\BibitemShut
  {NoStop}%
\bibitem [{\citenamefont {Yang}\ and\ \citenamefont
  {Leal}(1991)}]{yang1991note}%
  \BibitemOpen
  \bibfield  {author} {\bibinfo {author} {\bibfnamefont {S.~M.}\ \bibnamefont
  {Yang}}\ and\ \bibinfo {author} {\bibfnamefont {L.~G.}\ \bibnamefont
  {Leal}},\ }\bibfield  {title} {\bibinfo {title} {A note on memory-integral
  contributions to the force on an accelerating spherical drop at low
  {Reynolds} number},\ }\href@noop {} {\bibfield  {journal} {\bibinfo
  {journal} {Phys. Fluids A}\ }\textbf {\bibinfo {volume} {3}},\ \bibinfo
  {pages} {1822} (\bibinfo {year} {1991})}\BibitemShut {NoStop}%
\bibitem [{\citenamefont {Galindo}\ and\ \citenamefont
  {Gerbeth}(1993)}]{galindo1993note}%
  \BibitemOpen
  \bibfield  {author} {\bibinfo {author} {\bibfnamefont {V.}~\bibnamefont
  {Galindo}}\ and\ \bibinfo {author} {\bibfnamefont {G.}~\bibnamefont
  {Gerbeth}},\ }\bibfield  {title} {\bibinfo {title} {A note on the force on an
  accelerating spherical drop at {low-Reynolds} number},\ }\href@noop {}
  {\bibfield  {journal} {\bibinfo  {journal} {Phys. Fluids}\ }\textbf {\bibinfo
  {volume} {5}},\ \bibinfo {pages} {3290} (\bibinfo {year} {1993})}\BibitemShut
  {NoStop}%
\bibitem [{\citenamefont {Magnaudet}\ and\ \citenamefont
  {Legendre}(1998)}]{Magnaudet:1998.1}%
  \BibitemOpen
  \bibfield  {author} {\bibinfo {author} {\bibfnamefont {J.}~\bibnamefont
  {Magnaudet}}\ and\ \bibinfo {author} {\bibfnamefont {D.}~\bibnamefont
  {Legendre}},\ }\bibfield  {title} {\bibinfo {title} {The viscous drag force
  on a spherical bubble with a time-dependent radius},\ }\href@noop {}
  {\bibfield  {journal} {\bibinfo  {journal} {Phys. Fluids}\ }\textbf {\bibinfo
  {volume} {10}},\ \bibinfo {pages} {550} (\bibinfo {year} {1998})}\BibitemShut
  {NoStop}%
\bibitem [{\citenamefont {Maxey}\ and\ \citenamefont
  {Riley}(1983)}]{MaxeyRiley:1983.1}%
  \BibitemOpen
  \bibfield  {author} {\bibinfo {author} {\bibfnamefont {M.~R.}\ \bibnamefont
  {Maxey}}\ and\ \bibinfo {author} {\bibfnamefont {J.~J.}\ \bibnamefont
  {Riley}},\ }\bibfield  {title} {\bibinfo {title} {Equation of motion for a
  small rigid sphere in a nonuniform flow},\ }\href@noop {} {\bibfield
  {journal} {\bibinfo  {journal} {Phys. Fluids}\ }\textbf {\bibinfo {volume}
  {26}},\ \bibinfo {pages} {883} (\bibinfo {year} {1983})}\BibitemShut
  {NoStop}%
\bibitem [{\citenamefont {Leal}(2007)}]{Leal:2007}%
  \BibitemOpen
  \bibfield  {author} {\bibinfo {author} {\bibfnamefont {L.~G.}\ \bibnamefont
  {Leal}},\ }\href@noop {} {\emph {\bibinfo {title} {Advanced Transport
  Phenomena}}}\ (\bibinfo  {publisher} {Cambridge University Press},\ \bibinfo
  {address} {Cambridge},\ \bibinfo {year} {2007})\BibitemShut {NoStop}%
\bibitem [{\citenamefont {Jaroslawski}\ \emph {et~al.}(2025)\citenamefont
  {Jaroslawski}, \citenamefont {Jaganathan}, \citenamefont {Govindarajan},\
  and\ \citenamefont {McKeon}}]{Jaroslawski2025_StokesianSettling}%
  \BibitemOpen
  \bibfield  {author} {\bibinfo {author} {\bibfnamefont {T.}~\bibnamefont
  {Jaroslawski}}, \bibinfo {author} {\bibfnamefont {D.}~\bibnamefont
  {Jaganathan}}, \bibinfo {author} {\bibfnamefont {R.}~\bibnamefont
  {Govindarajan}},\ and\ \bibinfo {author} {\bibfnamefont {J.}~\bibnamefont
  {McKeon}},\ }\bibfield  {title} {\bibinfo {title} {Stokesian settling from
  quiescence: Experiments and theory on history effects and unsteady flow
  structures},\ }\href {https://doi.org/10.1103/PhysRevFluids.10.L062301}
  {\bibfield  {journal} {\bibinfo  {journal} {Phys. Rev. Fluids}\ }\textbf
  {\bibinfo {volume} {10}},\ \bibinfo {pages} {L062301} (\bibinfo {year}
  {2025})}\BibitemShut {NoStop}%
\bibitem [{\citenamefont {Pumir}\ and\ \citenamefont
  {Wilkinson}(2016)}]{pumir_wilkinson}%
  \BibitemOpen
  \bibfield  {author} {\bibinfo {author} {\bibfnamefont {A.}~\bibnamefont
  {Pumir}}\ and\ \bibinfo {author} {\bibfnamefont {M.}~\bibnamefont
  {Wilkinson}},\ }\bibfield  {title} {\bibinfo {title} {Collisional aggregation
  due to turbulence},\ }\href@noop {} {\bibfield  {journal} {\bibinfo
  {journal} {Annu. Rev. Condens. Matter Phys.}\ }\textbf {\bibinfo {volume}
  {7}} (\bibinfo {year} {2016})}\BibitemShut {NoStop}%
\bibitem [{\citenamefont {Bec}\ \emph {et~al.}(2024)\citenamefont {Bec},
  \citenamefont {Gustavsson},\ and\ \citenamefont
  {Mehlig}}]{bec_gustavsson_mehlig}%
  \BibitemOpen
  \bibfield  {author} {\bibinfo {author} {\bibfnamefont {J.}~\bibnamefont
  {Bec}}, \bibinfo {author} {\bibfnamefont {K.}~\bibnamefont {Gustavsson}},\
  and\ \bibinfo {author} {\bibfnamefont {B.}~\bibnamefont {Mehlig}},\
  }\bibfield  {title} {\bibinfo {title} {Statistical models for the dynamics of
  heavy particles in turbulence},\ }\href@noop {} {\bibfield  {journal}
  {\bibinfo  {journal} {Annu. Rev. Fluid Mech.}\ }\textbf {\bibinfo {volume}
  {56}} (\bibinfo {year} {2024})}\BibitemShut {NoStop}%
\bibitem [{\citenamefont {Shaw}(2003)}]{Shaw2003}%
  \BibitemOpen
  \bibfield  {author} {\bibinfo {author} {\bibfnamefont {R.~A.}\ \bibnamefont
  {Shaw}},\ }\bibfield  {title} {\bibinfo {title} {Turbulence and cloud
  microphysics},\ }\href
  {https://doi.org/10.1146/annurev.fluid.35.101101.161125} {\bibfield
  {journal} {\bibinfo  {journal} {Annu. Rev. Fluid Mech.}\ }\textbf {\bibinfo
  {volume} {35}},\ \bibinfo {pages} {183} (\bibinfo {year} {2003})}\BibitemShut
  {NoStop}%
\bibitem [{\citenamefont {Coimbra}\ and\ \citenamefont
  {Rangel}(2001)}]{Coimbra:2001.1}%
  \BibitemOpen
  \bibfield  {author} {\bibinfo {author} {\bibfnamefont {J.~F.~M.}\
  \bibnamefont {Coimbra}}\ and\ \bibinfo {author} {\bibfnamefont {R.~H.}\
  \bibnamefont {Rangel}},\ }\bibfield  {title} {\bibinfo {title} {Spherical
  particle motion in harmonic {Stokes} flow},\ }\href@noop {} {\bibfield
  {journal} {\bibinfo  {journal} {AIAA J.}\ }\textbf {\bibinfo {volume} {39}},\
  \bibinfo {pages} {1673} (\bibinfo {year} {2001})}\BibitemShut {NoStop}%
\bibitem [{\citenamefont {Grace}\ and\ \citenamefont
  {Weber}(1978)}]{Grace_Weber:1978}%
  \BibitemOpen
  \bibfield  {author} {\bibinfo {author} {\bibfnamefont {J.~R.}\ \bibnamefont
  {Grace}}\ and\ \bibinfo {author} {\bibfnamefont {M.~E.}\ \bibnamefont
  {Weber}},\ }\bibinfo {title} {{Bubbles, Drops, and Particles}}\ (\bibinfo
  {publisher} {Academic Press},\ \bibinfo {address} {New York},\ \bibinfo
  {year} {1978})\ Chap.~\bibinfo {chapter} {3}\BibitemShut {NoStop}%
\bibitem [{\citenamefont {Buchanan}\ \emph {et~al.}(1962)\citenamefont
  {Buchanan}, \citenamefont {Jameson},\ and\ \citenamefont
  {Oedjoe}}]{Buchanan1962}%
  \BibitemOpen
  \bibfield  {author} {\bibinfo {author} {\bibfnamefont {R.~H.}\ \bibnamefont
  {Buchanan}}, \bibinfo {author} {\bibfnamefont {G.}~\bibnamefont {Jameson}},\
  and\ \bibinfo {author} {\bibfnamefont {D.}~\bibnamefont {Oedjoe}},\
  }\bibfield  {title} {\bibinfo {title} {Cyclic migration of bubbles in
  vertically vibrating liquid columns},\ }\href@noop {} {\bibfield  {journal}
  {\bibinfo  {journal} {Ind. Eng. Chem. Fundam.}\ }\textbf {\bibinfo {volume}
  {1}},\ \bibinfo {pages} {82} (\bibinfo {year} {1962})}\BibitemShut {NoStop}%
\bibitem [{\citenamefont {Jameson}(1966)}]{JAMESON196635}%
  \BibitemOpen
  \bibfield  {author} {\bibinfo {author} {\bibfnamefont {G.}~\bibnamefont
  {Jameson}},\ }\bibfield  {title} {\bibinfo {title} {The motion of a bubble in
  a vertically oscillating viscous liquid},\ }\href@noop {} {\bibfield
  {journal} {\bibinfo  {journal} {Chem. Eng. Sci.}\ }\textbf {\bibinfo {volume}
  {21}},\ \bibinfo {pages} {35} (\bibinfo {year} {1966})}\BibitemShut {NoStop}%
\bibitem [{\citenamefont {Ellenberger}\ and\ \citenamefont
  {Krishna}(2007)}]{Ellenberger:2007.1}%
  \BibitemOpen
  \bibfield  {author} {\bibinfo {author} {\bibfnamefont {J.}~\bibnamefont
  {Ellenberger}}\ and\ \bibinfo {author} {\bibfnamefont {R.}~\bibnamefont
  {Krishna}},\ }\bibfield  {title} {\bibinfo {title} {Levitation of air bubbles
  in liquid under low frequency vibration excitement},\ }\href@noop {}
  {\bibfield  {journal} {\bibinfo  {journal} {Chem. Eng. Sci.}\ }\textbf
  {\bibinfo {volume} {62}},\ \bibinfo {pages} {5669} (\bibinfo {year}
  {2007})}\BibitemShut {NoStop}%
\bibitem [{\citenamefont {Sorokin}\ \emph {et~al.}(2012)\citenamefont
  {Sorokin}, \citenamefont {Blekhman},\ and\ \citenamefont
  {Vasilkow}}]{Sorokin:2012.1}%
  \BibitemOpen
  \bibfield  {author} {\bibinfo {author} {\bibfnamefont {V.~S.}\ \bibnamefont
  {Sorokin}}, \bibinfo {author} {\bibfnamefont {I.~I.}\ \bibnamefont
  {Blekhman}},\ and\ \bibinfo {author} {\bibfnamefont {V.~B.}\ \bibnamefont
  {Vasilkow}},\ }\bibfield  {title} {\bibinfo {title} {Motion of a gas bubble
  in fluid under vibration},\ }\href@noop {} {\bibfield  {journal} {\bibinfo
  {journal} {Nonlinear Dyn.}\ }\textbf {\bibinfo {volume} {67}},\ \bibinfo
  {pages} {147} (\bibinfo {year} {2012})}\BibitemShut {NoStop}%
\bibitem [{\citenamefont {Elbing}\ \emph {et~al.}(2016)\citenamefont {Elbing},
  \citenamefont {Still},\ and\ \citenamefont {Ghajar}}]{Elbing:2016.1}%
  \BibitemOpen
  \bibfield  {author} {\bibinfo {author} {\bibfnamefont {B.~R.}\ \bibnamefont
  {Elbing}}, \bibinfo {author} {\bibfnamefont {A.~L.}\ \bibnamefont {Still}},\
  and\ \bibinfo {author} {\bibfnamefont {A.~J.}\ \bibnamefont {Ghajar}},\
  }\href@noop {} {\bibfield  {journal} {\bibinfo  {journal} {Ind. Eng. Chem.
  Res.}\ }\textbf {\bibinfo {volume} {55}},\ \bibinfo {pages} {385} (\bibinfo
  {year} {2016})}\BibitemShut {NoStop}%
\bibitem [{\citenamefont {Berlemont}\ \emph {et~al.}(1990)\citenamefont
  {Berlemont}, \citenamefont {Desjonqu{\`e}res},\ and\ \citenamefont
  {Gouesbet}}]{Berlemont:1990.1}%
  \BibitemOpen
  \bibfield  {author} {\bibinfo {author} {\bibfnamefont {A.}~\bibnamefont
  {Berlemont}}, \bibinfo {author} {\bibfnamefont {P.}~\bibnamefont
  {Desjonqu{\`e}res}},\ and\ \bibinfo {author} {\bibfnamefont {G.}~\bibnamefont
  {Gouesbet}},\ }\bibfield  {title} {\bibinfo {title} {Particle {Lagrangian}
  simulation in turbulent flows},\ }\href@noop {} {\bibfield  {journal}
  {\bibinfo  {journal} {Int. J. Multiphase Flow}\ }\textbf {\bibinfo {volume}
  {16}},\ \bibinfo {pages} {19} (\bibinfo {year} {1990})}\BibitemShut {NoStop}%
\bibitem [{\citenamefont {Kim}\ \emph {et~al.}(1998)\citenamefont {Kim},
  \citenamefont {Elghobashi},\ and\ \citenamefont
  {Sirignano}}]{Sirignano:1998.1}%
  \BibitemOpen
  \bibfield  {author} {\bibinfo {author} {\bibfnamefont {I.}~\bibnamefont
  {Kim}}, \bibinfo {author} {\bibfnamefont {S.}~\bibnamefont {Elghobashi}},\
  and\ \bibinfo {author} {\bibfnamefont {W.~A.}\ \bibnamefont {Sirignano}},\
  }\bibfield  {title} {\bibinfo {title} {{On the equation for
  spherical-particle motion: effect of Reynolds and acceleration numbers}},\
  }\href@noop {} {\bibfield  {journal} {\bibinfo  {journal} {J. Fluid Mech.}\
  }\textbf {\bibinfo {volume} {367}},\ \bibinfo {pages} {221} (\bibinfo {year}
  {1998})}\BibitemShut {NoStop}%
\bibitem [{\citenamefont {Gladkov}(2022)}]{Gladkov2022}%
  \BibitemOpen
  \bibfield  {author} {\bibinfo {author} {\bibfnamefont {S.~O.}\ \bibnamefont
  {Gladkov}},\ }\bibfield  {title} {\bibinfo {title} {Stokes’ force for a
  stationary rotating sphere},\ }\href@noop {} {\bibfield  {journal} {\bibinfo
  {journal} {Russ. Phys. J.}\ }\textbf {\bibinfo {volume} {65}},\ \bibinfo
  {pages} {856} (\bibinfo {year} {2022})}\BibitemShut {NoStop}%
\bibitem [{\citenamefont {Ovseenko}\ and\ \citenamefont
  {Ovseenko}(1968)}]{Ovseenko1968}%
  \BibitemOpen
  \bibfield  {author} {\bibinfo {author} {\bibfnamefont {R.~I.}\ \bibnamefont
  {Ovseenko}}\ and\ \bibinfo {author} {\bibfnamefont {Y.~G.}\ \bibnamefont
  {Ovseenko}},\ }\bibfield  {title} {\bibinfo {title} {Drag of a rotating
  sphere},\ }\href@noop {} {\bibfield  {journal} {\bibinfo  {journal} {Fluid
  Dynamics}\ }\textbf {\bibinfo {volume} {3}},\ \bibinfo {pages} {523}
  (\bibinfo {year} {1968})}\BibitemShut {NoStop}%
\bibitem [{\citenamefont {Stone}(1994)}]{stone1994dynamics}%
  \BibitemOpen
  \bibfield  {author} {\bibinfo {author} {\bibfnamefont {H.~A.}\ \bibnamefont
  {Stone}},\ }\bibfield  {title} {\bibinfo {title} {Dynamics of drop
  deformation and breakup in viscous fluids},\ }\href
  {https://doi.org/10.1146/annurev.fluid.26.1.65} {\bibfield  {journal}
  {\bibinfo  {journal} {Annu. Rev. Fluid Mech.}\ }\textbf {\bibinfo {volume}
  {26}},\ \bibinfo {pages} {65} (\bibinfo {year} {1994})}\BibitemShut {NoStop}%
\bibitem [{\citenamefont {Russel}\ \emph {et~al.}(1989)\citenamefont {Russel},
  \citenamefont {Saville},\ and\ \citenamefont {Showalter}}]{Russel:89}%
  \BibitemOpen
  \bibfield  {author} {\bibinfo {author} {\bibfnamefont {W.~B.}\ \bibnamefont
  {Russel}}, \bibinfo {author} {\bibfnamefont {D.~A.}\ \bibnamefont
  {Saville}},\ and\ \bibinfo {author} {\bibfnamefont {W.~R.}\ \bibnamefont
  {Showalter}},\ }\href@noop {} {\emph {\bibinfo {title} {Colloidal
  Dispersions}}}\ (\bibinfo  {publisher} {Cambridge Univ. Press},\ \bibinfo
  {address} {Cambridge},\ \bibinfo {year} {1989})\BibitemShut {NoStop}%
\bibitem [{\citenamefont {Rybczynski}(1911)}]{Rybczynski:1911}%
  \BibitemOpen
  \bibfield  {author} {\bibinfo {author} {\bibfnamefont {W.}~\bibnamefont
  {Rybczynski}},\ }\bibfield  {title} {\bibinfo {title} {On the translatory
  motion of a fluid sphere in a viscous medium},\ }\href@noop {} {\bibfield
  {journal} {\bibinfo  {journal} {Bull. Acad. Sci. Cracow, Series A}\ }\textbf
  {\bibinfo {volume} {40}},\ \bibinfo {pages} {33} (\bibinfo {year}
  {1911})}\BibitemShut {NoStop}%
\bibitem [{\citenamefont {Hadamard}(1911)}]{Hadamard:1911}%
  \BibitemOpen
  \bibfield  {author} {\bibinfo {author} {\bibfnamefont {J.}~\bibnamefont
  {Hadamard}},\ }\bibfield  {title} {\bibinfo {title} {Mouvement permanent lent
  d'une sphère liquide et visqueuse dans un liquide visqueux},\ }\href@noop {}
  {\bibfield  {journal} {\bibinfo  {journal} {C. R. Acad. Sci. Paris}\ }\textbf
  {\bibinfo {volume} {152}},\ \bibinfo {pages} {1735} (\bibinfo {year}
  {1911})}\BibitemShut {NoStop}%
\bibitem [{\citenamefont {Takemura}\ and\ \citenamefont
  {Magnaudet}(2004)}]{Magnaudet:2004.1}%
  \BibitemOpen
  \bibfield  {author} {\bibinfo {author} {\bibfnamefont {F.}~\bibnamefont
  {Takemura}}\ and\ \bibinfo {author} {\bibfnamefont {J.}~\bibnamefont
  {Magnaudet}},\ }\bibfield  {title} {\bibinfo {title} {{The history force on a
  rapidely shrinking bubble rising at finite Reynolds number}},\ }\href@noop {}
  {\bibfield  {journal} {\bibinfo  {journal} {Phys. Fluids}\ }\textbf {\bibinfo
  {volume} {16}},\ \bibinfo {pages} {3247} (\bibinfo {year}
  {2004})}\BibitemShut {NoStop}%
\bibitem [{\citenamefont {Magnaudet}\ and\ \citenamefont
  {Eames}(2000)}]{Magnaudet:2000.1}%
  \BibitemOpen
  \bibfield  {author} {\bibinfo {author} {\bibfnamefont {J.}~\bibnamefont
  {Magnaudet}}\ and\ \bibinfo {author} {\bibfnamefont {I.}~\bibnamefont
  {Eames}},\ }\bibfield  {title} {\bibinfo {title} {The motion of
  {high-Reynolds-number} bubbles in inhomogeneous flows},\ }\href@noop {}
  {\bibfield  {journal} {\bibinfo  {journal} {Annu. Rev. Fluid Mech.}\ }\textbf
  {\bibinfo {volume} {32}},\ \bibinfo {pages} {659} (\bibinfo {year}
  {2000})}\BibitemShut {NoStop}%
\end{thebibliography}
 
ontrol: page (0) single
\providecommand{\noopsort}[1]{}\providecommand{\singleletter}[1]{#1}%
\end{document}